\documentclass[12 pt,a4paper, oneside, openany, notitlepage]{article}
\usepackage[inner=0.7in,outer=0.7in,bottom=1in,top=1in]{geometry}
\usepackage{setspace}
\usepackage{lmodern}
\usepackage{subcaption}
\usepackage{graphicx}
\usepackage[usenames,dvipsnames]{xcolor}
\usepackage{soul}
\usepackage{amsmath, amssymb, graphics}
\usepackage{mathtools}
\newcommand{\mathsym}[1]{{}}
\newcommand{\unicode}[1]{{}}

\usepackage{amsthm}
\usepackage{amsfonts}
\usepackage{multicol}
\usepackage{ulem}
\usepackage{multirow}
\usepackage{makecell}
\usepackage{float}
\usepackage{array, makecell} 
\newcommand{\PreserveBackslash}[1]{\let\temp=\\#1\let\\=\temp}
\newcolumntype{C}[1]{>{\PreserveBackslash\centering}p{#1}}
\usepackage{longtable}
\usepackage{tabu}
\usepackage{threeparttable}
\usepackage{threeparttablex}
\usepackage{booktabs}
\usepackage[nouppercase]{frontespizio}
\usepackage{epsfig}
\usepackage{geometry}
\usepackage{diagbox}
\usepackage{morefloats}
\usepackage{bbm}
\usepackage{changepage}
\usepackage{enumerate}
\usepackage[shortlabels]{enumitem}
\usepackage{xparse}
\usepackage{bigints}
\usepackage[font={footnotesize}]{caption}
\usepackage{chngcntr}
\usepackage{comment}
\usepackage[multiple]{footmisc}
\usepackage{xr}
\usepackage{mathrsfs}

\usepackage[gen]{eurosym}
\usepackage[T1]{fontenc}
\usepackage{pdflscape}
   
\usepackage[nolists,tablesfirst,nomarkers]{endfloat}
 \usepackage{fancyhdr}
\usepackage{float}
\restylefloat{table}
\restylefloat{figure}
\usepackage{changepage}
 \usepackage{bigints}
\DeclareDelayedFloatFlavour{ltable}{table}

 \newtheoremstyle{mystyle1}% name
  {\topsep}% Space above
  {\topsep}% Space below
  {\normalfont}% Body font
  {}% Indent amount
  {\bfseries \color{black}}% Theorem head font
  {.}%Punctuation after theorem head
  {.5em}%Space after theorem head
  {}% theorem head spec
  
\theoremstyle{mystyle1}

\newtheorem{theorem}{Theorem} 
\newenvironment{thm}{\begin{theorem}%
  \pushQED{\qed}}%
  {\popQED\end{theorem}}
  
\newtheorem{proposition}{Proposition}
\newenvironment{prop}{\begin{proposition}%
 \pushQED{\qed}}%
  {\popQED\end{proposition}}
  
  \newtheorem{lemma}{Lemma}
  {\popQED\end{lemma}}
  
    \newtheorem{definition}{Definition}
  {\popQED\end{definition}}
  
    \newtheorem{corollary}{Corollary}
  {\popQED\end{corollary}}
   
 \newtheoremstyle{mystyle}% name
  {\topsep}% Space above
  {\topsep}% Space below
  {\normalfont}% Body font
  {}% Indent amount
  {\bfseries \color{black}}% Theorem head font
  {.}%Punctuation after theorem head
  {.5em}%Space after theorem head
  {}% theorem head spec

\theoremstyle{mystyle}

  \newtheorem{ex}{Example}
\newenvironment{example}
  {\pushQED{\qed}\ex}
  {\popQED\endex}
 
 \renewenvironment{abstract}
 {\small
  \begin{center}
  \bfseries \abstractname\vspace{-.5em}\vspace{0pt}
  \end{center}
  \list{}{
    \setlength{\leftmargin}{.1cm}%
    \setlength{\rightmargin}{\leftmargin}%
  }%
  \item\relax}
 {\endlist}

\usepackage[colorlinks = true,
            allcolors  = black,  
            linkcolor=black,
            urlcolor=black
            ]{hyperref}           
            
\begin{document}
\title{{\scshape Identification in discrete choice models with imperfect information\thanks{\scriptsize We would like to thank Christian Bontemps, Andrew Caplin, Emmanuel Guerre, Christian Hellwig, Vishal Kamat,  Niccol\'o Lomys,  Thierry Magnac, and Lorenzo Magnolfi for their useful comments. We would also like to thank seminar participants at TSE,  UCL, Sciences Po Paris, Tinbergen Institute Amsterdam, University of Cambridge, University of Essex, Queen Mary University of London, Aarhus University, and UC Berkeley. We acknowledge funding from the French National Research Agency (ANR) under the Investments for the Future (Investissements d'Avenir) program, grant ANR-17-EURE-0010.}
}}
\author{Cristina Gualdani\thanks{\scriptsize Email: \href{mailto:c.gualdani@qmul.ac.uk}{c.gualdani@qmul.ac.uk}, Queen Mary University of London, London, UK.}
\text{ } \text{ } Shruti Sinha\thanks{\scriptsize Email: \href{mailto:shruti.sinha@tse-fr.eu}{shruti.sinha@tse-fr.eu}, Toulouse School of Economics, University of Toulouse Capitole, Toulouse, France.}
}
\date{\today}
\maketitle
\begin{abstract}
We study identification of preferences in static single-agent discrete choice models where decision makers may be imperfectly informed about the state of the world. We leverage the notion of one-player Bayes Correlated Equilibrium by \hyperlink{BM_2}{Bergemann and Morris (2016)} to provide a tractable characterization of the sharp identified set. We develop a procedure to practically construct the sharp identified set following a sieve approach, and provide sharp bounds on counterfactual outcomes of interest. We use our methodology and data on the 2017 UK general election to estimate a spatial voting model under weak assumptions on agents' information about the returns to voting. Counterfactual exercises quantify the consequences of imperfect information on the well-being of voters and parties.
\vspace{0.8cm} 
\vspace{0.8cm}  
\end{abstract}

\thispagestyle{empty}
\newpage
\setcounter{page}{1}

%%%%%%%%%%%%%%%%%%%%%%%%%%%%%%%
\section{Introduction}
\label{intro}
A fundamental issue in the empirical analysis of decision problems is the presence of frictions that prevent agents from learning the payoffs associated with the available alternatives. When estimating parameters and making counterfactual predictions, it is common to make strong assumptions about agents' beliefs, which can weaken the credibility of  the results. In particular, most of the applied  literature either  imposes perfect information, or incorporates   information frictions by fully specifying agents' beliefs, as in search models (\hyperlink{Mehta}{Mehta, Rajiv, and Srinivasan, 2003}; \hyperlink{Honka}{Honka and Chintagunta, 2016}; \hyperlink{Ursu}{Ursu, 2018}), models with rational inattention (\hyperlink{Caplin_Dean}{Caplin and Dean, 2015}; \hyperlink{Matejka_McKay}{Mat\u{e}jka and McKay, 2015};  \hyperlink{Fosgerau}{Fosgerau, Melo, de Palma, and Shum, 2020};   \hyperlink{Csaba}{Csaba, 2018}; \hyperlink{Caplin_Dean_Leahy}{Caplin, Dean, and Leahy, 2019}; \hyperlink{Brown}{Brown and Jeon, 2020}), and models with preferences for risk (for a review, see \hyperlink{Barseghyan}{Barseghyan, Molinari, O'Donoghue, and  Teitelbaum, 2018}). 
Instead, this paper develops a methodology to identify preferences and counterfactual outcomes from cross-sectional choice data that imposes weak restrictions on  the agents' beliefs and, hence, is robust to whether agents are perfectly or partially informed.

We consider a large class of static single-agent discrete choice models, where the decision maker (DM)  has to choose an alternative from a finite set. The payoff generated by each alternative depends on the state of the world, which is randomly determined by nature. The DM has a prior on the state of the world. Moreover, the DM can refine their prior upon reception of a private signal representing the   DM's information structure. This information structure can range from full revelation of the state of the world to no information whatsoever, depending on the latent frictions encountered by the DM in the learning process.
 The DM uses the acquired information structure to update their  prior and obtain a posterior through the Bayes' rule. Lastly, the DM chooses an alternative maximising their expected payoff, where the expectation is computed via the posterior. Under additional assumptions on information structures, this framework accommodates   additive random utility discrete choice models, discrete choice models with risk aversion, discrete choice models with rational inattention, and {some} discrete choice models with search. Our objective is (partially) identifying  preferences  and counterfactual outcomes while remaining agnostic about   information structures. 
 
 The model just described is a  game against nature (\hyperlink{Milnor}{Milnor, 1951}). That is, it is a 1-player game in which a single self-interested player must choose a strategy. The player's payoff depends on their own strategy and the realization of the state of the world, which is decided at random by a totally disinterested nature.
Thus, we can use results from the theoretical  and empirical literature on  $N$-player games with $N\geq 1$ and weak assumptions on information structures to characterize the sharp identified set for the payoff parameters. 
In particular, we revisit our framework through the lens of 1-player Bayes Correlated Equilibrium  (\hyperlink{Kamenica}{Kamenica and Gentzkow, 2011}; \hyperlink{BM_1}{Bergemann and Morris, 2013};  \hyperlink{BM_2}{2016}). A fundamental result in the theoretical literature on robust predictions (Theorem 1, \hyperlink{BM_2}{Bergemann and Morris, 2016}) is that the set of optimal strategies predicted by our model under a large range of possible information structures  is  {equivalent} to the collection of model-implied choice probabilities   under the notion of 1-player Bayes Correlated Equilibrium. Further, the latter collection is a convex set defined by linear equalities and inequalities. Therefore, as shown by \hyperlink{Tamer_auctions}{Syrgkanis,  Tamer, and Ziani (2021)}    and \hyperlink{Magnolfi_Roncoroni}{Magnolfi and Roncoroni (2023)}, determining whether a given parameter value belongs to the sharp identified set  consists of solving a linear program, which is a well-understood and computationally tractable problem.\footnote{Observe that the collection of model-implied choice probabilities   under the notion of 1-player Bayes Correlated Equilibrium can also be written as the Aumann expectation of the random set of 1-player Bayes Correlated Equilibria. Therefore, the above characterization of the sharp identified set   is equivalent to the one provided by  \hyperlink{BMM}{Beresteanu, Molchanov, and Molinari (2011)}. A distinctive feature of our framework is that this Aumann expectation is defined by linear equalities and inequalities and, therefore, can be computed by solving a linear program. }

 We make two methodological contributions. First, we develop a formal procedure to practically construct the sharp identified set for the payoff parameters when the state of the world is continuous. In such a case, a 1-player Bayes Correlated Equilibrium is an infinite-dimensional object, and, thus, the linear program to solve for each candidate parameter value is also infinite-dimensional. Previous papers simplify the analysis by assuming that the state of the world is discrete, or allowing for a continuous state of the world but, in practice, discretising its support in some arbitrary bins to operationalise the
linear programming procedure. Here, instead,  we propose a sieve approximation of the program along the lines of   \hyperlink{Han}{Han and Yang (2023)} in their study of treatment effects. We test this procedure in simulations and implement it in the empirical application.

Second, we characterize sharp bounds on the counterfactual choice probabilities when
agents receive information about the state of the world via a policy program. This is an important question in the empirical
literature on single-agent decision problems across different fields and is largely absent in the literature on many-player games. To cite just a few examples, see  \hyperlink{Hastings}{Hastings and Tejeda-Ashton (2008)} on retirement fund options in Mexico, \hyperlink{Hastings2}{Hastings and Weinstein (2008)} and \hyperlink{bettinger}{Bettinger, et al. (2012)} on school choice, and \hyperlink{Kling}{Kling, et al. (2012)} on  Medicare Part D prescription drug plans. This question is typically answered using field experiments which
can be very costly. Our characterization   represents a powerful result because it allows the analyst to assess the effect of programs of information provision before conducting such interventions. We also provide sharp bounds on the maximum
potential welfare cost of limited information which may keep agents from choosing their first best.

Lastly, for readers keen to dive deeper into the identification nuances of our framework, Section \ref{power} explores how it differs from traditional 2-player games commonly studied in the empirical literature.

% In traditional 2-player games, both players choose actions to maximise their own payoffs, and these choices are observed by the researcher. In our framework, however, nature randomly selects the state of the world, a value that remains unobserved by the researcher. These differences have significant implications for identification power. Specifically,  without variation from exogenous covariates or additional restrictions on the payoff function and prior, our framework lacks empirical content, similar to several many-player games. Conversely,  even when such restrictions are present, our framework is generally expected to provide less informative bounds on the payoff parameters compared to many-player games. This is due to the unobserved and random behaviour of nature, which needs to be integrated out and,  for instance, cannot be influenced by shifting covariates to extreme values, a technique often employed in many-player games to obtain tight or even unique values (also known as ``identification at infinity'').}

We use our methodology to study voting behaviour in the UK. 
We consider the spatial model of voting, which is an important framework in political economy to explain individual preferences for parties (\hyperlink{Downs}{Downs 1957}; \hyperlink{Black}{Black, 1958}). This model postulates that an agent has a most preferred policy and votes for the party whose position is closest to their ideal.  
In empirical analysis, it is typically implemented by estimating a classical parametric discrete choice model with perfect information (\hyperlink{Alvarez1995}{Alvarez and Nagler, 1995}; \hyperlink{Alvarez1998}{1998}; \hyperlink{Alvarez2000}{2000}; \hyperlink{Alvarez2000_2}{Alvarez, Nagler, and Bowler, 2000}). However, in reality, uncertainty pervades voting  (\hyperlink{Shepsle}{Shepsle, 1972};  \hyperlink{Weisberg}{Weisberg and Fiorina, 1980};   \hyperlink{Enelow}{Enelow and Hinich, 1981};   \hyperlink{Baron}{Baron, 1994};  \hyperlink{Matsusaka}{Matsusaka 1995};  \hyperlink{Carpini}{Carpini and Keeter, 1996};     \hyperlink{Lupia}{Lupia and McCubbins, 1998}; \hyperlink{Feddersen_Pesendorfer}{Feddersen and Pesendorfer, 1999};   \hyperlink{Tabellini2}{Mat\u{e}jka and Tabellini, 2019}). That is, voters may  be aware of their  own and  the parties' attitudes towards some popular issues, but they might be less informed on how they themselves and the parties stand towards more technical or less debated topics, and on the traits of the candidates other than those publicly advertised. Further, their competence on these matters is likely to be arbitrarily different depending on, for example, political sentiment, civic sense, attentional limits, media exposure, and candidates' candor.

Despite the acknowledgement of the central role played by the sophistication of voters in determining voting patterns, only a few empirical works have attempted to take it into account while estimating  a spatial voting framework (\hyperlink{Aldrich}{Aldrich and McKelvey, 1977}; \hyperlink{Bartels}{Bartels, 1986}; \hyperlink{Palfrey_Poole}{Palfrey and Poole, 1987}; \hyperlink{Franklin}{Franklin, 1991};  \hyperlink{Alvarez_solo}{Alvarez, 1998}; \hyperlink{Degan_Merlo2}{Degan and Merlo, 2011}). This has been done by exogenously and parametrically modelling
 how information frictions affect the perceptions of DMs about the returns to voting (for instance, via an additive, exogenous, and parametrically distributed evaluation error in the payoffs), or by parametrically specifying the probability  of being informed versus uninformed when voting. Instead, our methodology permits us to incorporate voter uncertainty under weak assumptions on the latent, heterogeneous, and potentially endogenous process followed by voters to gather and evaluate information. 

In particular, we focus on a setting where the state of the world consists of distances between the voters and the parties' ideological positions on a few popular policy issues, and of voter-party-specific taste variables capturing voter perception on candidates' qualities and parties' positions on obscure topics. We assume that each voter observes the realization of the former, but may be uncertain about the realization of the latter. We estimate the model using data from the British Election Study, 2017: Face-to-Face Post-Election Survey (\hyperlink{Fieldhouse}{Fieldhouse, et al., 2018}) on the UK general election held on 8 June 2017. We compare our findings with the results one gets under the standard assumption that all agents are perfectly informed about the returns to voting. 
Several conclusions on the payoff parameters achieved under the complete information assumption are not unambiguously  corroborated when we remain agnostic about information structures. 

We use our characterization of counterfactual bounds to robustly assess  to what extent imperfect information affects the well-being of voters and parties. 
We imagine an omniscient mediator implementing a policy that gives voters perfect information about the state of the world. We simulate the counterfactual vote shares and study how they change compared to the factual scenario. This question has been debated at length in the literature. Political scientists have often answered it by arguing that a large population composed of possibly uninformed citizens act as if it was perfectly informed (for a review, see \hyperlink{Bartels2}{Bartels, 1996}). \hyperlink{Carpini}{Carpini and Keeter (1996)}, \hyperlink{Bartels2}{Bartels (1996)}, and \hyperlink{Degan_Merlo2}{Degan and Merlo (2011)} provide quantitative evidence to disconfirm such claims; the first two by using auxiliary data on the level of information of the survey respondents as rated by the interviewers or assessed by test items, and the latter by parametrically specifying the probability that a voter is informed. We contribute to this literature by providing a way to construct counterfactual vote shares under perfect information, which neither requires the difficult task of measuring voters' knowledge level in the factual scenario, nor imposes parametric assumptions on the probability that a voter is informed. We find that  voters benefit from full information as it leads to a considerable drop in the abstention rate. We also find that  transparency harms the two historically dominant parties, i.e., the Conservative Party and the Labour Party, and favours the other minor parties, i.e., the Liberal Democrats and the Green Party.   This suggests that some payoff-relevant information is unobserved by voters, and the biggest parties in the British political scene benefit from such uncertainty. Moreover, we quantify the maximum voters' welfare cost of limited information and find that it is comparable in magnitude to reducing the left-right ideological distance from a given party by around three points. 

In addition to studying the impact of information provision, we investigate how the   parties' welfare changes when they modify their ideological positions on popular policy issues. 
We adapt to our setting Theorem 1 of \hyperlink{Bergemann_Brooks_Morris}{Bergemann, Brooks, and Morris (2022)}, which permits us to answer such a question while holding fixed the voters' information structures in the counterfactual scenario. In agreement with several post-election studies, we find that,  by holding a strong left ideological position about tax and social care, the Labour Party gained numerous votes during the election campaign.  

 The remainder of the paper is organised as follows. Section \ref{model} describes the model. Section \ref{identification} discusses identification. Section \ref{simulations} presents some simulations. Section \ref{empirical} illustrates the empirical application. Finally, section \ref{conclusions} concludes. The proofs are in Appendix \ref{proofs}.

  In what follows,  we shorten \hyperlink{BM_2}{Bergemann and Morris (2016)} as BM16, \hyperlink{Tamer_auctions}{Syrgkanis,  Tamer, and Ziani (2021)} as STZ21, and \hyperlink{Magnolfi_Roncoroni}{Magnolfi and Roncoroni (2023)} as MR23. We refer to 1-player   Bayes Correlated Equilibrium as 1BCE. Given a random variable $Z$ with support $\mathcal{Z}$, $P_Z$ denotes its probability mass function (if $Z$ is discrete) or density (if $Z$ is continuous). $\Delta(\mathcal{Z})$ denotes the set of all probability mass functions or densities with support contained in $\mathcal{Z}$.

%%%%%%%%%%%%%%%%%%%%%%%%%%%%%%%
\section{The model}
\label{model}
We consider a DM who faces the  problem of choosing an alternative from a  {finite} set, $\mathcal{Y}$. There is an unknown  state of the world, $V$, with support $\mathcal{V}$, that enters directly in the DM's utility, $u: \mathcal{Y}\times \mathcal{V} \rightarrow \mathbb{R}$. $u\in \mathcal{U}$, where $\mathcal{U}$  is the   set  of all functions mapping $\mathcal{Y}\times \mathcal{V}$ to $\mathbb{R}$. 
The DM does not observe the realization of $V$ and has a prior belief on it, $P_V$.  Before making a choice, the DM has an opportunity to learn more about the state of the world and resolve some uncertainty about the payoffs generated by the alternatives.
Formally, the DM can refine their prior on $V$ upon reception of a private signal, $T$, with support $\mathcal{T}$ and distribution $P_{T|V}(\cdot| v)$ conditional on $V=v$. We denote by $\mathcal{P}_{T| V}$ the family of the signal's conditional distributions for each $v\in \mathcal{V}$, i.e., $\mathcal{P}_{T| V}\coloneqq \{P_{T|V}(\cdot| v):  v \in \mathcal{V}\}$. 
 The DM uses $\mathcal{P}_{T| V}$ and the received signal realization, $t$,   to update their prior  on $V$ via Bayes' rule and obtains a posterior, $P_{V|T}(\cdot|t)$. The DM chooses alternative $y\in \mathcal{Y}$ maximising their expected utility computed under the posterior, $
\int_{\mathcal{V}} u(y, v) P_{V| T}(v|t)dv$. If there is more than one maximising alternative,   the DM applies some tie-breaking rule.

The informativeness of  $T$ about $V$ (in the Blackwell sense) is inherently related to the frictions potentially encountered by the DM while investigating the state of the world.\footnote{\hyperlink{blackwell51}{Blackwell (1951;} \hyperlink{blackwell53}{1953)} provides a rank-ordering of information structures in terms of their informativeness.} These frictions can stem from various sources, such as attentional and cognitive limits, financial constraints, spatial and temporal boundaries, and cultural and personal biases. If the DM faces no information frictions,  they may process a signal revealing the realization of $V$ and discover the payoffs with certainty.  
Instead, if the DM experiences considerable information frictions,  they may process a signal adding nothing to their prior on $V$. 
A signal whose informativeness is between such two extremes is plausible as well. In a typical empirical application, the information frictions the DM   encounters are not observed by the researcher. Hence, we proceed without assumptions on $\mathcal{T}$ and $\mathcal{P}_{T| V}$.

We now provide a more compact representation of our framework. Following the terminology of \hyperlink{BM_2}{BM16}, we define the {\it baseline decision problem} faced by the DM as $G\coloneqq \{\mathcal{Y},  \mathcal{V}, u, P_V\}$. 
We also define the {\it information structure} processed by the DM as $S\coloneqq\{\mathcal{T}, \mathcal{P}_{T|V}\}$. $G$ represents what the DM knows before processing any signal. $S$ consists of the additional information the DM  learns about $V$, together with the received  realization of the signal. $S$ belongs to $\mathcal{S}$, which is the set of all possible information structures processed by the DM. $\mathcal{S}$ contains the information structure giving complete information ({\it complete information structure}), the information structure giving no information in addition to the prior  ({\it null information structure}), and any information structure whose informativeness is between those two extremes. The pair $\{G, S\}$ constitutes the {\it augmented decision problem} faced by the DM. 

We denote by  $Y$   the DM's choice. 
An optimal (mixed) strategy  in the    augmented decision problem  $\{G, S\}$ is a  distribution of $Y$ conditional on $T$, $\mathcal{P}_{Y|T}\coloneqq \{P_{Y|T}(\cdot| t):  t \in \mathcal{T}\}$, such that, for each $t\in \mathcal{T}$, the DM maximises their expected utility by choosing  any alternative $y\in \mathcal{Y}$ featuring $P_{Y|T}(y| t)>0$.\footnote{A mixed strategy arises if there are ties.}

Observe that the model just described is a {\it game against nature} (\hyperlink{Milnor}{Milnor, 1951}). That is, it is a 1-player game in which a single self-interested player must choose a strategy. Nature is indifferent among outcomes, has no payoff, and chooses $V$ through   randomisation. 
The player's payoff depends on their own   strategy and the realization of $V$. 
 One can categorize our model as a specific example of 2-player games, where nature serves as the second player. However, our framework diverges from traditional 2-player games commonly studied in the empirical literature. 
  In those games, the researcher models the payoffs of both players, both players can affect each other's payoffs through their choices,   and such choices are observed by the researcher. In contrast, in our model, nature's payoff is not specified,   nature selects a value for $V$ randomly from $P_V$, and this value remains unobserved by the researcher. These differences have implications for identification power, as discussed in Section \ref{power}.

Our framework encompasses several settings of empirical interest, primitives of which are typically estimated under strong assumptions on $\mathcal{T}$ and $\mathcal{P}_{T| V}$. For example, it includes additive random utility discrete choice models (Logit, Nested Logit, Mixed Logit, Probit, etc.),  discrete choice models with preferences for risk (\hyperlink{Barseghyan_AER}{Barseghyan, Molinari, O' Donoghue, and Teitelbaum, 2013}; \hyperlink{Barseghyan_QE}{Barseghyan, Molinari, and Teitelbaum, 2016}), models with rational inattention (\hyperlink{Caplin_Dean}{Caplin and Dean, 2015}; \hyperlink{Matejka_McKay}{Mat\u{e}jka and McKay, 2015};    \hyperlink{Csaba}{Csaba, 2018}; \hyperlink{Caplin_Dean_Leahy}{Caplin, Dean, and Leahy, 2019}; \hyperlink{Brown}{Brown and Jeon, 2020}; \hyperlink{Fosgerau}{Fosgerau, Melo, de Palma, and Shum, 2020}), and some search models (\hyperlink{Hebert}{H\'ebert and Woodford, 2018};  \hyperlink{Morris}{Morris and Strack, 2019}). See Appendix \ref{example} for more details. Also, see Appendix \ref{optimal_appendix} for a connection with the consideration set literature.

%%%%%%%%%%%%%%%%%%%%%%%%%%%%%%%
\section{Identification}
\label{identification}

\subsection{Data generating process and characterization of the identified set}
\label{BM_application}
We assume that the utility function, $u$,  and prior, $P_V$, belong to parametric classes, $\{u(\cdot; \theta_u)\}_{\theta_u\in \Theta_U}$ and $\{P_V(\cdot; \theta_V)\}_{\theta_V\in \Theta_V}$, indexed by the finite-dimensional structural parameters $\theta_u\in \Theta_U$ and $\theta_V\in \Theta_V$, respectively. Let $\theta\coloneqq (\theta_u, \theta_V)\in \Theta\coloneqq\Theta_U\times \Theta_V $ denote a generic parameter vector  and $\theta_0\coloneqq (\theta_{0,u}, \theta_{0,V})$ denote the true parameter vector. 
Let $Y_1,\dots, Y_n$ be an i.i.d. sample of choices, where each choice is the outcome of the augmented decision problem, $\{G(\theta_0), S_1\},\dots, \{G(\theta_0), S_n\}$, respectively. 

We do not know the exact information structures, $S_1,\dots, S_n$, that were processed in each of these decision problems and remain agnostic about those. In particular, we allow $S_i$ to be different from $S_j$ for each $i\neq j$, implying that the empirical distribution of choices, $\mathbb{P}_Y$, is a mixture  of optimal strategies over various information structures. This heterogeneity embeds the fact that different agents could encounter different information frictions and hence process more or less informative signals. We treat the information structures processed by DMs as nuisance parameters and  study the question of identifying $\theta_0$ and counterfactuals of interest from $\mathbb{P}_Y$.\footnote{It is implicit in our discussion that we also remain agnostic about tie-breaking rules and allow them to be heterogenous across DMs.} 

All DMs are assumed to rely on a common prior, $P_V(\cdot; \theta_{0,V})$. Some heterogeneity of priors  across DMs can be introduced by including discrete payoff-relevant variables, $(X,\epsilon)$, that are observed by  DMs together with the signal and correlated with $V$. In that case, each DM has a prior $P_{V|X,\epsilon}(\cdot;| x,e; \theta_{0,V})$ conditional on $(X,\epsilon)=(x,e)$. See Section \ref{computation} and Appendix   \ref{epsilon} on how to add $(X,\epsilon)$ to our framework.

In certain settings, some or all the components of  $V$ are observed by the researcher.   For example, in models of insurance plans, the researcher often has data on the ex-post claim experience of the agents in the sample (see Example \ref{risk} in Appendix \ref{example}).  In those cases,  $\theta_{0,V}$ could be identified directly from such additional data.  In our general discussion below, we focus on the  scenario where $V$ is unobserved to the researcher. This is the case considered in our empirical application on voting behaviour.

The  identified set of $\theta_0$ can be characterised according to Result 3 of \hyperlink{Tamer_auctions}{STZ21} which encompasses any $N$-player games with $N\geq 1$. We explain how this result can be adapted to our specific case for the sake of completeness in our exposition.
 Intuitively, the  identified set of $\theta_0$ is the set of $\theta$s for which the model predicts a distribution of $Y$  that matches $\mathbb{P}_Y$. 
Let $\mathcal{R}(\theta,S)$ be the set of optimal strategies of    $\{G(\theta), S\}$ and  $\mathcal{R}(\theta)$ be the set of model-implied choice probabilities  while remaining agnostic about information structures.   That is,
\begin{equation}
\label{convex}
\begin{aligned}
 \mathcal{R}(\theta) \coloneqq \text{Conv}\Big \{  P_{Y} \in \Delta(\mathcal{Y}):    P_Y(y)=\int_{\mathcal{T}} \int_{\mathcal{V}}     P_{Y|T}(y|t)   & P_{T|V}(t|v)   P_{V}(v; \theta_V)  dv dt\text{ } \forall y \in \mathcal{Y}, \\
& \mathcal{P}_{Y| T} \in\mathcal{R}(\theta,S), S\coloneqq \{\mathcal{T}, \mathcal{P}_{T| V}\} \in \mathcal{S}\Big \},
\end{aligned}
\end{equation}
where we have used the fact that $Y$ is independent of $V$ conditional on $T$.
Convexification (via the convex hull operator, $\text{Conv}\{\cdot\}$) allows us to include in $\mathcal{R}(\theta)$ distributions of $Y$  that are mixtures of optimal strategies over various information structures. This ensures   the heterogeneity of information structures   in the cross-section, as discussed above.\footnote{To understand the convexification step better, note that the information structures in our framework  are econometrically similar to  the equilibrium selection mechanisms in incomplete many-player games (\hyperlink{Tamer}{Tamer, 2003}; \hyperlink{CT}{Ciliberto and Tamer, 2009}). In the former, convexification allows  the information structures to differ across DMs. In the latter, convexification allows  the equilibrium selection mechanisms   to differ across markets. 
See also  \hyperlink{Tamer_auctions}{STZ21} and \hyperlink{Magnolfi_Roncoroni}{MR23} about  convexification.}
The  identified set of $\theta_0$ is defined as,
$$
\Theta^*\coloneqq \{\theta\in \Theta: \mathbb{P}_Y\in \mathcal{R}(\theta) \}.
$$

The above definition of $\Theta^*$ is not helpful in practice. This is because constructing $\mathcal{R}(\theta)$ following (\ref{convex}) is infeasible due to the necessity of exploring the large class $\mathcal{S}$, which contains infinite-dimensional objects. We overcome this issue by recalling that our decision problem is a   1-player game (game against nature), as discussed in Section \ref{model}. Hence, we can use results from the theoretical literature on  $N$-player games with $N\geq 1$ and weak assumptions on information  to give a simpler characterization of $\mathcal{R}(\theta)$ and, in turn, $\Theta^*$. 

In particular, we consider the notion of 1BCE (\hyperlink{Kamenica}{Kamenica and Gentzkow, 2011}; \hyperlink{BM_1}{Bergemann and Morris, 2013}; \hyperlink{BM_2}{BM16}). This notion refers to a theoretical setting where an omniscient mediator makes incentive-compatible recommendations to the DM as a function of the state of the world and consistent with the DM's prior. If the DM follows such recommendations, the resulting distribution of choices is a 1BCE. A fundamental result in the theoretical literature on robust predictions (Theorem 1, \hyperlink{BM_2}{BM16}) is that the set of optimal strategies that arise from adding an arbitrary information structure   to the baseline decision problem $G(\theta)$ is equivalent to the set of 1BCE of $G(\theta)$. Therefore, to obtain $\mathcal{R}(\theta)$, we do not need to explore  the collection of all possible information structures, but instead, we can calculate the set of 1BCE of $G(\theta)$.    In the remainder of the section, we formalise the equivalent characterization of $\mathcal{R}(\theta)$ and $\Theta^*$ based on 1BCE. In Section \ref{computation}, we zoom into the computational part. In Section \ref{power}, we discuss the identification power of our model.   In Section \ref{count}, we characterize bounds on counterfactuals of interest.
 
 First, we give the definition of 1BCE of $G(\theta)$. In what follows, $P_{Y,V}$ denotes the joint distribution of $Y$ and $V$.
 \begin{definition}{\normalfont({\itshape 1BCE of  $G(\theta)$})}
\label{1BCE}
Given $\theta\in \Theta$,  $P_{Y,V} $ is a 1BCE of   $G(\theta)$ if:
\begin{enumerate}[1.]
\item It is {\it consistent}, i.e., the marginal of $P_{Y,V} $ on $Y$ is equal to  the  DM's prior, $P_V(\cdot; \theta_V)$:
$$
\sum_{y\in \mathcal{Y}} P_{Y,V}(y,v)=P_V(v; \theta_V) \text{ }\forall v\in \mathcal{V}.
$$
\item It is {\it obedient}, i.e., the DM who is recommended alternative $y\in \mathcal{Y}$ by an omniscient mediator  has no incentive to deviate: 
$$
\begin{aligned}
\int_{\mathcal{V}} P_{Y,V}(y, v )(u(y,v; \theta_u) - u(y', v; \theta_u ))dv \geq   0,\text{ } \forall y'\in \mathcal{Y}\setminus \{y\},\forall y\in \mathcal{Y}.
\end{aligned}
$$
\end{enumerate}
\end{definition} 

We now state Theorem 1 of \hyperlink{BM_2}{BM16}.  

\begin{thm}{\normalfont({\itshape Theorem 1 BM16})}
\label{main}
Given $\theta\in \Theta$, $P_{Y,V}$ is a 1BCE of   $G(\theta)$  if and only if there exists an information structure $S \coloneqq \{\mathcal{T}, \mathcal{P}_{T|V}\}\in \mathcal{S}$ and  an optimal strategy $\mathcal{P}_{Y| T}$ of   $\{G(\theta), S\}$ such that $P_{Y,V }$ arises from $\mathcal{P}_{Y| T}$.\footnote{We say that $P_{Y,V }$ arises from $\mathcal{P}_{Y| T}$ if $P_{Y,V}(y,v)=\int_{t\in \mathcal{T}} P_{Y| T}(y|t)P_{T|V}(t|v) P_{V}(v; \theta_V)dt $, for every $ y \in \mathcal{Y}$ and $  v\in \mathcal{V}$.}
\end{thm}

We use Theorem \ref{main} to equivalently rewrite $\mathcal{R}(\theta)$ and  $\Theta^*$. 
For each $\theta\in \Theta$, let $\mathcal{W}(\theta)$ be the set of  1BCEs of   $G(\theta)$.  Let $\mathcal{Q}(\theta)$ be the set of distributions of $Y$ that arise from the  1BCEs of   $G(\theta)$:
\begin{equation*}
\label{Q}
\begin{aligned}
 \mathcal{Q}(\theta) \coloneqq \Big \{ & P_{Y} \in \Delta(\mathcal{Y}): \text{ }P_{Y }(y)=\int_{\mathcal{V}} P_{Y,V}(y,v )  dv\text{ } \forall y \in \mathcal{Y}, P_{Y,V }  \in \mathcal{W}(\theta)\Big \}.
\end{aligned}
\end{equation*}
Theorem \ref{main} implies that 
$
\mathcal{R}(\theta)=\mathcal{Q}(\theta)$. This can be seen in three steps. 
First, observe that the {\it Consistency} and {\it Obedience} requirements defining 1BCE  are  linear in $P_{Y,V}$. Therefore,  $\mathcal{Q}(\theta)$ is a convex set. Second, $\mathcal{R}(\theta)$ contains distributions of $Y$ that are mixture  of optimal strategies over various information structures. Hence, by Theorem \ref{main}, each $P_Y\in \mathcal{R}(\theta)$  maps into a mixture of    1BCEs. Third, since $\mathcal{Q}(\theta)$ is convex, any mixture of elements from the set is itself an element of $\mathcal{Q}(\theta)$. Therefore, $
\mathcal{R}(\theta)=\mathcal{Q}(\theta)$ and we can use $\mathcal{Q}(\theta)$ in place of $\mathcal{R}(\theta)$ to characterize $\Theta^*$,  as shown by Result 3 of \hyperlink{Tamer_auctions}{STZ21} and stated in our Proposition \ref{main_ident}.
\begin{prop}{\normalfont({\itshape Result 3 STZ21})}
\label{main_ident}
Let $
\Theta^{**}\coloneqq  \{\theta\in \Theta: \mathbb{P}_Y\in \mathcal{Q}(\theta)\}$. 
It holds that $\Theta^*=\Theta^{**}$. 
\end{prop}

%Proposition  \ref{main_ident} provides an analytically convenient characterization of the identified set. Note that the notion of 1BCE is not used as an alternative equilibrium assumption. In fact, under the assumed data generating process, the DMs choose actions based on optimal strategies.

 %%%%%%%%%%%%%%%%%%%%%%%%%%%%%%%%%%%%%%%%%%%%%%%%%%%%%%%%%%%%%%%%
%%%%%%%%%%%%%%%%%%%%%%%%%%%%%%%%%%%%%%%%%%%%%%%%%%%%%%%%%%%%%%%%
%%%%%%%%%%%%%%%%%%%%%%%%%%%%%%%%%%%%%%%%%%%%%%%%%%%%%%%%%%%%%%%%
\subsection{Construction of the identified set}
\label{computation}
To see how to use Proposition \ref{main_ident} in practice,  we first rewrite it in a more explicit way using Definition \ref{1BCE}. By Proposition \ref{main_ident} and Definition \ref{1BCE}, $\theta$ belongs to $ \Theta^*$ if and only if there exists a function $P_{Y,V}: \mathcal{Y}\times \mathcal{V}\rightarrow \mathbb{R}$ which satisfies the following constraints:
\begin{equation}
\label{lin_pr0}
%\hspace{-0.7cm}
\begin{alignedat}{3}
& \text{Consistency:} \quad && \sum_{y\in \mathcal{Y}}  {P_{Y,V}(y,v)}={P_{V}(v; \theta_V)} \text{ }\forall v\in \mathcal{V},\\
&\text{Obedience:}  \quad&&\int_{\mathcal{V}} {P_{Y,V}(y,v)}({u(y,v; \theta_u)}-{u(y',v; \theta_u)})dv \geq 0 \text{ } \forall y\in \mathcal{Y}, \forall y'\in \mathcal{Y}\setminus\{y\},\\
&\text{Probability requirements:}  \quad&&P_{Y,V}(y,v)\geq 0 \text{ } \forall y\in \mathcal{Y}, \forall v \in \mathcal{V}, \text{ }\sum_{y\in \mathcal{Y}}\int_{\mathcal{V}} P_{Y,V}(y,v)dv=1,\\
&\text{Data match:} \quad &&\int_{\mathcal{V}} P_{Y,V }(y,v )dv =\mathbb{P}_Y(y) \text{ }\forall y \in \mathcal{Y}.
\end{alignedat}
\end{equation}

Observe that the above constraints are linear in $P_{Y,V}$.
Therefore, when $\mathcal{V}$ is a finite set, (\ref{lin_pr0}) reduces to a {\it finite-dimensional} linear program:
\begin{equation}
\label{lin_pr}
%\hspace{-0.7cm}
\begin{alignedat}{3}
& \text{Consistency:} \quad && \sum_{y\in \mathcal{Y}}  {P_{Y,V}(y,v)}={P_{V}(v; \theta_V)} \text{ }\forall v\in \mathcal{V},\\
&\text{Obedience:}  \quad&&\sum_{v\in \mathcal{V}} {P_{Y,V}(y,v)}({u(y,v; \theta_u)}-{u(y',v; \theta_u)})\geq 0 \text{ } \forall y\in \mathcal{Y}, \forall y'\in \mathcal{Y}\setminus\{y\},\\
&\text{Probability requirements:}  \quad&&P_{Y,V}(y,v)\geq 0 \text{ } \forall y\in \mathcal{Y}, \forall v \in \mathcal{V}, \text{ }\sum_{y\in \mathcal{Y}}\sum_{v\in \mathcal{V}} P_{Y,V}(y,v)=1,\\
&\text{Data match:} \quad &&\sum_{v\in  \mathcal{V}} P_{Y,V }(y,v ) =\mathbb{P}_Y(y) \text{ }\forall y \in \mathcal{Y}.
\end{alignedat}
\end{equation}

In turn, we  can construct $\Theta^*$  following this procedure: first, generate a grid of points covering $\Theta$ as precisely as possible, depending on the available computational resources; second, for each $\theta$ in such a grid, check if (\ref{lin_pr}) has a solution with respect to $P_{Y,V}$; third, any $\theta$ for which (\ref{lin_pr}) has a solution belongs to the identified set.\footnote{Note that there is no need to recover the entire set of solutions of (\ref{lin_pr}) for a given $\theta$. Existence of at least one solution of (\ref{lin_pr}) is sufficient to include such a $\theta$ in the identified set.}

Observe that, when  $\mathcal{V}$ is finite, we can  dispense with the parameterisation of $u$ and $P_V$ via $\theta$, as these functions can be fully and flexibly characterised by a finite number of parameters, one for each combination of values of $(Y,V)$. In this case, we can also add nonparametric restrictions on $P_V$ to the linear program, such as monotonicity, concavity/convexity, and Lipschitz restriction, which can be written as linear constraints. 

Further, we  remark that (\ref{lin_pr}) is linear in $P_{Y,V}$ for a given $\theta$, but it is {\it not} linear  in {\it both} $P_{Y,V}$  and $\theta$. This is why we grid over $\Theta$  to construct $\Theta^*$. Linearity of (\ref{lin_pr}) in   $P_{Y,V}$ {\it and} the parameters is achieved when $u$ is {\it known} and $P_V$  is treated as a  finite-dimensional vector of parameters without indexing it by $\theta_V$. In this case, we can   obtain the identified set of  moments of $P_V$ by solving a {\it unique}  linear program, without gridding, as proposed by \hyperlink{Tamer_auctions}{STZ21}  for an auction framework.  In single-agent decision problems and other types of games, $u$ is typically unknown and, therefore,  we have linearity {\it only} in  $P_{Y,V}$. 

The setting with finite $\mathcal{V}$ is considered by \hyperlink{Tamer_auctions}{STZ21} and \hyperlink{Magnolfi_Roncoroni}{MR23}. We refer to those papers for a discussion on the computational burden of solving (\ref{lin_pr}) as the cardinalities of $\mathcal{Y}$ and $\mathcal{V}$ increase. 

When $\mathcal{V}$ is not a finite set, the simple finite-dimensional linear programming approach is no longer applicable as $P_{Y,V }$ is an {\it infinite-dimensional} object. This case has not been  addressed in the econometric literature on games, where assuming  a finite $\mathcal{V}$ or discretising a non-finite $\mathcal{V}$ in a few arbitrary bins to operationalise (\ref{lin_pr0}) are standard practices.  Here  we propose a formal  procedure to approximate $\Theta^*$ when $\mathcal{V}$ is not finite, along the lines of   \hyperlink{Han}{Han and Yang (2023)}. As a preliminary step, it is useful to equivalently rewrite (\ref{lin_pr0}) using the distribution of $Y$ conditional on $V$ as unknown, $\mathcal{P}_{Y|V}\coloneqq \{P_{Y|V}(\cdot |v): v\in \mathcal{V}\}$:\footnote{Note that  the {\it Consistency} constraint is redundant in (\ref{lin_pr_2}) as $\sum_{y\in \mathcal{Y}}  P_{Y|V}(y|v)P_{V}(v; \theta_V)=P_{V}(v; \theta_V)$ becomes $\sum_{y\in \mathcal{Y}}  P_{Y|V}(y|v)=1$, which is one of  the probability requirements.}
\par\nobreak
\vspace{-0.7cm}
{\small \begin{equation}
\label{lin_pr_2}
\begin{alignedat}{3}
&\text{Obedience:}  \quad&&\int_{\mathcal{V}} P_{Y|V}(y|v)P_V(v; \theta_V)({u(y,v; \theta_u)}-{u(y',v; \theta_u)}) dv \geq 0 \text{ } \forall y\in \mathcal{Y}, \forall y'\in \mathcal{Y}\setminus\{y\},\\
&\text{Probability requirements:}  \quad&&P_{Y|V}(y|v)\geq 0 \text{ } \forall y\in \mathcal{Y}, \forall v \in \mathcal{V}, \text{ }\sum_{y\in \mathcal{Y}} P_{Y|V}(y|v)=1 \text{ }\forall v \in \mathcal{V},\\
&\text{Data match:} \quad &&\int_{\mathcal{V}} P_{Y|V }(y|v ) P_V(v; \theta_V) dv =\mathbb{P}_Y(y) \text{ }\forall y \in \mathcal{Y}.
\end{alignedat}
\end{equation}}%

 We start from the simple case where $V$ is a random variable and $\mathcal{V}\coloneqq [v_\ell, v_u]$. 
Consider the following sieve approximation of $P_{Y|V}(y|v)$ using Bernstein polynomials of order $K$:
\begin{equation}
\label{approx}
P_{Y|V}(y|v)\approx \sum_{k\in \mathcal{K}} \lambda_{k,K}^y a_{k,K}(v),
\end{equation}
where $a_{k,K}(v)\coloneqq {K \choose k} \frac{(v-v_\ell)^k (v_u-v)^{K-k}}{(v_u-v_\ell)^K}$ is a univariate Bernstein basis, $\lambda_{k,K}^y \coloneqq P_{Y|V}(y| v_\ell+(v_u-v_\ell)\frac{k}{K})$ is its coefficient, $\mathcal{K}\coloneqq \{0,1,\dots,K\}$, and $K$ is finite. It can be shown that this approximation tends to $P_{Y|V}(y|v)$ uniformly over $\mathcal{V}$ as $K$ goes to infinity.  Observe that $a_{k,K}(v)\geq 0$. Hence,  for each $y$, $P_{Y|V}(y|v)\geq 0$ for each $v$ if and only if $\lambda_{k,K}^y\geq 0$ for each $k$.  Further, $\sum_{y\in \mathcal{Y}}  P_{Y|V}(y|v) =1$   is approximately equal to $\sum_{y\in \mathcal{Y}} \lambda_{k,K}^y= 1$ for each $k$ (\hyperlink{Coolidge}{Coolidge, 1949}).\footnote{See also \hyperlink{Mogstad}{Mogstad, Santos, and Torgovitsky (2018)} for using Bernstein polynomials in their study of treatment effects to approximate the marginal treatment effect function.}
Motivated by this result, a finite-dimensional linear program approximating (\ref{lin_pr_2}) can be obtained with respect to $\lambda\coloneqq (\lambda_{k,K}^y: (k,y)\in \mathcal{K}\times \mathcal{Y})$:
 \begin{equation}
\label{lin_pr_3}
\begin{alignedat}{3}
&\text{Obedience:}  \quad&&\sum_{k\in \mathcal{K}} \lambda_{k,K}^y \gamma^{y,y'}_{1,k,K}(\theta) \geq 0 \text{ } \forall y\in \mathcal{Y}, \forall y'\in \mathcal{Y}\setminus\{y\},\\
&\text{Probability requirements:}  \quad&&\lambda_{k,K}^y\geq 0 \text{ }\forall k\in \mathcal{K}, \forall y \in \mathcal{Y},  \text{ }\sum_{y\in \mathcal{Y}} \lambda_{k,K}^y= 1\text{ } \forall k\in \mathcal{K},\\
&\text{Data match:} \quad && \sum_{k\in \mathcal{K}} \lambda_{k,K}^y \gamma_{2,k,K}(\theta_V) =\mathbb{P}_Y(y) \text{ }\forall y \in \mathcal{Y},
\end{alignedat}
\end{equation} 
where 
$$ 
\begin{aligned}
&\gamma^{y,y'}_{1,k,K}(\theta) \coloneqq \int_{\mathcal{V}}  a_{k,K}(v) P_V(v; \theta_V)({u(y,v; \theta_u)}-{u(y',v; \theta_u)}) dv,\\
& \gamma_{2,k,K}(\theta_V) \coloneqq 
\int_{\mathcal{V}} a_{k,K}(v)  P_V(v; \theta_V) dv.
\end{aligned}
$$
We can then approximate $\Theta^*$ by verifying if (\ref{lin_pr_3}) has a solution with respect to $\lambda$ for each $\theta\in \Theta$.

The same logic applies when $V$ is a $D\times 1$ random vector  and $\mathcal{V}\coloneqq \times _{d=1}^D [v_{\ell_d}, v_{u_d}] $. In this case, we can  use (\ref{approx}), where $\mathcal{K}\coloneqq \times_{d=1}^D\{0,1,\dots, K_d\}$,  $K_d$ is finite for each $d=1,\dots, D$, $K$ is the cardinality of $\mathcal{K}$,  $a_{k,K}(v)\coloneqq \prod_{d=1}^D{K_d \choose k_d} \frac{(v_d-v_{\ell_d})^{k_d} (v_{u_d}-v_d)^{K_d-k_d}}{(v_{u_d}-v_{\ell_d})^{K_d}}$ is a $D$-variate Bernstein basis, and $\lambda_{k,K}^y \coloneqq P_{Y|V}(y| v_{\ell_1}+(v_{u_1}-v_{\ell_1})\frac{k_1}{K_1},\dots ,v_{\ell_D}+(v_{u_D}-v_{\ell_D})\frac{k_D}{K_D} )$ is its coefficient. We can further generalise Bernstein polynomials to the case where $\mathcal{V}$ is the real line (\hyperlink{Szasz}{Szasz, 1950};  \hyperlink{Butzer}{Butzer, 1954}).

%Instead of applying the above procedure, we could approximate (\ref{lin_pr_2}) using an i.i.d. sample, $\{V_j\}_{j=1}^J$, from $P_V(\cdot; \theta_V)$ (\hyperlink{calafiore}{Calafiore and Campi, 2005}). This method is similar to the discretisation in bins  by \hyperlink{Magnolfi_Roncoroni}{MR23}. However, as discussed by \hyperlink{Han2}{Han and Xu (2022)} in their Remark 6.1, the linear program (\ref{lin_pr_3})  may be more stable. %This is because, by writing down the dual of the problem, the randomisation approach can be shown to be a crude approximation to ... %that involves a uniform kernel. 

 Suppose there are other {\it discrete} payoff-relevant variables which enter the DM's information set together with the signal. In this case, one should solve  (\ref{lin_pr_3}) for each value of such variables. For instance, $u$ could depend on covariates $X$ observed by the DM and the researcher. $u$ could also depend on some variable $\epsilon$ observed by the DM but unobserved by the researcher. 
 Our procedure allows $(X,V,\epsilon)$ to be correlated, so that DMs can have heterogenous priors, $P_{V|X,\epsilon}(\cdot;| x,e; \theta_{V})$ conditional on $(X,\epsilon)=(x,e)$. See  Appendix   \ref{epsilon} on how to include $(X,\epsilon)$ in (\ref{lin_pr_3}).

 Verifying if (\ref{lin_pr_3}) has a solution for a given $\theta$ is computationally easy using standard algorithms, such as  the simplex optimizer
or the interior-point optimizer. In Section \ref{simulations}, we discuss how to choose $K$  in practice and the computing time.

 %%%%%%%%%%%%%%%%%%%%%%%%%%%%%%%%%%%%%%%%%%%%%%%%%%%%%%%%%%%%%%%%
%%%%%%%%%%%%%%%%%%%%%%%%%%%%%%%%%%%%%%%%%%%%%%%%%%%%%%%%%%%%%%%%
%%%%%%%%%%%%%%%%%%%%%%%%%%%%%%%%%%%%%%%%%%%%%%%%%%%%%%%%%%%%%%%%
\subsection{On the identification power of the model}
\label{power}
Our identification procedure imposes weak assumptions on the DMs' behavior by allowing for any information structures through the 1BCE characterization. One might wonder whether being agnostic about information structures strips discrete choice models of any empirical content. Proposition \ref{id_power} shows that, {\it even under the restrictive assumption of complete information}, discrete choice models do not retain identification power with respect to $(u,P_V)$, unless further assumptions  are made, such  as  exogenous covariates   and (non)parametric restrictions on $(u,P_V)$. Therefore,  to generate informative bounds, our framework necessitates additional assumptions on $(u,P_V)$, mirroring the requirements of traditional ``complete-information'' discrete choice models. In particular, in our simulations and empirical application, we show that our model can generate relatively tight bounds by introducing exogenous covariates and parameterising $u$ and $P_V$.

\begin{prop}{\normalfont({\itshape Identification power})}
\label{id_power}
Let $V\coloneqq (V_y: y\in \mathcal{Y})$ be a $|\mathcal{Y}|\times 1$ vector whose distribution $P_V$ has full support and is absolutely continuous   with respect to the Lebesgue measure. Let  $\tilde{\Delta}(\mathbb{R}^{|\mathcal{Y}|})$ be the set of all such $P_V$. Let $u(y, v)=\tilde{u}(y)+v_y$ for each $(y,v)\in \mathcal{Y}\times \mathcal{V}$, where $\tilde{u}:\mathcal{Y}\rightarrow \mathbb{R}$.  Let  $\tilde{\mathcal{U}}$ be the set of all  such $\tilde{u}$. Assume that all DMs process the complete information structure and denote by $P_{Y; \tilde{u}, P_V}$ the (unique) model-implied choice distribution for any given $(\tilde{u}, P_V)\in \tilde{\mathcal{U}}\times \tilde{\Delta}(\mathbb{R}^{|\mathcal{Y}|})$. Let $\Theta^*\coloneqq \{(\tilde{u},P_V)\in \tilde{\mathcal{U}}\times \tilde{\Delta}(\mathbb{R}^{|\mathcal{Y}|}): \mathbb{P}_Y= P_{Y; \tilde{u}, P_V}\}$ be the identified set under these restrictions. Then, for every $\mathcal{Y}$: (a)  the projection of $\Theta^*$ on $ \tilde{\Delta}(\mathbb{R}^{|\mathcal{Y}|}) $ is equal to $\tilde{\mathcal{U}}$; 
(b) the projection of $\Theta^*$ on $\tilde{\mathcal{U}} $ is equal to $ \tilde{\Delta}(\mathbb{R}^{|\mathcal{Y}|}) $.
\end{prop}

Nevertheless, even in the presence of exogenous covariates and parametric restrictions, we expect  our  framework to generally provide less informative bounds on the primitives  than those in 2-player games and no assumptions on information structures. To see why, consider a 2-player entry game with $\mathcal{Y}\coloneqq \{0,1\}$. Let the  payoff of player $i\in \{1,2\}$ be $Y_i(X_i \beta+ \delta Y_j+\epsilon_i)$, where $Y_j$ is the competitor's action, $X_i$ represents  $i$'s covariates   (scalar, for simplicity), and $\epsilon_i$ captures $i$'s characteristics unobserved by the researcher. Each player $i$ is assumed to observe $(X_i, X_j, \epsilon_i)$. The researcher remains agnostic about  $i$'s knowledge of $\epsilon_j$ and, hence, their ability to accurately predict $Y_j$.  Note that this framework resembles our model (a game against nature) by letting player $i$ be the DM, assigning player $j$'s role   to nature,  and setting $Y_j\coloneqq V$. However, while in the aforementioned 2-player  game player $j$ chooses $Y_j$  to maximise their own payoff, in our setting nature is indifferent to the outcomes and selects $V$ randomly from $P_V$. Additionally, this value is unobserved by the researcher. This leads to a generic reduction in the identification power of our framework compared to 2-player games, given the fewer data and model restrictions available for analysis. For example, a common way to achieve point identification of $\beta$ and the parameters governing the distribution of $(\epsilon_1,\epsilon_2)$  in the above 2-player  game is to use  ``at infinity'' arguments (Theorem 1, \hyperlink{Tamer}{Tamer, 2003}; Proposition 3, \hyperlink{Magnolfi_Roncoroni}{MR23}).  
These involve finding extreme values of $X_j$ that induce player $j$ always to choose one action, so that player $i$'s problem turns to a single agent parametric discrete choice problem with complete information that we know to be point identified (\hyperlink{Manski}{Manski, 1988}). Such a strategy is clearly not implementable in our setting because the model lacks assumptions regulating nature's behaviour, let alone covariates influencing nature's latent choice of $V$, which needs to be integrated out.

 %%%%%%%%%%%%%%%%%%%%%%%%%%%%%%%%%%%%%%%%%%%%%%%%%%%%%%%%%%%%%%%%
%%%%%%%%%%%%%%%%%%%%%%%%%%%%%%%%%%%%%%%%%%%%%%%%%%%%%%%%%%%%%%%%
%%%%%%%%%%%%%%%%%%%%%%%%%%%%%%%%%%%%%%%%%%%%%%%%%%%%%%%%%%%%%%%%
\subsection{Counterfactual identification}
\label{count}
A key objective in the empirical analysis of single-agent decision problems is predicting how individual choices and social welfare change in counterfactual decision settings deployed in the same environment. 
Our framework allows us to consider two types of policy experiments, illustrated in what follows.

\subsubsection{Information provision}
\label{count1_sec}
 In the first policy experiment, we study how the choice probabilities change in response to changes in the availability of information to agents about the state of the world. Suppose the policy maker implements some intervention that urges all agents to process information structure $S$. Some agents may settle for this information structure, while others might prefer to collect {\it further} information. In this new environment, the DM  selects their favourite alternative from $\mathcal{Y}$ by solving the augmented decision problem $\{G(\theta_0), S^\dagger\}$, where $S^\dagger$ is an unknown information structure that is  {\it at least as} informative as $S$. $S^\dagger$ is also called an {\it expansion} of $S$. Formally,  $S^\dagger\coloneqq \{\mathcal{T}^\dagger, \mathcal{P}_{T^\dagger|V}\}$ is an expansion of ${S}\coloneqq \{{\mathcal{T}}, {\mathcal{P}}_{T|V}\}$ if there exists ${S}^\diamond \coloneqq \{{\mathcal{T}}^\diamond, {\mathcal{P}}_{T^\diamond |V}\} $ such that $S^\dagger$ is the {\it combination} of $S$ and ${S}^\diamond$. That is, 
 $\mathcal{T}^\dagger \coloneqq \mathcal{T}\times \mathcal{T}^\diamond$, 
 $\int_{\mathcal{T}^\diamond} P_{T^\dagger|V}( t, t^\diamond |v) d t^\diamond= P_{T|V}(t|v)$ for each $t\in \mathcal{T}$ and $v\in \mathcal{V}$, and
  $\int_{\mathcal{T}} P_{T^\dagger|V}( t, t^\diamond |v) d t = P_{T^\diamond |V}(t^\diamond |v)$ for each $t^\diamond \in \mathcal{T}^\diamond $ and $v\in \mathcal{V}$ (Definition 5, \hyperlink{BM_2}{BM16}). 
  
We can characterize sharp bounds on the counterfactual choice probabilities while fixing the policy-implemented information structure $S$ and remaining agnostic about $S^\diamond$ and, hence, $S^\dagger$. To do this, we   use once again Theorem 1 of \hyperlink{BM_2}{BM16}, which generically applies to any 1-player game whose {\it minimal} information structure $S$ has been set by the researcher. Namely,  by Theorem 1 of \hyperlink{BM_2}{BM16}, the set of optimal strategies that arise from arbitrarily expanding $S$ is equivalent to the set of 1BCE of $\{G(\theta_0), S\}$. Therefore,   the collection of counterfactual choice probabilities is simply   the set of  1BCE of $\{G(\theta_0), S\}$. Proposition \ref{count1} formalises these arguments.

\begin{prop}{\normalfont({\itshape Counterfactual bounds - information provision})}
\label{count1}
  Let ${S}\coloneqq \{{\mathcal{T}}, {\mathcal{P}}_{T|V}\}\in \mathcal{S}$ be the information structure implemented by a policy program.  Let $\mathcal{P}^*_{Y^\dagger}(h)$ be  the identified set of the   counterfactual probability of choosing alternative $h\in \mathcal{Y}$ when  DMs face the augmented decision problem $\{G(\theta_0), S^\dagger\}$, where  $S^\dagger\coloneqq \{\mathcal{T}^\dagger, \mathcal{P}_{T^\dagger|V}\}\in \mathcal{S}$  is some unknown expansion of $S$, potentially heterogenous across DMs. Then, 
$
 \mathcal{P}^*_{Y^\dagger}(h)= \cup_{\theta \in \Theta^*} [ \Psi^{\ell}(h;\theta),  \Psi^{u}(h;\theta)]
$, 
  where 
$$
\begin{aligned}
&  \Psi^{\ell}(h;\theta)\coloneqq     \min_{P_{Y^\dagger, V,T}: \mathcal{Y}\times \mathcal{V}\times \mathcal{T}\rightarrow \mathbb{R}}  \int_{\mathcal{V}}\int_{\mathcal{T}}  P_{Y^\dagger, V,T}(h, v, t) dt dv  \text{ s.t. } (\ref{con1})-(\ref{con2}),
\end{aligned}
$$
$$
\begin{aligned}
&  \Psi^{u}(h;\theta)\coloneqq     \max_{P_{Y^\dagger, V,T}: \mathcal{Y}\times \mathcal{V}\times \mathcal{T}\rightarrow \mathbb{R} } \int_{\mathcal{V}}\int_{\mathcal{T}}  P_{Y^\dagger, V,T}(h, v, t) dt dv  \text{ s.t. } (\ref{con1})-(\ref{con2}),
\end{aligned}
$$
and
\begin{align} 
 & \sum_{y^\dagger \in \mathcal{Y}}  {P_{Y^\dagger,V,T}(y^\dagger,v,t)}={P_{V}(v; \theta_{V}) {P}_{T|V}(t|v)} \text{ }\forall v\in \mathcal{V}, \forall t \in {\mathcal{T}}, \label{con1}\\
&\int_{\mathcal{V}} {P_{Y^\dagger,V,T}(y^\dagger,v,t)}({u(y^\dagger,v; \theta_{u})}-{u(k^{\dagger},v; \theta_{u})})dv \geq 0 \text{ } \forall y^\dagger\in \mathcal{Y}, \forall k \in \mathcal{Y}\setminus\{y^\dagger\}, \forall t \in {\mathcal{T}}, \label{con1a}\\
&P_{Y^\dagger, V, T}(y^\dagger ,v, t)\geq 0 \text{ }\forall y^\dagger \in \mathcal{Y},  \forall v \in \mathcal{V}, \forall t \in \mathcal{T} \text{, } \sum_{y^\dagger \in \mathcal{Y}} \int_{\mathcal{V}} \int_{\mathcal{T}} P_{Y^\dagger, V, T}( y^\dagger ,v, t) dt dv =1.\label{con2}
\end{align}
\end{prop}
In Proposition \ref{count1}, (\ref{con1})-(\ref{con2}) define a 1BCE of  $\{G(\theta), S\}$. In particular, (\ref{con1}) is the {\it Consistency} constraint,  (\ref{con1a}) is the {\it Obedience} constraint, and (\ref{con2}) ensures that $P_{Y^\dagger,V,T}$ is a proper distribution. Observe that computing $\Psi^{\ell}(h;\theta)$ and $\Psi^{u}(h;\theta)$   for a given $\theta$ requires us to solve two linear programs with respect to $P_{Y^\dagger,V,T}$. Further,  Proposition \ref{count1} presumes that the identified set, $\Theta^*$, has been constructed in a pre-step. One could also append the program of Proposition \ref{count1} to (\ref{lin_pr0}), solve a unique program,  and thus find $ \mathcal{P}^*_{Y^\dagger}(h)\times \Theta^*$ in one step. However, this strategy would not bring notable computational advantages because the programs of Proposition \ref{count1} and (\ref{lin_pr0}) are not linear in  $\theta$.

Understanding how information changes  behaviour is a key question in the literature on single-agent decision problems (see, for example, \hyperlink{Athey}{Athey, 2002}) and is largely absent in the literature on many-player games. In particular, as discussed in Section \ref{intro}, several empirical papers are concerned with assessing the impact on choices of sending agents information about payoff-relevant variables. These papers typically exploit field experiments which can be very costly.  Proposition \ref{count1} represents a powerful result because it allows the analyst to assess the effect of programs of information provision {\it prior} to conducting such interventions. In the empirical application of Section \ref{empirical}, we apply Proposition \ref{count1} to the benchmark case where the policy intervention fully reveals the state of the world  to agents, i.e., $S$ is the complete information structure and, therefore, $S^\dagger=S$ for each DM.\footnote{Any expansion $S^\dagger$ of the complete information structure $S$ is equal to $S$.}

The complete information benchmark is  also useful to evaluate the welfare cost of limited information which may keep agents from choosing their first best. Specifically, given the true parameter vector $\theta_0\in \Theta$, we consider the average gains in the model-implied  {\it ex-post} utility  from expanding every DM's information structure from the null information structure to the complete information structure:  
\begin{equation}
\label{welfare}
\begin{aligned}
\Delta \mathbb{E}_{\theta_0}\coloneqq &  \displaystyle \bigintss_{\mathcal{V}} u\Big(\text{argmax}_{y\in \mathcal{Y}} u(y,v; \theta_{0,u}),v\Big)  P_V(v;\theta_{0,V}) dv\\
& - \displaystyle \bigintss_{\mathcal{V}}  u\Big(\text{argmax}_{y\in \mathcal{Y}} \int_{\mathcal{V}} u(y,v; \theta_{0,u}) P_V(v; \theta_{0,V}) dv, v\Big)  P_V(v;\theta_{0,V}) dv,
\end{aligned}
\end{equation}
where the first term is the average ex-post utility when all agents process the complete information structure and the second term is the average ex-post utility when all agents process the null information structure. 
Observe that, in the complete information scenario, each DM  earns an ex-post payoff that is, on average, greater or equal than the ex-post payoff under the null information structure. Therefore, $\Delta \mathbb{E}_{\theta_0}$ captures the maximum welfare cost of limited information. 
The identified set of $\Delta \mathbb{E}_{\theta_0}$ is 
$ \cup_{\theta\in \Theta^*} \Delta \mathbb{E}_{\theta}$.

\subsubsection{Changes in covariates}
The second policy experiment considers a more traditional, yet important, question that interests the literature on both single-agent decision problems and games.
In particular, we  study how the choice probabilities  change in response to changes
 in  covariates $X$ entering the utility function and   observed by the researcher and the DMs.   
 Suppose the policy maker implements some intervention which shifts the realization of $X$ assigned to agents. In this new environment, the DM   processes the same information structure as in the factual scenario (i.e., differently from Proposition \ref{count1}, information structures are now held fixed), but has to account for the new realization of $X$ in evaluating their payoffs. Proposition \ref{count2}  characterises   sharp  bounds on the  counterfactual choice probabilities.

Before presenting Proposition \ref{count2}, we introduce some useful notation and provide the intuition behind the result. Let $\mathcal{X}$ be the finite support of $X$. Let $X^\dagger$ be  the  vector of covariates after the intervention, with finite support $\mathcal{X}^\dagger$. 
To compute the counterfactual choice probabilities, imagine a hypothetical scenario where the DM simultaneously chooses  an alternative from $ \mathcal{Y}$ for the augmented decision problem $\{G(\theta_0, X), S\}$  and an alternative from $   \mathcal{Y}$ for the augmented decision problem $\{G(\theta_0, X^\dagger), S\}$. The choices  for $\{G(\theta_0, X), S\}$ and $\{G(\theta_0, X^\dagger), S\}$ are denoted by $Y$ and $Y^\dagger$, respectively.  $Y^\dagger$ and $S$ are not observed by the researcher. There is no interaction between the two choices except for the common information structure.  $\{G(\theta_0, X), G(\theta_0, X^\dagger), S\}$ is called the {\it linked augmented decision problem}. $\{G(\theta_0, X), G(\theta_0, X^\dagger)\}$ is called the {\it linked baseline decision problem}.  Let $\mathcal{P}_{Y,Y^\dagger,V| X, X^\dagger}$ be a  1BCE of the linked  baseline decision problem and consider the set of such 1BCEs. By Theorem \ref{main} above, the marginal of these distributions on $(Y^\dagger, V)$ is precisely the set of counterfactual choice probabilities that we are looking for. These steps are formalised by Theorem 1  in \hyperlink{Bergemann_Brooks_Morris}{Bergemann, Brooks, and Morris (2022)} and readapted to our case by Proposition \ref{count2}.

\begin{prop}{\normalfont({\itshape Counterfactual bounds - change in covariates})}
\label{count2}
Let $ \mathcal{Q}^*_{Y^\dagger|x, x^\dagger}(h)$ denote the identified set of the   counterfactual probability of choosing alternative $h\in \mathcal{Y}$ when   DMs face the augmented decision problem $\{G(\theta_0, x^\dagger), S\}$, where  $S$  is the unknown information structure processed  in the factual scenario and potentially heterogenous across DMs, and $x\in \mathcal{X}$ and $x^\dagger\in \mathcal{X}^\dagger$ are the covariate realizations before and after the policy intervention, respectively. Then, 
$
 \mathcal{Q}^*_{Y^\dagger|x, x^\dagger}(h)= \cup_{\theta \in \Theta^*} [ \Phi^{\ell}(h| x, x^\dagger; \theta),  \Phi^{u}(h| x, x^\dagger; \theta)]
$, 
where 
$$
\begin{aligned}
&  \Phi^{\ell}(h| x, x^\dagger; \theta)\coloneqq     \min_{P_{Y,Y^\dagger,V| X, X^\dagger}(\cdot| x, x^\dagger): \mathcal{Y}^2\times \mathcal{V}\rightarrow \mathbb{R}}  \sum_{y\in \mathcal{Y}}\int_{\mathcal{V}}P_{Y,Y^\dagger,V| X, X^\dagger}(y,h, v| x, x^\dagger) dv.    \text{ s.t. } (\ref{con3})-(\ref{con4}),\\
& \Phi^{u}(h| x, x^\dagger; \theta) \coloneqq     \max_{P_{Y,Y^\dagger,V| X, X^\dagger}(\cdot| x, x^\dagger): \mathcal{Y}^2\times \mathcal{V}\rightarrow \mathbb{R}}  \sum_{y\in \mathcal{Y}}\int_{\mathcal{V}}P_{Y,Y^\dagger,V| X, X^\dagger}(y,h, v| x, x^\dagger) dv.    \text{ s.t. } (\ref{con3})-(\ref{con4}),
\end{aligned}
$$
and
\par\nobreak
\vspace{-0.7cm}
{\footnotesize
\begin{align} 
&   \sum_{y\in \mathcal{Y}} \sum_{y^\dagger\in \mathcal{Y}}  {P_{Y,Y^\dagger,V| X, X^\dagger}(y,y^\dagger, v| x, x^\dagger)}={P_{V}(v| x, x^\dagger; \theta_{0,V})} \text{ }\forall v\in \mathcal{V}, \label{con3}\\
& \int_{\mathcal{V}} {P_{Y,Y^\dagger,V| X, X^\dagger}(y,y^\dagger, v | x, x^\dagger)}({u(y,v, x; \theta_{0,u})}-{u(y',v, x; \theta_{0,u})})dv \geq 0,  \forall y\in \mathcal{Y}, \forall y'\in \mathcal{Y}\setminus\{y\}, \forall y^\dagger \in \mathcal{Y}, \label{con3a}\\
& \int_{\mathcal{V}} {P_{Y,Y^\dagger,V| X, X^\dagger}(y,y^\dagger, v | x, x^\dagger)}({u(y^\dagger,v, x^\dagger; \theta_{0,u})}-{u(k^\dagger,v, x^\dagger; \theta_{0,u})})dv \geq 0,  \forall y^\dagger\in \mathcal{Y}, \forall k^\dagger\in \mathcal{Y}\setminus\{y^\dagger\}, \forall y \in \mathcal{Y}, \label{con3b}\\
& P_{Y,Y^\dagger,V| X, X^\dagger}(y,y^\dagger, v| x, x^\dagger)\geq 0 \text{ } \forall y\in \mathcal{Y}, \forall y^\dagger\in \mathcal{Y}, \forall v \in \mathcal{V},\sum_{y\in \mathcal{Y}}\sum_{y^\dagger\in \mathcal{Y}} \int_{\mathcal{V}}   P_{Y,Y^\dagger,V| X, X^\dagger}(y,y^\dagger, v| x, x^\dagger)  dv=1, \label{con3c}\\
& \sum_{y^\dagger\in \mathcal{Y}} \int_{\mathcal{V}} P_{Y,Y^\dagger,V | X, X^\dagger}(y,y^\dagger,v|x, x^\dagger)dv =\mathbb{P}_Y(y|x) \text{ }\forall y \in \mathcal{Y}.\label{con4}
\end{align}}
\end{prop}
In proposition \ref{count2}, (\ref{con3})-(\ref{con4}) define a 1BCE of  the linked baseline decision problem $\{G(\theta, X), G(\theta, X^\dagger)\}$. In particular,  (\ref{con3}) is the {\it Consistency} constraint,  (\ref{con3a}) and  (\ref{con3b}) are the {\it Obedience} constraints for each decision problem, (\ref{con3c}) ensures that $P_{Y, Y^\dagger,V|x, X^\dagger}$ is a proper distribution, and (\ref{con4}) imposes that the factual choice probabilities are equal to the empirical ones. 

In the empirical application of Section \ref{empirical}, we apply Proposition \ref{count2} to study how the well-being of UK parties changes when they modify their ideological positions on popular policy issues.

%%%%%%%%%%%%%%%%%%%%%%%%%%%%%%%%%%%%%%%%%%%%%%%%%%%%%%%%%%%%%%%%%%%%%%%%%%%%%%%%%%%%
\section{Simulations}
\label{simulations}
We consider a model specification close to the one used in the empirical application of Section \ref{empirical}. The payoff function is
\begin{equation}
\label{probit_model}
u(y, X_i, V_i; \beta)\coloneqq 
\beta X_{iy}+  V_{iy},
\end{equation}
where $i$ indexes a generic DM,  $\mathcal{Y}\coloneqq \{0,1,\dots, D\}$, $0$ denotes the outside option and its utility is normalised to zero, $X_{iy}$ and $V_{iy}$ are DM-alternative specific features. DM $i$ observes the realization of $X_i\coloneqq (X_{i1},..., X_{iD})$ but may be uncertain about the realization of $V_i\coloneqq (V_{i1},..., V_{iD})$.  $X_i$ and $V_i$ are assumed to be independent. DM $i$ has a prior on $V_i$, which is assumed to be standard Normal. The researcher observes the choice made by DM $i$ and the realization of   $X_i$ for a large sample of DMs, without knowing their information structures.  Note that this framework reduces to a standard multinomial Probit model under the additional assumption that each DM processes the complete information structure. 

In the first simulation exercise, we illustrate how to choose the order of the Bernstein polynomials, $K$. As developing data-driven procedures to choose  
$K$ is still an open question in nonparametric frameworks with point identification, here we follow the heuristic approach developed by   \hyperlink{Han}{Han and Yang (2023)} for partially identified settings. We simulate the data from  (\ref{probit_model}) with $|\mathcal{Y}|=3$, $D=2$, and $\beta=1.3$. Each $X_{iy}$ is randomly drawn from
a probability mass function constructed by taking a bivariate normal and then discretising it to have support $\mathcal{X}\coloneqq \{-2.4, -0.4, 0.3\}$.\footnote{That is, $\Pr(X=x)=\frac{\exp(-(x-\mu)^2/\sigma^2)}{\sum_{x\in \mathcal{X}} \exp(-(x-\mu)^2/\sigma^2)}$ for each $x\in \mathcal{X}$.} The data is generated assuming all DMs process the complete information structure.  
The first column of Table \ref{sim1} reports the identified set    as $K$  increases. To obtain such identified set, we explored a grid of candidate values of $\beta$ between -30 and 30 equally distanced at 0.001. The second and third columns  report  $K_d$, which is taken to be constant across $d=1,\dots, D$,  and $K\coloneqq(K_d+1)^{D}$, respectively.  For a given covariate realization $x$ and parameter value $\beta$, the fourth column computes the number of unknowns of the linear program (\ref{lin_pr_3}), which is $(D+1) K$. Observe that the width of the bounds tends to increase weakly with $K$. When $K_d=3$  ($K =64$), the bounds are narrow, but this may be due to our misspecification of the smoothness of the family of functions to be approximated, $\mathcal{P}_{Y|V}$. When $K_d$ is sufficiently large, the width increases at a slower rate, and the bounds start to converge. In particular, the bounds become stable from $K_d=10$    ($K =1,331$) onwards. Therefore,  we set $K_d=10$ for $d=1,\dots, D$   in  the  next simulations.\footnote{\label{0_footnote}When the data are generated from (\ref{probit_model}) with $X_i$ exogenous and  $V_i$ distributed as a standard Normal, then $\beta=0$ is expected to be part of the identified set. To see why, suppose $\beta=0$ and each DM processes the null information structure, so that the posterior is equal to the prior. Then, given the standard Normal prior, each DM gets the same expected payoff, $\mathbb{E}(V_{iy})=0$, from every alternative $y\in \mathcal{Y}$ and chooses according to some tie-breaking rule. Clearly, there will be a tie-breaking rule that reproduces the data. Hence, $\beta=0$, coupled with the null information structure for all DMs and some tie-breaking rule,  generates the data. Nevertheless, the model maintains enough identification power to exclude negative values of $\beta$, as shown in Table \ref{sim1}.}

 The fifth column of Table \ref{sim1} shows the average CPU time to assess if (\ref{lin_pr_3}) has a solution for   a given $(x,\beta)$, using the MOSEK solver for Matlab. The CPU time includes the calculation of the integral $\gamma^{y,y',x}_{1,k,K}(\beta)\coloneqq \int_{\mathcal{V}} a_{k,K}(v)P_V(v)(u(y,x, v;\beta)-u(y', x, v; \beta)) dv$  for each $k\in \mathcal{K}\coloneqq \times_{d=1}^D \{0,1,\dots, K_d\}$ and $y,y'\in \mathcal{Y}$.\footnote{Recall that $P_V$ is assumed to be standard normal, with $V_i$ independent of $X_i$. Hence, $\gamma^{y,y'}_{2,k,K}\coloneqq \int_{\mathcal{V}} a_{k,K}(v) P_v(v)dv$ does not vary across values of the covariates and parameters and can be computed once.} These integrals are computed by Monte Carlo integration taking $10^4$ random draws from the standard Normal distribution.
The number of unknowns of  (\ref{lin_pr_3})  increases  linearly in $K$ and exponentially in $K_d$. Hence, the   CPU time increases approximately linearly in $K$ and exponentially in $K_d$.
The total time   to construct the  identified set   depends on the possibility of parallelising across $(x,\beta)$. Using a computing cluster, we could exploit 600 parallel workers by coding our procedure as an SGE array job. Based on those, the rough total CPU time is listed in the last column of Table \ref{sim1}.

\begin{table}[h!]
  \captionsetup{font=normal}
\caption{Identified set   in the first simulation exercise.}
\label{sim1}
\centering
{\footnotesize
\begin{tabular}[t]{c c c c c c}
\toprule
Identified set & $K_d$, $d=1,\dots, D$  & Order polynomial  & Unknowns in (\ref{lin_pr_3}) & CPU time per $(x,\beta)$ & Total CPU time  \\
&  &$K\coloneqq(K_d+1)^{D}$  & $(D+1) K$ &  &  \\
\midrule
$[0, 1.321]$ & 3 & 16 & 48& 0.046 s & 42 s  \\
$[0, 1.402]$ & 5 & 36 & 108 &0.128 s & 115 s \\
$[0, 1.504]$ & 7 & 64 &192 &0.226  s & 204 s\\
$[0, 1.565]$ & 10 & 121  &363& 0.463 s & 7 min \\
$[0, 1.565]$ & 15 & 256 &7688 &0.923  s & 14 min \\
$[0, 1.565]$ & 20 & 441 & 1,323&1.539 s & 23 min \\
$[0, 1.565]$ & 25 & 676 & 2,028&2.762 s & 41 min \\
$[0, 1.565]$ & 30 & 961 & 2,883&3.561 s & 53 min \\
\bottomrule
\end{tabular}
		\begin{tablenotes}
			\item   \footnotesize \textit{Note}:     The true value of $\beta$ is 1.3. $D=2$. The CPU time per $(x,\beta)$ includes the calculation of $\{\gamma^{y,y',x}_{1,k,K}(\beta)\}$ using $10^4$  random draws from the standard Normal distribution. The total CPU time is based on 600 parallel workers, a grid of candidate values of $\beta$ between -30 and 30 equally distanced at 0.001, and 9 possible realizations of $X_i$.
		\end{tablenotes}}
\end{table}
% \vspace{1cm}

% \begin{figure}[!htbp]
	%\centering
	%	\includegraphics[scale=0.3]{choice_K.jpg}
	%\caption{Una figura.}
	%\label{choice_K}
%\end{figure}

 In the second simulation exercise, we investigate how the identified set   varies as the cardinality of the support of  $X_i$ increases. We generate data   from  (\ref{probit_model}) with $|\mathcal{Y}|=3$, $D=2$,  $\beta=1.3$,  and all DMs processing the complete information structure. We distinguish two scenarios regarding the dependence between $X_i$ and $V_i$. In the first scenario, $X_i$ and $V_i$ are independent and the prior on $V_i$ is a standard normal, as in the first  simulation exercise. In the second scenario, $X_i$ and $V_i$ are allowed to be correlated and the prior on $V_i$ conditional on $X_i=x$ is a normal distribution with mean and variance that vary with $x$. 
For each scenario, we study three cases. In the first case, each $X_{iy}$ is randomly drawn from a probability mass function  constructed by taking a bivariate normal and then discretising it to have support $\mathcal{X}\coloneqq\{-2.4, -0.4, 0.3\}$,   as in the first   simulation exercise.  In the second case, $\mathcal{X}\coloneqq\{-2.4, -0.5, -0.4, 0.1, 0.3\}$.  In the third case, $\mathcal{X}\coloneqq\{-2.4, -0.7, -0.5, -0.4, 0.1, 0.2, 0.3\}$.  We set $K_d=10$ for $d=1,\dots, D$ to construct the identified set.
   For each $\mathcal{X}$, the second and third columns of Table \ref{sim3} show the identified set    in the first and second scenarios, respectively.   When $X_i$ is exogenous, we find that the identified set shrinks as the cardinality of $\mathcal{X}$ increases. This is because variation in  exogenous covariates induces variation in agents' choices, which   helps the identification of $\beta$, as in standard parametric analysis. Conversely, when $X_i$ and $V_i$ are allowed to be correlated, 
 the identified set  becomes larger as the cardinality of $\mathcal{X}$ increases. This is because, as the cardinality of $\mathcal{X}$ increases,  there are smaller groups
of  DMs with the same prior, which worsens the identification of $\beta$ by ``weakening'' the {\it Consistency} requirement of 1BCE.\footnote{When the data are generated from (\ref{probit_model}) and the prior varies across realizations of $X_i$, $\beta=0$ may not be part of the identified set. To see why the logic of Footnote \ref{0_footnote} does not apply, suppose $\beta=0$ and each DM processes the null information structure, so that the posterior is equal to the prior. If $X_i=x$, then DM $i$ gets an expected payoff equal to $\mathbb{E}(V_{iy}|X_i=x)\coloneqq \mu_{y,x}$ from every alternative $y\in \mathcal{Y}\setminus \{0\}$, and a payoff equal to 0 from the outside option. Hence, the DMs' choices will differ depending on the realization of the covariates and  the tie-breaking rule adopted. In turn, the resulting distribution of choices may not coincide with the empirical one and, so, $\beta=0$, may not belong to the identified set.}

   \begin{table}[h!]
     \captionsetup{font=normal}
\caption{Identified set   in the second simulation exercise.}
\label{sim3}
\centering
\begin{tabular}[t]{ l c c}
\toprule
Support of $X_{iy}$ &  Identified set &   Identified set     \\
& $(X_i, V_i)$ independent & $(X_i, V_i)$ correlated \\
\midrule
$\mathcal{X}\coloneqq\{-2.4, -0.4, 0.3\}$   & $[0, 1.565]$ & [0.548, 1.317]   \\
$\mathcal{X}\coloneqq\{-2.4, -0.5, -0.4, 0.1, 0.3\}$   & $[0, 1.505]$ &  [0.494, 1.448]\\
$\mathcal{X}\coloneqq\{-2.4, -0.7, -0.5, -0.4, 0.1, 0.2, 0.3\}$ & $[0, 1.474]$ &[0.479, 1.627] \\
\bottomrule
\end{tabular}
		\begin{tablenotes}
			\item \hspace{0.2cm} \footnotesize \textit{Note}:     The true value of $\beta$ is 1.3.
		\end{tablenotes}
\end{table}

 In the third simulation exercise, we investigate how the identified set  changes as the information structures processed by DMs in the underlying data generating process vary.  We generate data   from  (\ref{probit_model}) with  $|\mathcal{Y}|=3$, $D=2$,     $\beta=1.3$, $X_i$ independent of $V_i$,  each $X_{iy}$   randomly drawn from a probability mass function  constructed by taking a bivariate normal and then discretising it to have support $\mathcal{X}\coloneqq\{-2.4, -0.4, 0.3\}$, and $V_i$ distributed as a standard Normal. We set $K_d=10$ for $d=1,2$ to construct the identified set. 
 We consider three scenarios. In the first scenario,  all DMs process the complete information structure, as in the first simulation exercise. In the second scenario,  5/6 of the DMs process the complete information structure, and 1/6 process the null information structure. In the third scenario, all DMs process the null information structure. For each scenario, Table \ref{sim2} shows the identified set   (second column) and the value of $\beta$ that would be identified if the researcher assumed  that all DMs process the complete information structure (third column). Assuming that all DMs process the complete information structure, as in a standard multinomial Probit model, leads to
recovering one parameter value contained in the identified set. When this assumption is
misspecified (second and third scenario), the recovered parameter value differs from the truth. The model has the least identifying power in the third scenario, when all DMs process the null information structure. As soon as a significant
proportion of DMs process the complete information structure (first and second scenarios), the identifying power of the model
improves. Note, in fact, that, under the first scenario, the DMs take their decisions
based on the actual payoffs. Instead, under the third scenario, the DMs  choose based on the expected
payoffs which are relatively homogenous in the population because computed using the same posterior. Hence, under the third scenario, there is less variation in the
DMs' choices, which leads to wider bounds.

 \begin{table}[h!]
   \captionsetup{font=normal}
\caption{Identified set   in the third simulation exercise.}
\label{sim2}
\centering
\begin{tabular}[t]{ l l  l}
\toprule
Information structure & Identified set   & $\beta^\text{com}$ \\
\midrule
Complete  & $[0, 1.565]$   & 1.3 \\
Mixture & $[0, 1.706]$   & 1.632 \\
Null & $[0, \infty)$   & 15.186 \\
\bottomrule
\end{tabular}
		\begin{tablenotes}
			\item \hspace{3.5cm} \footnotesize \textit{Note}:     The true value of $\beta$ is 1.3.
		\end{tablenotes}
\end{table}

%%%%%%%%%%%%%%%%%%%%%%%%%%%%%%%%%%%%%%%%%%%%%%%%%%%%%%%%%%%%%%%%%%%%%%%%%%%%%%%%%%%%
\section{Empirical application}
\label{empirical}
 
In this section, we use our methodology to study the determinants of voting behaviour during the UK general election held on 8 June 2017 and perform some counterfactual exercises aiming to evaluate the impact of uncertainty on the well-being of voters and parties. 

\subsection{Setting and model specification}
\label{model_spec}
The spatial model of voting is a dominant framework in political economy to explain individual preferences for parties and, in turn, how such preferences shape the policies implemented by democratic societies (\hyperlink{Downs}{Downs, 1957};  \hyperlink{Black}{Black, 1958}).
This model posits that an agent has a most preferred policy (also called ``bliss point'') and casts their vote in favour of the party whose position is closest to their ideal (i.e., she votes ``ideologically''). 
In empirical analysis, it is typically implemented by estimating a classical parametric discrete choice model with perfect information. That is, it is assumed that each DM $i$ processes the complete information structure and votes for party $y\in \mathcal{Y}$ maximising their utility, 
\begin{equation*}
\label{appl}
u(y, X_i, V_i; \theta_u)\coloneqq\beta_y^\top Z_{iy} +\gamma_y^ \top W_i+V_{iy}, 
\end{equation*}
where $Z_{iy}\coloneqq|Z_{i}-Z_{y}|$ is an $M\times 1$ vector observed by the researcher, representing the distance between DM $i$'s opinion ($Z_i$) and party $y$'s opinion ($Z_y$) on $M$  issues, as measured in some common $M$-dimensional ideological metric space. $W_i$ is a   vector of individual-specific covariates observed by the researcher. $X_i\coloneqq(W_i, Z_{iy}:  y \in \mathcal{Y})$ collects the ideological distances and the individual-specific covariates. $V_i \coloneqq (V_{iy}:  y \in \mathcal{Y})$ is a vector of tastes of DM $i$ for each party/candidate that is unknown to the researcher and independent of $X_i$, whose distribution belongs to a parametric family, thereby outlining a (Multinomial) Logit model, Probit model, Nested Logit model, etc.  If voters vote ideologically, then each $\beta_y$ is expected to be negative so that  DM $i$'s utility declines with increasing distance between $Z_i$ and $Z_y$.  $\theta_u\coloneqq (\beta_y, \gamma_y: y \in \mathcal{Y})$ is the vector of payoff parameters.

The above framework is scientifically appealing because of its elegance and simplicity but it has limitations. Importantly,  uncertainty affects voting (\hyperlink{Tabellini2}{Mat\u{e}jka and Tabellini, 2021}, and other references in Section \ref{intro}). That is, voters may be unsure about their  own and  the parties' ideological positions and, more generally, about the qualities of the candidates.  This is because of the inevitable difficulty of making precise political judgments and understanding associated returns, or because the parties deliberately obfuscate information   to attract voters with different preferences and expand electoral support. More plausibly, in the wake of election campaigns, voters are conscious of their  own and the parties' attitudes towards some popular issues, but might be uncertain about how they themselves and the parties stand towards more technical or less debated topics, and about the traits of the candidates other than those publicly advertised. Further, they may attempt to fill such gaps in information with various degrees of success and in different ways,  depending on a priori inclination for certain parties, political sentiments, interest in specific issues, civic sense,  attentional limits, participation in  townhall debates, candidates'  transparency, opinion makers, and media exposure.
In turn, some individuals might become much more informed, others less, giving rise to heterogeneity in the public understanding of politics.

Despite the acknowledgement of the central role played by the sophistication of voters in determining voting patterns, only a few empirical works have attempted to take it into account while estimating  a spatial voting framework. 
This has been done, for instance, through an additive, exogenous, and parametrically distributed error in the payoffs representing the evaluation mistakes made by voters,  a parametric specification of the variance of the perceived party position across voters, or a parametric specification of the probability  of being informed versus uninformed when voting  (\hyperlink{Degan_Merlo2}{Degan and Merlo, 2011}, and other references in Section \ref{intro}). By contrast, our methodology permits us to incorporate uncertainty under weak assumptions on the latent, heterogeneous, and potentially endogenous process followed by voters to gather and evaluate information.

In particular, we   assume that, when assessing the returns to voting for   party $y\in \mathcal{Y}$, DM $i$ is:
\begin{itemize}[leftmargin=*]
\item[-] Aware of the distances between their position and party $y$'s  position  on highly debated topics (for instance, EU integration). These distances are captured by the vector $Z_{iy}\coloneqq|Z_i - Z_{y}|$. 
\item[-] Potentially uncertain about their tastes towards party $y$'s opinion on more complicated and less media-covered issues (for instance,  public expenditure management and reactions to pandemics) and towards the candidates' qualities that  have been less advertised (for instance, disclosure of assets, liabilities, and any conflict of interests). These tastes are captured, in some aggregate way, by the vector  $V_i \coloneqq (V_{iy}:  y \in \mathcal{Y})$. DM $i$ has a prior on $V_{i}$. Further, DM $i$ has access to a learning technology that allows them to become more informed about $V_{i}$. 
\end{itemize}

We follow the literature on voting under uncertainty which typically models priors as normal distributions (\hyperlink{Knight}{Knight and Schiff, 2010};    \hyperlink{Tabellini2}{Mat\u{e}jka and Tabellini, 2021}; \hyperlink{Yuksel}{Yuksel, 2022}). In particular, we assume that $V_i$ is distributed as a standard normal, independent of  $X_i$.  See also \hyperlink{Feddersen}{Feddersen and Pesendorfer (1997)} and \hyperlink{McMurray}{McMurray (2013)} on the use of the common prior assumption and Bayesian rationality in models of voting. Lastly,   we consider abstention as the base category and normalise its payoff to zero as, for example, in \hyperlink{Knight}{Knight and Schiff (2010)}. 

%%%%%
\subsection{Data}
We estimate our model by using data on the UK general election held on 8 June 2017. Specifically, we use data from the British Election Study, 2017: Face-to-Face Post-Election Survey (\hyperlink{Fieldhouse}{Fieldhouse, et al., 2018}). The survey took place immediately after the election. It asks questions concerning key contemporary problems about political representation, accountability, and engagement, and aims to  explain changes in party support.  The interviewees constitutes an address-based random probability sample of eligible voters living in 468 wards in 234 Parliamentary Constituencies across England, Scotland, and Wales.  

We believe that such data fit  the framework  described in Section \ref{model_spec} for four main reasons. First, the UK parties were clearly focused on the topic of Brexit, along with issues of public health and austerity, thus inducing potential uncertainty among voters with respect to many other factors (\hyperlink{Hutton}{Hutton, 2017}; \hyperlink{Snowdon}{Snowdon and Demianyk, 2017}). Second, the UK political scene is dominated by historical parties. Hence, past election outcomes and consequent behaviour of parties can justify the common prior assumption on $V_i$.  Third, the survey reports  the positions of the  respondents on topics that were debated at length before the election, which we discuss more precisely below. The survey also asks respondents to state the parties's positions with respect to those topics and the answers provided are substantially aligned. This suggests that there was no uncertainty among voters on those topics. Hence, they can be used to construct the vector $Z_{iy}$ for each party $y \in \mathcal{Y}$, whose realization is assumed to be in the information set of voters. The survey does not contain data on other relevant factors that might have induced uncertainty among voters.
Hence, it is natural to treat the realization of $V_i$ as unobserved by the researcher.  
Fourth, the survey asks respondents to declare if they voted tactically. Only $2.16\%$ of the respondents answer affirmatively. We drop them from our final sample, in order for the assumption that voters vote ideologically to apply.

To limit the impact of Scottish and Welsh independentist fronts on our results, we focus on the respondents who reside in England. We consider the answers  of respondents on which party they have voted for among the Conservative Party, Labour Party, Liberal Democrats, United Kingdom Independence Party (UKIP), Green Party, and none.

%\footnote{The original questionnaire includes among the possible answers also the Scottish National Party (i.e, the Scottish nationalist social-democratic party in Scotland), Plaid Cymru (i.e., the Welsh nationalist social-democratic party in Wales), other unspecified minor parties, and ``Refused to declare''. None of the respondents who reside in England have voted for the Scottish National Party. Only one of the respondents who reside in England has voted for Plaid Cymru. $4$ respondents who reside in England have voted for other unspecified minor parties. Lastly, $3.52\%$ of the respondents who reside in England have refused to declare who they have voted for. We have dropped all these observations. } 

We collect in $Z_{iy}$ the distances
 between DM $i$'s position  and party $y$'s  position on four dimensions: EU integration, taxation and social care, income inequality, and left-right political orientation.   More precisely, we select the answers of the respondents to the following questions (summarised with respect to the original version, for brevity):
\begin{enumerate}[1.]
\item\text{}[EU integration]: On a scale from $0$ to $10$, do you think that Britain should do all it can to unite fully with the European Union ($0$), or do all it can to protect its independence from the European Union ($10$)? Also provide the positions of the parties on the same scale.
\item\text{}[Taxation and social care]: On a scale from $0$ to $10$, do you think that government should cut taxes a lot and spend much less on health and social services ($0$), or that government should raise taxes a lot and spend much more on health and social services ($10$)? Also provide the positions of the parties on the same scale.
\item\text{}[Income inequality]: On a scale from $0$ to $10$, do you think that government should make much greater efforts to make people's incomes more equal ($0$), or that  government should be much less concerned about how equal people's incomes are ($10$)? Also provide the positions of the parties on the same scale.
\item\text{}[Left-right political orientation] Where would you place yourself on a scale from $0$ to $10$ where $0$ denotes  left political attitudes and $10$ denotes right political attitudes? Also provide the positions of the parties on the same scale.
\end{enumerate}
Following the literature (\hyperlink{Alvarez1995}{Alvarez and Nagler, 1995}; \hyperlink{Alvarez1998}{1998}; \hyperlink{Alvarez2000}{2000}; \hyperlink{Alvarez2000_2}{Alvarez, Nagler, and Bowler, 2000}), we set party $y$'s position on dimensions 1-4 equal to the  median placement of the party on each dimension across the sample, although as noticed above there is substantial alignment among the respondents' answers.

We collect in $W_i$ some demographic characteristics of respondents. In particular, we focus on gender, socio-economic class, and  total income before tax. 
Recall that $\gamma_y$ captures the impact  of $W_i$ on the vote shares.  We allow this impact to be heterogeneous across the parties. To be parsimonious on the number of parameters to estimate, we further parameterise $\gamma_y$ by requiring that $\gamma_y\coloneqq\gamma Z^{\text{LR}}_y$ for every party $y$, where $Z^{\text{LR}}_y$ is the position of party $y$ with regards to left-right orientation. In other words, we assume that the aforementioned heterogeneity  is driven by the position of each party in the left-right political spectrum. Similarly, to reduce dimensionality, we impose $\beta_y\coloneqq \beta$ for each party $y$.

In our final sample, $36.48\%$ of people have voted for the Labour Party,  $36.65\%$ for the Conservative Party,  $6.41\%$ for the Liberal Democrats,  $1.73\%$ for UKIP, $1.56\%$ for the Green Party, and $17.17\%$ did not vote. Table \ref{descriptive} presents some descriptive statistics. The second column refers to the positions of the respondents on dimensions 1-4 and reports  the mean (rounded to the nearest integer), median, and standard deviation across the sample.  The remaining columns reports $Z_y$ for each party $y$. As expected, the Conservative Party and UKIP are more right-wing, less concerned with income inequality, more Eurosceptic, and stronger supporters of low taxes and a minimal welfare state, than the Labour Party and the Green Party.  The Liberal Democrats are more centrist. 

%\footnote{Before omitting observations with missing data, the percentages are: $34.74\%$ for the Labour Party,  $33.78\%$ for the Conservative Party,  $5.39\%$ for the Liberal Democrats,  $1.91\%$ for UKIP, $1.63\%$ for the Green Party, and $22.56\%$ did not vote. 
%The survey seems  slightly skewed towards supporters of the Labour Party. The actual vote shares in England were
%$41.9\%$ for the Labour Party,  $45.6\%$ for the Conservative Party,  $7.8\%$ for the Liberal Democrats,  $2.1\%$ for UKIP, $1.9\%$ for the Green Party.  The election results led to a hung parliament and the Conservative Party formed a minority government supported by an agreement with the Northern Ireland's Democratic Unionist Party. The vote shares of each party (including minorities) can be  obtained, for example,  here \url{https://www.bbc.co.uk/news/election/2017/results/england}. See also \url{https://en.wikipedia.org/wiki/Opinion_polling_for_the_2017_United_Kingdom_general_election\#2017}   for opinion polls organised by various organisations to gauge voting intentions.} 

 \begin{table} [ht]
   \captionsetup{font=normal}
   \captionsetup{font=normal}
\centering
{\footnotesize \begin{tabular}{c c c c c c c}
\toprule
                        & Self                                                                   & Conservative & Labour & Lib. Dem. & UKIP & Green\\
                       & (Mean, Median, St.Dev.)                                   &  &  &  &  & \\
\midrule
EU                 &   $5$\text{ }\text{ } $5$ \text{ }\text{ }$3.355$     &$7$    &$4 $     & $3 $   & $10$        & $3$  \\

Social care     &  $7$\text{ }\text{ } $7$ \text{ }\text{ }$2.051$       &$5$     &$7$    &$6$     &$4$           & $6$  \\

Inequality     &   $4$\text{ }\text{ } $4$ \text{ }\text{ }$2.743$       &$6$    &$ 3 $    &$4$     & $5$            &$3$\\

Left-right         &  $5$\text{ }\text{ } $5$ \text{ }\text{ }$2.059$      &$8$     & $2$     &$5$     & $9$          &  $3$  \\
\bottomrule
 \end{tabular}}
\caption{Descriptive statistics on the ideological positions.}
\label{descriptive}
\end{table}

The sample is gender balanced, with $48.97\%$ of males and $51.03\%$ of females. We assign label $1$ to females and $0$ to males. In the original data, the socio-economic class is divided into seven categories, following the Standard Occupation Classification 2010: professional occupations; managerial and technical occupations; skilled occupations - non-manual; skilled occupations - manual; partly skilled occupations; unskilled occupations; armed forces. To lessen the computational burden, we reorganise these categories into three groups. The first group is assigned label $0$ and collects professional occupations, managerial and technical occupations, skilled occupations - non-manual, and armed forces ($68.04\%$ of the sample). The second group is assigned label $1$ and collects skilled occupations - manual and partly skilled occupations ($29.42\%$ of the sample). The third group is assigned label $2$ and collects unskilled occupations ($2.54\%$ of the sample). Similarly, in the original data, the total income before tax is bracketed into $14$ categories. We reorganise these categories into four groups, which we construct by approximately  following the UK income tax rates. The first group is for income between $\textsterling 0$ and $\textsterling15,599$ ($21.78\%$ of the sample).  The second group is for income between $\textsterling15,600$ and $\textsterling49,999$ ($51.68\%$ of the sample). The third group is for income between $\textsterling50,000$ and $ \textsterling99,999$ ($21.45\%$ of the sample). The fourth group is for income above $\textsterling 100,000$ ($5.09\%$ of the sample). To each of the four groups, we assign as value the logarithm of the median income across the respondents belonging to that group ($9.4727$, $10.4282$, $11.1199$, and $12.6115$, respectively).  We summarise these numbers in Table \ref{descriptive2}.\\
 \begin{table} [ht]
   \captionsetup{font=normal}
\centering
{\footnotesize \begin{tabular}{c c c}
\toprule
Gender & Socio-economic class & Income (in log)\\
\midrule
Males (0): $48.97\%$      & First group ($0$):  $68.04\%$         & First group ($9.4727$): $21.78\%$\\
Females (1): $51.03\%$  & Second group ($1$):  $29.42\%$    & Second group ($10.4282$): $51.68\%$\\ 
                                   & Third group ($2$):  $2.54\%$          & Third group ($11.1199$): $21.45\%$\\
                                   &                                                        & Fourth group ($12.6115$): $5.09\%$\\
   \bottomrule                            
 \end{tabular}}
\caption{Descriptive statistics on the demographic characteristics.}
\label{descriptive2}
\end{table}

%%%%%%%%%%%%%%%%%%%%%%%%%%%%%%%%%%%%%%%%%%%%%%%%%
%%%%%%%%%%%%%%%%%%%%%%%%%%%%%%%%%%%%%%%%%%%%%%%%%
%%%%%%%%%%%%%%%%%%%%%%%%%%%%%%%%%%%%%%%%%%%%%%%%%
%%%%%%%%%%%%%%%%%%%%%%%%%%%%%%%%%%%%%%%%%%%%%%%%%
%%%%%%%%%%%%%%%%%%%%%%%%%%%%%%%%%%%%%%%%%%%%%%%%%
\subsection{Results}
\label{result}
We estimate the identified set of the $1\times 7$ vector $\theta_0 \coloneqq (\beta_0, \gamma_0)$ by evaluating a grid of 100,000  parameter values. The grid is constructed by exploring the parameter space, $\Theta\subseteq \mathbb{R}^7$, via the simulated annealing algorithm.\footnote{In the case of a high-dimensional vector of parameters, it is common in the partial identification literature to construct the grid of parameter values  by using the simulated annealing algorithm   (\hyperlink{CT}{Ciliberto and Tamer, 2009}). In particular, we proceed in four steps. First, we estimate  $\theta_0$ by maximum likelihood under the assumption that all DMs process the complete information structure and obtain  the estimate $\hat{\theta}^\text{com}$. Second, we construct an Halton set of $10^6$ points around $\hat{\theta}^\text{com}$. We draw 100 points at random from this set. We stack these points, together with $\hat{\theta}^\text{com}$, in an $101\times 7$ matrix, $A$. Third, we minimise the test statistic $\text{TS}_n(\theta)$ defined in Appendix \ref{inference} with respect to $\theta$ by running the simulated annealing algorithm from each row of $A$ as starting point and experimenting at different temperatures. We save every  parameter value encountered in the course of the algorithm.  We stack all the saved parameter values in a matrix $G$. Fourth, we draw 100,000 rows at random from $G$.  Such 100,000 rows constitute our final grid of candidate parameter values.}

To decide the order of the Bernstein polynomials, we take $K_d$ constant across $d=1,\dots, D$ and consider $K_d=3,5,7,10$. $D=5$  because there are five parties and the payoff from abstention is normalised to zero. For each of these four values of $K_d$, we construct an estimate of $\Theta^*$, $\widehat{\Theta}^*$, by replacing the empirical choice probabilities in (\ref{lin_pr_3}) with their sample analogues and solving (\ref{lin_pr_3}) for every realization $x$ of $X_i$ and parameter value $\theta$ in the grid. 
Table \ref{res0} reports the projections of $\widehat{\Theta}^*$. As seen in Section \ref{simulations}, the width of the bounds tends to increase weakly with $K_d$. The bounds become approximately stable from $K_d=5$      onwards. Therefore, in our next computations,  we set $K_d=5$ for $d=1,\dots, D$.  

\begin{table}[h!]
  \captionsetup{font=normal}
\caption{Projections of $\widehat{\Theta}^*$ for different orders of the Bernstein polynomials.}
\label{res0}
\centering
{\scriptsize 
\begin{adjustwidth}{-1.2cm}{}
\begin{tabular}[t]{c c c c c c c c}
\toprule
$K_d$ & $\beta_1$   & $\beta_2$   &   $\beta_3$   & $\beta_4$    & $\gamma_1$    & $\gamma_2$    & $\gamma_3$  \\
$d=1,\dots,D$ & EU                              & Social care                     &  Inequality         &  Left-right                & Gender                   & Class                        &  Income\\
\midrule
3                     &$[-0.3512,0.0055]$     & $[-0.0823,0.0172]$         & $[-0.1078,0.0000]$    &$[-0.3221,0.0003]$  &$[-0.3876,0.3952]$  &$[-0.2483,0.1134]$      &$[-0.1965, 0.4387]$\\ 
5                      &$[-0.3844,0.0827]$     & $[-0.1105,0.0818]$         & $[-0.1345,0.0000]$    &$[-0.3948,0.0084]$  &$[-0.4103,0.4142]$  &$[-0.2895,0.1712]$      &$[-0.2151, 0.4614]$\\ 
7                    &$[-0.3844,0.0829]$     & $[-0.1105,0.0820]$         & $[-0.1345,0.0000]$    &$[-0.3949,0.0084]$  &$[-0.4103,0.4142]$  &$[-0.2897,0.1712]$      &$[-0.2151, 0.4614]$\\ 
10                      &$[-0.3844,0.0829]$     & $[-0.1105,0.0823]$         & $[-0.1345,0.0000]$    &$[-0.3949,0.0084]$  &$[-0.4105,0.4142]$  &$[-0.2897,0.1712]$      &$[-0.2151, 0.4614]$\\ 
\bottomrule
\end{tabular}
\end{adjustwidth}}
\end{table}

Table \ref{res1} provides some computational details, as in Table \ref{sim1} of Section \ref{simulations}. In particular, the fourth column shows the average CPU time to assess if (\ref{lin_pr_3}) has a solution for    a given $(x,\theta)$, using the MOSEK solver for Matlab. The CPU time includes the calculation of the integrals $\{\gamma^{y,y',x}_{1,k,K}(\theta)\}$   by Monte Carlo integration, taking $10^2$ random draws from the standard Normal distribution. The last column of Table \ref{res1} reports the  rough total CPU time to construct $\widehat{\Theta}^*$ based on exploiting 800 parallel workers from a computing cluster.

\begin{table}[h!]
  \captionsetup{font=normal}
\caption{CPU time across different orders of the Bernstein polynomial.}
\label{res1}
\centering
{\small
\begin{tabular}[t]{c c c c c}
\toprule
$K_d$, $d=1,\dots,D$  & Order polynomial                    & Unknowns in (\ref{lin_pr_3}) & CPU time per $(x,\theta)$ & Total CPU time  \\
                                    &$K\coloneqq(K_d+1)^{D}$       & $(D+1) K$                             &                                           &  \\
\midrule
3 & 1,024 & 6,144&  0.036 s & 2 h  \\
5 & 7,776& 46,656&  0.281 s & 12 h  \\
7 & 32,768 & 196,608&  1.191 s & 2 d  \\
10 & 161,051 & 966,306& 6.025 s & 10 d  \\
\bottomrule
\end{tabular}
		\begin{tablenotes}
			\item   {\footnotesize \textit{Note}:  $D=5$.  The CPU time per $(x,\theta)$ includes the calculation of $\{\gamma^{y,y',x}_{1,k,K}(\theta)\}$ using $10^2$  random draws from the standard Normal distribution. The total CPU time is based on 800 parallel workers, a grid of 100,000 candidate values of $\theta$, and 1,174 possible realizations of $X_i$.}
		\end{tablenotes}}
\end{table}

 \begin{table} [h]
  \captionsetup{font=normal}
\centering
{\small \begin{tabular}{c  c  c c c c }
\toprule
                 &                                                 & $\hat{\theta}^\text{com}$ &$C^\text{com}_{0.95}$   & $\widehat{\Theta}^*$ &$C_{0.95}$ \\
\midrule
$\beta_1$  & EU                                            & $ -0.0770$    & $[-0.0998,-0.0541] $               &$[-0.3844,0.0827] $         &$[-0.5858,  0.0827] $ \\
                  &                                                  &$(0.0117)$      &                &                               & \\                                         
$\beta_2$  & Social care                               & $-0.0064$      &$[-0.0402,0.0274] $              & $[-0.1105,0.0818]$          & $[-2.1726,  0.6404]$\\       
                  &                                                  &$(0.0173)$     &                 &                               & \\ 
$\beta_3$  & Inequality                                 &$ -0.0342 $      &$[-0.0613,-0.0070] $              &$[-0.1345,0.0000] $           & $[-0.6414,  0.0000]$\\
                  &                                                 &$(0.0138)$     &                  &                               &\\                               
$\beta_4$  & Left-right                                  &$  -0.1507$     &$[-0.1782,-0.1232] $               &$[-0.3948,0.0084]$            &$[-3.7394, 0.0086]$\\   
                  &                                                 &$  (0.0140) $   &                &                               &\\                                                                
$\gamma_1$  & Gender                              &$ -0.0071$       &$[-0.0248,0.0107] $               & $[-0.4103,0.4142]$          &$[-0.4142,   0.9715]$ \\                  
                  &                                               &$(0.0091)$        &                &                                 &\\                                       
$\gamma_2$  & Class                                 &$ -0.0461$      &$[-0.0626,-0.0296] $               &   $[-0.2895,0.1712] $        &$[-0.2901, 0.1869 ]  $ \\
                  &                                               &$(0.0084)$       &                 &                                  &  \\                                                           
$\gamma_3$  & Income                              &$ 0.0044  $      &$[0.0028,0.0060]$               & $ [-0.2151, 0.4614]$         & $[-0.2172, 0.4618]$\\
                  &                                                &$ (0.0008) $    &                                               &\\       
   \bottomrule                            
 \end{tabular}}
\caption{Inference results.}
\label{res2}
\end{table}

 Table \ref{res2} presents the inference results. In particular, the second and third columns report the maximum likelihood estimate of $\theta_0$ ($\hat{\theta}^\text{com}$) and 95\% confidence intervals ($C^\text{com}_{0.95}$), respectively, under the assumption that all DMs process the complete information structure. The third and fourth columns report  the projections of $\widehat{\Theta}^*$ and of the 95\% confidence region for any $\theta\in \Theta^*$  ($C_{0.95}$), respectively.  $C_{0.95}$ is constructed following \hyperlink{Andrews_Shi}{Andrews and Shi (2013)}, as outlined in Appendix \ref{inference}, based on 50 bootstrap samples.
 
 %\footnote{It took us around 12 days to obtain $C_{0.95}$ with $K_d=5$ for $d=1,\dots, D$.} 

 Under the assumption that voters are fully informed, all the $\beta$ coefficients, except $\beta_2$, are statistically different from zero at $5\%$. This suggests that DMs vote ideologically on the EU,  inequality, and left-right dimensions. That is, the smaller the distance between DM $i$ and party $y$'s ideological positions on those dimensions, the more likely DM $i$ votes for party $y$, ceteris paribus. Further, $\beta_4$ has the highest absolute value magnitude among the $\beta$ coefficients. That is, voters particularly disvalue casting their votes in favour of a party ideologically distant on the left-right axis. A one-unit increase in the ideological distance on the left-right axis produces a payoff decrease that is roughly 2, 23, and 4 times bigger than the payoff decrease produced by a one-unit increase in the ideological distance on the EU, social care, and inequality dimensions, respectively.

When we remain agnostic about voter sophistication, all the projections of $\widehat{\Theta}^*$ for the $\beta$ coefficients include zero. Therefore, differently from above, we cannot reject the possibility that the election outcomes have been generated under some combinations of information structures under which the ideological distances on the EU,  social care, inequality, and left-right dimensions are irrelevant for voter preferences. Nevertheless, the model maintains enough identification power to completely exclude positive values of $\beta_3$ and almost entirely positive values of $\beta_4$. This is roughly confirmed by the projections of  $C_{0.95}$. In particular, $\beta_4$ can have the highest absolute value magnitude among the $\beta$ coefficients, in agreement with the maximum likelihood results. The fact that this finding on $\beta_4$ is robust to the restrictions on the information environment reflects several post-election descriptive studies  run by political experts, which emphasise that the traditional left-right values, rather than specific policy issues,  have been the main driver of the British electoral behaviour in 2017 (\hyperlink{Hobolt}{Hobolt, 2018}). 
 
The upper bounds of the projections for $\beta_1$ and $\beta_2$ are non-negligibly positive. While this aligns with the non-significance of $\beta_2$  under the complete information assumption ($C^\text{com}_{0.95}$ includes positive values), it is in contrast with the  significance of $\beta_1$ under the complete information assumption and     the  2017 election being often referred  to as the ``Brexit election'' (\hyperlink{Mellon}{Mellon, et al., 2018}). Upon closer inspection, however, the inconclusive results on $\beta_1$ reflect that the 2017 pre-election period saw a substantial increase in the relationship between EU referendum choice and Labour versus Conservative vote choice, with a sort of alignment of the remain-leave axis with the traditional left-right axis. The parties with the clearest positions against Brexit (the Liberal Democrats) and in favour of Brexit (UKIP) lost many supporters. These switched {\it en masse} to the Labour Party, offering a ``soft Brexit'' and the Conservative Party, offering a ``hard Brexit'', respectively (\hyperlink{Mellon}{Mellon, et al., 2018}; \hyperlink{Heath}{Heath and Goodwin, 2017}). Such a tendency may have dampened the role of the distance in the Brexit sentiment in determining the preferences of voters, an insight that is not picked up under the complete information assumption.
%\footnote{The projections of $\widehat{\Theta}^*$   contain $\hat{\theta}^\text{com}$. This is in line with our identification results, according to which the sets of parameter values recovered under specific information structures belong to $\Theta^*$ when they are non-empty.}  

            %%%%%%%%%%%%%%%%%%%%%%%%%%%%%%%%%%%%%%%%%%%%%%%%%
%%%%%%%%%%%%%%%%%%%%%%%%%%%%%%%%%%%%%%%%%%%%%%%%%
%%%%%%%%%%%%%%%%%%%%%%%%%%%%%%%%%%%%%%%%%%%%%%%%%
%%%%%%%%%%%%%%%%%%%%%%%%%%%%%%%%%%%%%%%%%%%%%%%%%
%%%%%%%%%%%%%%%%%%%%%%%%%%%%%%%%%%%%%%%%%%%%%%%%%
\subsection{Counterfactuals}
\label{counterfactuals}   

\paragraph{Information provision.} 
The uncertainty about the payoffs resulting from voting can occur due to deliberate strategies of the candidates who ``{\it becloud}'' their characteristics and opinions ``{\it in a fog of ambiguity}''  (\hyperlink{Downs}{Downs, 1957}, p.136), in order to expand the electoral support by attracting groups of voters with different political preferences (\hyperlink{Campbell}{Campbell, 1983}; \hyperlink{Dahlberg}{Dahlberg, 2009}; \hyperlink{Tomz}{Tomz and van Houweling, 2009}; \hyperlink{Somer}{Somer-Topcu, 2015}). It remains unclear, however, to what extent such uncertainty affects the vote shares and, in turn,   the election results. A better understanding is important for designing transparency policies that can improve citizens' welfare and parties' well-being. We investigate this question by imagining an omniscient mediator who implements a policy that gives voters complete information. This can be achieved, for instance, by organising school campaigns that develops  political literacy; forcing candidates to publicly disclose their assets, liabilities, and criminal records; and enforcing a strict regulation regarding campaign spending and airtime.\footnote{See, for example,  \hyperlink{Niemi}{Niemi and Junn (1998)}, \hyperlink{Hooghe}{Hooghe and Wilkenfeld (2007)}, and  \hyperlink{Pontes}{Pontes, Henn, and Griffiths (2019)} on the impact of civic education on political engagement.} 
We simulate the counterfactual vote shares under complete information and study how they change compared to the factual scenario.

This question has been largely debated in the literature. As explained by \hyperlink{Bartels2}{Bartels (1996)}, political scientists have often answered it by arguing that a large population composed of possibly uninformed citizens acts as if it was fully informed, either because each voter uses cues and information shortcuts helping them to figure out what she needs to know about the political world; or because individual deviations from fully informed voting cancel out in a large election, producing the same aggregate election outcome as if voters were fully informed. \hyperlink{Carpini}{Carpini and Keeter (1996)} and  \hyperlink{Bartels2}{Bartels (1996)} are the first studies to use quantitative evidence to disconfirm such claims. They simulate counterfactual vote shares under complete information using data on the level of information of the survey respondents as rated by the interviewers or assessed by test items. 
\hyperlink{Degan_Merlo2}{Degan and Merlo (2011)} propose an alternative approach, which is closer to ours. They consider a spatial model of voting with latent uncertainty. Differently from us, they estimate such a model by parametrically and exogenously specifying the probability that a voter is informed. They use their estimates to obtain counterfactual vote shares under complete information and find that making citizens more informed about electoral candidates decreases abstention. We contribute to this thread of the literature by providing a way to construct counterfactual vote shares under complete information, which neither requires the difficult task of measuring voters' level of information in the factual scenario, nor imposes parametric assumptions on the probability that a voter is informed. 

Recall from Proposition \ref{count1} that,  given a realization $x$ of $X_i$ and a parameter value $\theta$, $\Psi^{\ell}(y|x;\theta)$ and $\Psi^{\ell}(y|x;\theta)$ are the minimum and maximum counterfactual probabilities of  choosing alternative $y\in \mathcal{Y}$ when the DMs face the augmented decision problem $\{G(\theta,x), S^\dagger\}$, where $S^\dagger$ is some unknown expansion of the information structure $S$ implemented by the policy program, possibly heterogeneous across DMs. In the case analysed, $S$ is the complete information structure and, hence, $S^\dagger=S$ for each DM. Moreover, observe that, when all DMs process the complete information structure, there is a unique model-implied choice distribution. Therefore, $\Psi^{\ell}(y|x;\theta)=\Psi^{u}(y|x;\theta)\coloneqq \Psi(y|x;\theta)$. In light of all this, we compute 
$$
\widehat{\Delta}^*_y\coloneqq \cup_{\theta \in \widehat{\Theta}^*}  \sum_{x}(\Psi(y|x;\theta)-\widehat{\mathbb{P}}_Y(y|x))\widehat{\mathbb{P}}_X(x) \quad \text{and} \quad \widehat{\Delta}^*_{y,0.95} \coloneqq \cup_{\theta \in C_{0.95}}  \sum_{x}(\Psi(y|x;\theta)-\widehat{\mathbb{P}}_Y(y|x))\widehat{\mathbb{P}}_X(x),
$$ 
for each $y\in \mathcal{Y}$, where $\widehat{\mathbb{P}}_Y(y|x)$ and $\widehat{\mathbb{P}}_X(x)$ are  the sample   probability of choosing $y$ conditional on $x$ and the sample  probability of $x$, respectively. $\widehat{\Delta}^*_y$ and $\widehat{\Delta}^*_{y,0.95}$ are the estimated gains/losses in  vote shares under complete information   compared to the  factual scenario. 

Table \ref{count1_table} reveals that, when voters are fully informed, abstention drops with respect to the  factual scenario. This shows that voters are more confident  in choosing a party and aligns with the empirical results in \hyperlink{Degan_Merlo2}{Degan and Merlo (2011)}.  We also find that the ``losers'' from the policy intervention are the two biggest parties, i.e., the Conservative Party and the Labour Party. Conversely, the ``winners'' from the policy intervention are the other minor parties, i.e., the Liberal Democrats and the Green Party.   This suggests that there exists some payoff-relevant information unobserved by voters, and the historically dominant parties in the British political scene benefit the most from such uncertainty.\footnote{The observed drop in abstention is not  a mechanical effect of the normalisation to zero of the payoff from not voting. This is because the observed realization of $V_{iy}$ could add to or subtract from the  component of the payoff ``observed pre-signal'', $\beta_y^\top Z_{iy} +  \gamma_y^\top W_i$. Further, note that it could be that many/all voters in the population already  observe the realization of $V_i$, in which case we should expect no significant change in the abstention share, regardless of the normalisation adopted.}

  \begin{table} [ht]
\centering
  \captionsetup{font=normal}
 \begin{tabular}{l c c}
 \toprule
                                                             &  $\widehat{\Delta}^*_y$ &      $ \widehat{\Delta}^*_{y,0.95}$\\ 
\midrule
 Abstention                                        &      $[-0.1591,   -0.0350$]  &  $[-0.1654,  -0.0020$]  \\

Conservative                                   &      $[  -0.1722,  -0.1293$]  &  $[-0.1722 ,-0.0190$]  \\

Labour                                                &  $[-0.2316,   -0.1716$]    &  $[-0.3317, -0.1211$]   \\

Lib. Dem                                           &   $[0.1234,  0.1673$]   &     $[0.0857, 0.2507$]  \\

UKIP                                                  &  $[-0.0173,  0.2358$]  &  $[ -0.1389,  0.3559$] \\

Green                                               &  $[0.1458,0.1781$]   & $[0.0469 , 0.2056$] \\
\bottomrule
 \end{tabular}
\caption{Gains/losses in  vote shares under complete information   compared to the  factual scenario.}
\label{count1_table}
\end{table}

Lastly, we quantify the voters' maximum   welfare cost of limited information, based on (\ref{welfare}). In particular, we compute
$$
\widehat{W}^*\coloneqq \cup_{\theta \in \widehat{\Theta}^*}  \Delta \mathbb{E}_{\theta} \quad \text{and} \quad \widehat{W}^*_{0.95}\coloneqq \cup_{\theta \in C_{0.95}}  \Delta \mathbb{E}_{\theta},
$$
where
$$
\begin{aligned}
\Delta \mathbb{E}_{\theta}\coloneqq  \sum_{x} \Big[ \int_{\mathcal{V}} u\Big(\text{argmax}_{y\in \mathcal{Y}} & u(y,x,v; \theta),v\Big)  P_V(v) dv\\
& - \int_{\mathcal{V}}  u\Big(\text{argmax}_{y\in \mathcal{Y}} \int_{\mathcal{V}}u(y,x,v; \theta) P_V(v) dv, v\Big)  P_V(v) dv\Big] \widehat{\mathbb{P}}_X(x),
\end{aligned}
$$
Table \ref{count1_table2} highlights that up to 1.1697 utility points would be gained, on average, if all voters were perfectly informed. We interpret this utility increase by translating it into the corresponding change in a specific covariate. For example,  consider $Z_{i4}$ and take the estimated lower bound of its coefficient, $\beta_4$,  which is $ -0.3948$, as shown in Table \ref{res2}. A utility increase of 1.1697 is equivalent to the utility gain achieved by reducing the ideological distance from a given party on the left-right dimension by approximately $1.1697/0.3948\approx 3$ points.  

   \begin{table} [ht]
\centering
  \captionsetup{font=normal}
 \begin{tabular}{ l c}
 \toprule
 $\widehat{W}^*$ & $[0, 1.1697]$ \\
$\widehat{W}^*_{0.95}$ & $[0, 1.2387]$ \\
\bottomrule
 \end{tabular}
\caption{Maximum  welfare cost of limited information.}
\label{count1_table2}
\end{table}

\paragraph{Changes in covariates.}
Various political experts sustain that, while at the beginning of the 2017 election campaign the Conservative Party had a sizeable lead in the opinion polls over the Labour Party, as the campaign progressed the Labour Party recovered ground because  it strengthened its left ideological position on social spending and nationalization of key public services (for example,  \hyperlink{Heath}{Heath and Goodwin, 2017}; \hyperlink{Mellon}{Mellon, at al., 2018}).  To evaluate this, we reset the Labour Party's  placement on dimension 2 (social care) to be two points less (i.e., $5$ instead of $7$) and study if the Labour Party's well-being worsens by looking at the change in its vote shares.

Recall from Proposition \ref{count2} that,  given a realization $x$ of $X_i$ and a parameter value $\theta$, $\Phi^{\ell}(y|x, x^\dagger;\theta)$ and $\Phi^{u}(y|x, x^\dagger;\theta)$ are the minimum and maximum counterfactual probabilities of  choosing alternative $y\in \mathcal{Y}$ when the DMs face the augmented decision problem $\{G(\theta,x^\dagger), S\}$, where $S$ is the unknown   information structure $S$ processed in the factual scenario, possibly heterogeneous across DMs, and $x$ and $x^\dagger$ are the covariate realizations before and after the intervention, respectively.
Given $y$ denoting the Labour Party, we compute 
$$
\widehat{\Lambda}_y\coloneqq \cup_{\theta \in \widehat{\Theta}^*} \Big[ \sum_{(x, x^\dagger)} (\Phi^{\ell}(y| x, x^\dagger; \theta)-\widehat{\mathbb{P}}_Y(y|x))\widehat{\mathbb{P}}_X(x),   \sum_{(x, x^\dagger)} (\Phi^{u}(y| x, x^\dagger; \theta)-\widehat{\mathbb{P}}_Y(y|x))\widehat{\mathbb{P}}_X(x)\Big],
$$
and
$$
\widehat{\Lambda}_{y,0.95}\coloneqq \cup_{\theta \in C_{0.95}} \Big[ \sum_{(x, x^\dagger)} (\Phi^{\ell}(y| x, x^\dagger; \theta)-\widehat{\mathbb{P}}_Y(y|x))\widehat{\mathbb{P}}_X(x),   \sum_{(x, x^\dagger)} (\Phi^{u}(y| x, x^\dagger; \theta)-\widehat{\mathbb{P}}_Y(y|x))\widehat{\mathbb{P}}_X(x)\Big].
$$
$\widehat{\Lambda}^*_y$ and $\widehat{\Lambda}^*_{y,0.95}$ are the estimated gains/losses in the Labour Party's vote share compared to the  factual scenario. We also calculate the difference between the counterfactual and factual choice probabilities under the assumption that all voters process the complete information structure:
$$
\widehat{\Lambda}^{\text{com}}_y\coloneqq  \sum_{x} (P_Y(y| x^\dagger; \hat{\theta}^{\text{com}}) -\widehat{\mathbb{P}}_Y(y|x))\widehat{\mathbb{P}}_X(x),
$$
where $P_Y(y| x^\dagger; \hat{\theta}^{\text{com}})$ is the counterfactual probability of choosing $y$ conditional on $x^\dagger$ under complete information and $\theta=\hat{\theta}^{\text{com}}$, obtained using the standard Multivariate Probit formulas.

$\widehat{\Lambda}^{\text{com}}_y$ in Table \ref{count2_table} reveals that, when voters are assumed to be fully informed, weakening the social care position to 5 leads to a decrease in the Labour Party's vote share. Our results partly confirm this finding, as $\widehat{\Lambda}_y$ and $\widehat{\Lambda}_{y,0.95}$ mostly lie on the negative real line. This supports the claim that, by strengthening its left ideological position on the social care dimension, the Labour Party gained some votes during the election campaign.
  \begin{table} [ht]
\centering
  \captionsetup{font=normal}
 \begin{tabular}{ l c}
 \toprule
 $\widehat{\Lambda}^{\text{com}}_y$ & $-0.2283$ \\
$\widehat{\Lambda}_y$ & $[-0.3648, 0.0158]$ \\
$\widehat{\Lambda}_{y,0.95}$ & $[-0.5682, 0.0163]$\\
\bottomrule
 \end{tabular}
\caption{Gains/losses in the Labour Party's vote share compared to the  factual scenario.}
\label{count2_table}
\end{table}

            %%%%%%%%%%%%%%%%%%%%%%%%%%%%%%%%%%%%%%%%%%%%%%%%%
%%%%%%%%%%%%%%%%%%%%%%%%%%%%%%%%%%%%%%%%%%%%%%%%%
%%%%%%%%%%%%%%%%%%%%%%%%%%%%%%%%%%%%%%%%%%%%%%%%%
%%%%%%%%%%%%%%%%%%%%%%%%%%%%%%%%%%%%%%%%%%%%%%%%%
%%%%%%%%%%%%%%%%%%%%%%%%%%%%%%%%%%%%%%%%%%%%%%%%%
\section{Conclusions}                 
 \label{conclusions}              
In this paper, we study identification of preferences in static single-agent discrete choice models where decision makers may be imperfectly informed about the state of the world. We leverage the notion of 1BCE by \hyperlink{BM_2}{BM16} to provide a tractable characterization of the sharp identified set.
We make three main methodological contributions. First, by reinterpreting our framework as a 1-player game against nature, we provide insightful comparisons between the identification power of our framework and several many-player games. Second, we develop a formal procedure to practically construct the sharp identified set for the payoff parameters when the state of the world is continuous. Third, we characterize sharp bounds on the counterfactual choice probabilities when
agents receive information about the state of the world via a policy program, which is an important question in the empirical
literature on single-agent decision problems across different fields.   The method developed in this paper is used to estimate a spatial voting model under weak assumptions on agents' information about the returns to voting for the 2017 UK general election. We show the usefulness of our methodology to quantify the consequences of imperfect information on the welfare of voters and parties.

                %%%%%%%%%%%%%%%%%%%%%
\newpage
\begin{center}
\section*{References}
\end{center}

\begin{description}
\setlength\itemsep{-0.5em}
\footnotesize{
\item \hypertarget{Abaluck_Adams}{ } Abaluck, J., and A. Adams (2021): ``What Do Consumers Consider Before They Choose? Identification from Asymmetric Demand Responses,'' {\it The Quarterly Journal of Economics}, 136(3), 1611--1663. 

\item \hypertarget{Aldrich}{ }   Aldrich, J.H., and R.D. McKelvey (1977): ``A Method of Scaling with Applications to the 1968 and 1972 Presidential Elections,'' {\it American Political Science Review}, 71(1), 111--130.

\item \hypertarget{Alvarez_solo}{ } Alvarez, R.M. (1998): {\it Information and Elections}, University of Michigan Press, Ann Arbor, MI.

\item \hypertarget{Alvarez1995}{ } Alvarez, R.M., and J. Nagler (1995): ``Economics, Issues and the Perot Candidacy: Voter Choice in the 1992 Presidential Election,'' {\it American Journal of Political Science}, 39(3), 714--744.

\item \hypertarget{Alvarez1998}{ } Alvarez, R.M., and J. Nagler (1998): ``When Politics and Models Collide: Estimating Models of
Multiparty Elections,'' {\it American Journal of Political Science}, 42(1), 55--96.

\item \hypertarget{Alvarez2000}{ } Alvarez, R.M., and J. Nagler (2000): ``A New Approach for Modelling Strategic Voting in Multiparty Elections,'' {\it British Journal of Political Science}, 30(1), 57--75.

\item \hypertarget{Alvarez2000_2}{ }  Alvarez, R.M., J. Nagler, and S. Bowler (2000):  ``Issues, Economics, and the Dynamics of Multiparty Elections: the British 1987 General Election,'' {\it American Political Science Review}, 94(1), 131--149.
 
\item \hypertarget{Andrews_Shi}{ } Andrews, D.W.K., and X. Shi (2013): ``Inference Based on Conditional Moment Inequalities,'' {\itshape Econometrica}, 81(2), 609--666.

\item \hypertarget{Athey}{ }  Athey, S. (2002): ``Monotone Comparative Statics under Uncertainty,'' {\it The Quarterly Journal of Economics}, 117(1), 187Ð223.

\item \hypertarget{Baron}{ } Baron, D.P. (1994): ``Electoral Competition with Informed and Uniformed Voters,'' {\it American Political Science Review}, 88(1), 33--47.

\item \hypertarget{Molinari_2}{ } Barseghyan, L., M. Coughlin, F. Molinari, and J.C. Teitelbaum (2021): ``Heterogeneous Choice Sets and Preferences,'' 	{\it Econometrica}, 89(5), 2015--2048.

\item \hypertarget{Barseghyan_AER} Barseghyan, A., F. Molinari, T. O'Donoghue, and J.C. Teitelbaum (2013): ``The Nature of Risk Preferences: Evidence from Insurance Choices,'' {\it American Economic Review}, 103(6), 2499--2529.

\item \hypertarget{Barseghyan}{ }  Barseghyan, A., F. Molinari, T. O'Donoghue, and J.C. Teitelbaum (2018): ``Estimating Risk Preferences in the Field,'' {\itshape Journal of Economic Literature}, 56(2), 501--564.

\item \hypertarget{Barseghyan_QE} Barseghyan, A., F. Molinari, and J.C. Teitelbaum (2016): ``Inference under Stability of Risk Preferences,'' {\it Quantitative Economics}, 7(2), 367--409.

\item \hypertarget{Molinari_1}{ } Barseghyan, L., F. Molinari, and M. Thirkettle (2021): ``Discrete Choice Under Risk with Limited Consideration,'' {\it American Economic Review}, 111(6), 1972--2006.

\item \hypertarget{Bartels}{ } Bartels, L.M. (1986): ``Issue Voting Under Uncertainty: An Empirical Test,'' {\it American Journal of Political Science}, 30(4), 709--728.

\item \hypertarget{Bartels2}{ } Bartels, L.M. (1996): ``Uninformed Votes: Information Effects in Presidential Elections,'' {\it American Journal of Political Science}, 40(1), 194--230.

\item \hypertarget{BMM}{ }  Beresteanu, A., I. Molchanov, and F. Molinari (2011): ``Sharp Identification in Models with Convex Moment Predictions,'' {\itshape Econometrica}, 79(6), 1785--1821.

\item \hypertarget{Bergemann_Brooks_Morris}{ } Bergemann, D., B. Brooks, and S. Morris (2022): ``Counterfactuals with Latent Information,''  {\it American Economic Review}, 112(1), 343--368.

\item \hypertarget{BM_1}{ } Bergemann, D., and S. Morris (2013): ``Robust Predictions in Games With Incomplete Information,'' {\it Econometrica}, 81(4), 1251--1308.

\item \hypertarget{BM_2}{ } Bergemann, D., and S. Morris (2016): ``Bayes Correlated Equilibrium and the Comparison of Information Structures in Games,'' {\it Theoretical Economics}, 11(2), 487--522.

\item \hypertarget{Berry}{ } Berry, S. (1994): ``Estimating Discrete-Choice Models of Product Differentiation,'' {\it The RAND Journal of Economics}, 25(2), 242--262.

\item \hypertarget{bettinger}{ }  Bettinger, E.P., B.T. Long, P. Oreopoulos, L. Sanbonmatsu (2012): ``The Role of Application Assinstance and Information in College Decisions: Results from the H\&R Block Fafsa Experiment,'' {\it Quarterly Journal of Economics}, 127(3), 1205--1242.

\item \hypertarget{Black}{} Black, A. (1998): {\it Information and Elections}, Michigan University Press, Ann Arbor, MI.

\item \hypertarget{blackwell51}{ } Blackwell, D. (1951): ``Comparison of Experiments,'' {\it Proceedings of the Second Berkeley Symposium on Mathematical Statistics and Probability}, 2, 93--102.

\item \hypertarget{blackwell53}{ } Blackwell, D. (1953): ``Equivalent Comparisons of Experiments,'' {\it The Annals of Mathematical Statistics},  24(2), 265--272.

\item \hypertarget{Brown}{ } Brown, Z.Y., and J. Jeon (2020): ``Endogenous Information and Simplifying Insurance Choice,'' Working Paper.

\item \hypertarget{Butzer}{ } Butzer, P.L. (1954): ``On the Extensions of Bernstein Polynomials to the Infinite Interval,'' {\it Proceedings of the American Mathematical Society}, 5(4), 547--553. 

%\item \hypertarget{Calafiore}{ } Calafiore, G., and M.C. Campi (2005): ``Uncertain Convex Programs:
%Randomized Solutions
%and Confidence Levels,'' {\it Mathematical Programming}, 102, 25--46. 

\item \hypertarget{Campbell}{ } Campbell, J.E. (1983): ``The Electoral Consequences of Issue Ambiguity: An Examination of the Presidential
Candidates' Issue Positions from 1968 to 1980,'' {\it Political Behavior}, 5(3), 277-291.

\item \hypertarget{Caplin_Dean}{ } Caplin, A., and M. Dean (2015): ``Revealed Preference, Rational Inattention, and Costly Information Acquisition,'' {\it American Economic Review}, 105(7), 2183--2203.

\item \hypertarget{Caplin_Dean_Leahy}{ } Caplin, A., M. Dean, and J. Leahy (2019): ``Rational Inattention, Optimal Consideration Sets and Stochastic Choice,''  {\it The Review of  Economic Studies}, 86(3), 1061--1094.

\item \hypertarget{Caplin_Martin}{ } Caplin, A., and D. Martin (2015): ``A Testable Theory of Imperfect Perception,'' {\it The Economic Journal}, 125(582), 184--202.

\item \hypertarget{Cardell}{ } Cardell, N.S. (1997): ``Variance Components Structures for the Extreme-Value and Logistic Distributions with Application to Models of Heterogeneity,'' {\it Econometric Theory}, 13(2), 185--213. 

\item \hypertarget{Carpini}{ }Carpini Delli, M.X., and S. Keeter (1996): {\it What Americans Know about Politics and Why It Matters}, Yale University Press.

\item \hypertarget{Cattaneo}{ } Cattaneo, M., X. Ma, Y. Masatlioglu, and E. Suleymanov (2020): ``A Random Attention Model,'' 	{\it Journal of Political Economy}, 128(7), 2796--2836.

\item \hypertarget{CT}{ } Ciliberto, F., and E. Tamer (2009): ``Market Structure and Multiple Equilibria in Airline
Markets,'' {\itshape Econometrica}, 77(6), 1791--1828.

\item \hypertarget{Coolidge}{ } Coolidge, J.L. (1949): ``The Story of the Binomial Theorem,'' {\it The American Mathematical Monthly}, 56(3), 147--157.

\item \hypertarget{Csaba}{ } Csaba, D. (2018): ``Attentional Complements,'' Working Paper.

\item \hypertarget{iaria}{ } Crawford, G.S., R. Griffith, and A. Iaria (2021): ``A Survey of Preference Estimation with Unobserved Choice Set
Heterogeneity,'' {\it Journal of Econometrics}, 222(1), 4--43.

\item \hypertarget{Dahlberg}{} Dahlberg, S. (2009): ``Political Parties and Perceptual Agreement: The Influence of Party Related Factors on Voters' Perceptions in Proportional Electoral Systems,'' {\it Electoral Studies}, 28(2), 270--278.

\item \hypertarget{Manzini1}{} Dardanoni, V., P. Manzini, M. Mariotti, and C.J. Tyson (2020): ``Inferring Cognitive Heterogeneity From Aggregate Choices,'' {\it Econometrica}, 88(3), 1269--1296.

\item \hypertarget{Manzini2}{} Dardanoni, V., P. Manzini, M. Mariotti, H. Petri, and and C.J. Tyson (2022): ``Mixture Choice Data: Revealing
Preferences and Cognition,'' forthcoming at the {\it Journal of Political Economy}.

\item \hypertarget{Degan_Merlo2}{ }  Degan, A., and A. Merlo (2011): ``A Structural Model of Turnout and Voting in Multiple Elections,'' {\it Journal of the European Economic Association}, 9(2), 209--245.

\item \hypertarget{Downs}{ } Downs, A. (1957): {\it An Economic Theory of Democracy}, Harper and Row, New York.

\item \hypertarget{Enelow}{} Enelow, J., and M.J. Hinich (1981): ``A New Approach to Voter Uncertainty in the Downsian Spatial Model,''  {\it American Journal of Political Science}, 25(3), 483--493.

\item \hypertarget{Feddersen}{} Feddersen, T., and W. Pesendorfer (1997): ``Voting Behavior and Information Aggregation in Elections With Private Information,'' {\it Econometrica}, 65(5), 1029--1058.

\item \hypertarget{Feddersen_Pesendorfer}{ }  Feddersen J., and W. Pesendorfer (1999): ``Abstention in Elections with Asymmetric Information and Diverse Preferences,'' {\it American Political Science Review}, 93(2), 381--398.

\item \hypertarget{Fieldhouse}{} Fieldhouse, E., Green, J. Evans G., Schmitt H., van der Eijk C., Mellon J., and Prosser C. (2019): {\it British Election Study, 2017: Face-to-Face Post-Election Survey}, available at \text{https://www.britishelectionstudy.com/.}

\item \hypertarget{Fosgerau}{ } Fosgerau, M., E.  Melo, A. de Palma, and M. Shum (2020): ``Discrete Choice and Rational Inattention: a General Equivalence Result,'' {\it International Economic Review}, 61(4), 1569--1589.

\item \hypertarget{Franklin}{ } Franklin, C.H. (1991): ``Eschewing Obfuscation? Campaigns and the Perception of U.S. Senate Incumbents,'' {\it American Political Science Review}, 85(4), 1193--1214. 

\item \hypertarget{GS2021}{ } Galichon, A., and B. Salani\'e (2022):  ``Cupid's Invisible Hand: Social Surplus and Identification in Matching Models,''   {\it The Review of  Economic Studies}, 89(5), 2600-2629.

\item \hypertarget{HHK}{} Haile, P.A., A. Horta\c{c}su,  and G. Kosenok (2008): ``On the Empirical Content of Quantal Response Equilibrium,'' {\it American Economics Review}, 98(1), 180--200.

\item \hypertarget{Han}{ } Han, S., and and S. Yang (2023): ``A Computational Approach to Identification of Treatment Effects for Policy Evaluation,'' arXiv:2009.13861. 

%\item \hypertarget{Han2}{ } Han, S., and and H. Xu (2022): ``On Quantile Treatment Effects, Rank Similarity, and the
%Variation of Instrumental Variables,'' Working Paper.

\item \hypertarget{Hastings}{ }  Hastings, J.S., and L. Tejeda-Ashton (2008): ``Financial Literacy, Information, and Demand Elasticity: Survey and Experimental Evidence from Mexico,'' NBER Working Paper 14538. 

\item \hypertarget{Hastings2}{ } Hastings, J.S., and J.M. Weinstein (2008): ``Information, School Choice, and Academic Achievement: Evidence from Two Experiments,'' {\it The Quarterly Journal of Economics}, 123(4), 1373--1414.

\item \hypertarget{Heath}{ } Heath, O., and M. Goodwin (2017):  ``The 2017 General Election, Brexit and the Return to Two- Party Politics: An Aggregate-level Analysis of the Result,'' {\it Political Quarterly}, 88(3), 345--358.

\item \hypertarget{Hebert}{ } H\'ebert, B., and M. Woodford (2018): ``Information Costs and Sequential Information Sampling,'' NBER Working Paper 25316.

\item \hypertarget{Hobolt}{} Hobolt, S.B. (2018): ``Brexit and the 2017 UK General Election,'' {\it Journal of Common Market Studies}, 56(S1), 39--50.

\item \hypertarget{Honka}{ }  Honka, E., and P. Chintagunta (2016): ``Simultaneous or Sequential? Search Strategies in the US Auto Insurance Industry,'' {\it Marketing Science}, 36(1), 21--42.

\item \hypertarget{Hooghe}{} Hooghe, M., and B. Wilkenfeld (2007): ``The Stability of Political Attitudes and Behaviors Across Adolescence and Early Adulthood: A Comparison of Survey Data on Adolescents and Young Adults in Eight Countries,'' {\it Journal of Youth and Adolescence}, 37(2), 155--167.

\item \hypertarget{Hutton} Hutton, R. (2017): ``General Election 2017: What We've Learned So Far,'' Bloomberg, accessed 
  \href{https://www.bloomberg.com/news/articles/2017-05-19/general-election- 2017-what-we-ve-learned-so-far on 14 September 2017}{{\it here}}.
  
\item \hypertarget{Kamenica}{ } Kamenica, E., and M. Gentzkow (2011): ``Basyesian Persuasion,'' {\it American Economic Review}, 101(6), 2590--2615. 

\item \hypertarget{Kling}  Kling, Jeffrey R., S. Mullainathan, E. Shafir, L.C. Vermeulen, M.V. Wrobel (2012): ``Comparison Friction: Experimental Evidence from Medicare Drug Plans,'' {\it Quarterly Journal of Economics}, 127(1), 199--235. 

\item \hypertarget{Knight}{} Knight, B., and N. Schiff (2010): ``Momentum and Social Learning in Presidential Primaries,'' {\it Journal of Political Economy}, 118(6), 1110--1150.

\item \hypertarget{Lin}{ } Lin, Y.-H. (2022): ``Stochastic Choice and Rational Inattention,'' {\it Journal of Economic Theory}, 202, 105450. 

\item \hypertarget{Lu}{ } Lu, J. (2016): ``Random Choice and Private Information,'' {\it Econometrica}, 84(6), 1983--2027. 

\item \hypertarget{Lupia}{ } Lupia, A., and M.D. McCubbins (1998): {\it The Democratic Dilemma. Can Citizens Learn What They Need to Know},  Cambridge University Press.

\item \hypertarget{Magnolfi_Roncoroni}{ } Magnolfi, L., and C. Roncoroni (2023): ``Estimation of Discrete Games with Weak Assumptions on Information,''  {\it The Review of Economic Studies}, 90(4), 2006--2041.

\item \hypertarget{Manski} Manski, C.F. (1988): ``Identification of Binary Response Models,'' {\it Journal of the American Statistical
Association}, 83(403), 729--738.

\item \hypertarget{Matejka_McKay}{ } Mat\u{e}jka, F., and A. McKay (2015): ``Rational Inattention to Discrete Choices: A New Foundation for the Multinomial Logit
Model,'' {\itshape American Economic Review}, 105(1), 272--298.

\item \hypertarget{Tabellini2}{ } Mat\u{e}jka, F., and G. Tabellini (2021): ``Electoral Competition with Rationally Inattentive Voters,'' {\it Journal of the European Economic Association}, 19(3), 1899-1935.

\item \hypertarget{Matsusaka}{} Matsusaka, J.G. (1995): ``Explaining Voter Turnout Patterns: An Information Theory,'' {\it Public Choice}, 84, 91--117.

 \item \hypertarget{McMurray}{} McMurray, J.C. (2013): ``Aggregating Information by Voting: The Wisdom of the Experts versus the Wisdom of the Masses,'' {\it The Review of  Economic Studies}, 80(1), 277--312. 
 
 \item \hypertarget{Mehta}{ }  Mehta, N., S. Rajiv, and K. Srinivasan (2003): ``Price Uncertainty and Consumer Search: A Structural Model of Consideration Set Formation,'' {\it Marketing Science}, 22(1), 58--84.
  
 \item \hypertarget{Mellon}{ } Mellon, J., G. Evans, E. Fieldhouse, J. Greeb, and C. Prosser (2018): ``Brexit or Corbyn? Campaign and Inter-Election Vote Switching in the 2017 UK General Election,'' {\it  Parliamentary Affairs}, 71(4), 719--737.

\item \hypertarget{Milnor}{ } Milnor, H. (1951): {\it Games against Nature}, Research Memorandum RM-679, The Rand Corporation.

%\item \hypertarget{Mogstad}{} Mogstad, M., A. Torgovitsky, and C.R. Walters (2021): ``The Causal Interpretation of Two-Stage Least Squares with Multiple Instrumental Variables,''  {\it American Economic Review}, 111(11), 3663--3698.

\item \hypertarget{Mogstad}{} Mogstad, M., A. Santos, and A. Torgovitsky  (2018): ``Using Instrumental Variables for Inference about Policy Relevant Treatment Parameters,''  {\it Econometrica}, 86(5), 1589--1619.

\item \hypertarget{Morris}{ } Morris, S., and P. Strack (2019): ``The Wald Problem
and the Relation of Sequential Sampling and Ex-Ante Information Costs,'' SSRN Working Paper.

\item \hypertarget{Niemi}{} Niemi, R.G., and J. Junn (1998): {\it Civic Education: What Makes Students Learn}, New Haven, CT: Yale University
Press.

\item \hypertarget{Palfrey_Poole}  Palfrey, T.R., and K.T. Poole (1987): ``The Relationship between Information, Ideology, and Voting Behavior,'' {\it American Journal of Political Science}, 31(3), 511--530.

\item \hypertarget{Pontes}{} Pontes, A.I., M. Henn, and M.D. Griffiths (2019): ``Youth Political (Dis)engagement and the Need for Citizenship Education: Encouraging Young People's Civic and Political Participation through the Curriculum,'' {\it Education, Citizenship and Social Justice}, 14(1), 3--21.

\item \hypertarget{Shepsle}{ } Shepsle, A.K. (1972): ``The Strategy of Ambiguity: Uncertainty and Electoral Competition,'' {\it American Political Science Review}, 66(2), 555--568.

\item \hypertarget{Sims_2003}{ } Sims, C.A. (2003): ``Implications of Rational Inattention,'' {\itshape Journal of Monetary Economics }, 50(3), 665--690.

\item \hypertarget{Snowdon}{} Snowdon, K. (2017): ``General Election 2017: 10 Forgotten Issues During The Campaign,'' HuffPost UK, accessed  \href{http://www.huffingtonpost.co.uk/entry/general- election-2017-10-forgotten-issues-during-the-campaign_uk_593688d7e4b013c4816a d4ef on 14 September 2017}{{\it here}}.

\item \hypertarget{Somer}{} Somer-Topcu (2015): ``Everything to Everyone: The Electoral Consequences of the Broad-Appeal Strategy in
Europe,'' {\it American Journal of Political Science}, 59(4), 841--854.

\item \hypertarget{Szasz}{ } Szasz, O. (1950): ``Generalisation of S. Bernstein's Polynomials to the Infinite Interval,'' {\it Journal of Research of the National Bureau of Standards}, 45(3), 239--245.

\item \hypertarget{Tamer_auctions}{ } Syrgkanis, V., E. Tamer, and J. Ziani (2021): ``Inference on Auctions with Weak Assumptions on Information,'' arXiv:1710.03830. 

\item \hypertarget{Tamer}{ } Tamer, E. (2003): ``Incomplete Simultaneous Discrete Response Model with Multiple Equilibria,'' {\it The Review of  Economic Studies}, 70(1), 147--165. 

\item \hypertarget{Tomz}{} Tomz, M., and R. P. van van Houweling (2009): ``The Electoral Implications of Candidate Ambiguity,'' {\it American Political Science Review}, 103(1), 83--98.

\item \hypertarget{Ursu}{ } Ursu, R. M. (2018): ``The Power of Rankings: Quantifying the Effect of Rankings on Online Consumer Search and Purchase Decisions,'' {\it Marketing Science}, 37(4), 530--552.

\item \hypertarget{Weisberg}{} Weisberg, H.F., and M.P. Fiorina (1980): ``Candidate Preference under Uncertainty:
An Expanded View of Rational Voting,'' in John C. Pierce and John L. Sullivan, eds., {\it The
Electorate Reconsidered}, Beverly Hills: Sage. 

\item \hypertarget{Yuksel}{} Yuksel, S. (2022): ``Specialized Learning and Political Polarization,'' International Economic Review, 63(1), 457--474. 
}
\end{description}
          
%%%%%%%%%%%%%%%%%%%%%%%%%%%%%%%%%%%%%%%%%%%%%%%%%%%%%%%%%%%%%%%%%%%%%%%%%%%%%%%%%%%%
%%%%%%%%%%%%%%%%%%%%%%%%%%%%%%%%%%%%%%%%%%%%%%%%%%%%%%%%%%%%%%%%%%%%%%%%%%%%%%%%%%%%%
\newpage
\begin{appendix}
\counterwithin{figure}{section}
\counterwithin{table}{section}
\counterwithin{equation}{section}
\counterwithin{theorem}{section}
\counterwithin{corollary}{section}
\counterwithin{table}{section}
\counterwithin{proposition}{section}
\counterwithin{lemma}{section}

%%%%%%%%%%%%%%%%%%%%%%%%%%%%%%%%%%%%%%%%%%%%%%%%%%%%%%%%%%%%%%%%%%
%%%%%%%%%%%%%%%%%%%%%%%%%%%%%%%%%%%%%%%%%%%%%%%%%%%%%%%%%%%%%%%%%%
%%%%%%%%%%%%%%%%%%%%%%%%%%%%%%%%%%%%%%%%%%%%%%%%%%%%%%%%%%%%%%%%%%
%%%%%%%%%%%%%%%%%%%%%%%%%%%%%%%%%%%%%%%%%%%%%%%%%%%%%%%%%%%%%%%%%%
\section{Examples}
\label{example}

\begin{example}{\normalfont({\itshape Additive random utility discrete choice models})}
\label{nested}
Consider the Nested Logit model with one nest collecting all goods but the outside option. The payoff function, $u$, is 
\begin{equation}
\label{nested_model}
u(y, X_i,   V_i,\epsilon_i(\lambda); \beta, \lambda)\coloneqq \begin{cases}
%\beta^\top{Z}_{iy}+ \lambda \log(\xi_i)+ \lambda \eta_{iy} & \text{if } y\in \mathcal{Y}\setminus\{0\},\\
\beta^\top{X}_{iy}+ \epsilon_i(\lambda)+ \lambda V_{iy} & \text{if } y\in \mathcal{Y}\setminus\{0\},\\
V_{i0} & \text{if } y=0, 
\end{cases}
\end{equation}
where $i$ indexes a generic DM,  $\mathcal{Y}\coloneqq \{0,1,\dots, D\}$, $0$ denotes the outside option, $X_{iy}$ are DM-alternative specific covariates, and $(\epsilon_i(\lambda), V_{iy})$ are DM $i$'s tastes i.i.d. across $y$ and independent of $X_i\coloneqq (X_{i1},..., X_{iD-1})$. The densities of $\epsilon_i(\lambda)$ and $V_i\coloneqq (V_{i0},..., V_{iD-1})$ are chosen to yield the familiar Nested Logit market share function, with $\lambda\in (0,1]$.  The researcher observes the choice made by DM $i$ and the realization of $X_i$ for a large sample of DMs. 

Suppose that DM $i$ observes the realization of $(X_i,\epsilon_i(\lambda))$ but might be uncertain about the realization of  $V_i$.   DM $i$ has a prior on $V_i$ conditional on $(X_i,  \epsilon_i(\lambda))$, which is assumed to obey the Nested Logit parameterisation above. Further, DM $i$ processes an information structure, $S_i$, to update their prior. Note that this framework reduces to the traditional Nested Logit model under the additional assumption that each agent in the population processes the complete information structure. 
Similar considerations can be made for the (Multinomial) Logit model, Mixed Logit model, Probit model, etc.
\end{example}
\begin{example}{\normalfont({\itshape Risk aversion})}
\label{risk}
Consider a discrete choice model of  insurance plans. Specifically, DM $i$ faces an underlying risk of a loss (for example,  a car accident) and can choose among $L$ insurance plans. The loss event is denoted by $C_i$, where it takes value $1$  if the loss event occurs and $0$ otherwise. Each insurance plan $y\in \mathcal{Y}$ is characterised by a deductible, $D_y$, and a premium, $P_{iy}$.  Further, DM $i$ is endowed with some wealth ($\text{Wealth}_i$).  
The payoff function, $u$, belongs to the CARA family:
\begin{equation}
\label{model_risk}
u(y, P_i, D, \text{Wealth}_i, r_i, C_i)\coloneqq \begin{cases}
\frac{1-\exp[-r_i\times (\text{Wealth}_i-P_{iy}-D_{y})]}{r_i} & \text{if } C_i=1, r_i\neq 0,\\
\frac{1-\exp[-r_i \times (\text{Wealth}_i-P_{iy})]}{r_i}  & \text{if } C_i=0, r_i\neq 0,\\
\text{Wealth}_i-P_{iy}-D_{y}   & \text{if } C_i=1, r_i=0,\\
\text{Wealth}_i-P_{iy}  & \text{if } C_i=0, r_i=0,\\
\end{cases}
\end{equation}
where  $P_i\coloneqq (P_{i1},..., P_{iL})$, $D\coloneqq (D_1,..., D_L)$, and $r_i$ is the coefficient of absolute risk aversion. $r_i$ is often assumed distributed according to some parametric distribution such as the Beta distribution. The researcher observes the choice made by DM $i$ and the realization of $(P_i, D, \text{Wealth}_i)$. In some cases, the researcher also observes  the realization of $C_i$ from ex-post data on claims.

Before choosing an insurance plan, DM $i$ is aware of the realization of $(P_i, D, \text{Wealth}_i, r_i)$. However, DM $i$ does not observe the realization of $C_i$ because it is realised after  the insurance plan choice has been made. Hence, following our general notation, ${V_i\coloneqq  C_i}$  and $(P_i, D, \text{Wealth}_i, r_i)$  are the variables in  DM $i$'s information set in addition to the signal.   DM $i$ has a prior on $V_i$ conditional on $(P_i, D, \text{Wealth}_i, r_i)$, which can be assumed to belong to some parametric family. For instance, one can use a simple Probit model or a more sophisticated Poisson-Gamma model (for an example of the latter see \hyperlink{Barseghyan_AER}{Barseghyan, Molinari, O' Donoghue, and Teitelbaum, 2013}; \hyperlink{Barseghyan_QE}{Barseghyan, Molinari, and Teitelbaum, 2016}). 
Further, DM $i$ processes an information structure, $S_i$, to update their prior. $S_i$ incorporates any extra private information on the risky event at the disposal of DM $i$, other than their level of risk aversion, and can arbitrarily depend on DM $i$'s risk aversion.

Under the additional restriction that each agent processes the null information structure, note that this framework collapses to the standard risk aversion setting considered in the empirical literature, where individuals have no extra private information on the risky event.
\end{example}
\begin{example}{\normalfont({\itshape Rational inattention})}
\label{sensitivity} Consider the  rational inattention framework by \hyperlink{Caplin_Dean}{Caplin and Dean (2015)} and \hyperlink{Matejka_McKay}{Mat\u{e}jka and McKay (2015)}. In that setting, the decision problem has two stages. In the first stage, DM $i$ optimally chooses an information structure to update their prior. Although DM $i$ is free to choose  any information structure, attention is a scarce resource and there is a cost of processing information. As a result, more informative signals are more costly. Such attentional costs are parameterised in various ways, for example,  using the Shannon entropy (\hyperlink{Sims_2003}{Sims, 2003}) and the posterior-separable function (\hyperlink{sep}{Caplin, Dean, and Leahy, 2019a}). Formally, in the first stage, the DM   chooses an information structure $S^*\in \mathcal{S}$ such that  
$$
\begin{aligned}
S^*\in & \text{argmax}_{S\coloneqq \{\mathcal{T}, \mathcal{P}_{T|V}\}\in \mathcal{S}} \\
&\hspace{2cm}\int_{\mathcal{T}} \int_{\mathcal{V}}\Big[\max_{y\in \mathcal{Y}}\mathbb{E}_{S,t}u(y, V) \Big] P_{T| V}(t|v) P_{V}(v)dv dt- C(S), 
\end{aligned}
$$
where $\mathbb{E}_{S,t}u(y, V)$ is the expected payoff from choosing $y\in \mathcal{Y}$ under the posterior induced by   information structure $S$ and   signal realization $t$,  and $C(S)$ represents the parameterised attentional costs associated with the information structure $S$.
Then, in the second stage, the DM  observes a signal realization, $t$, randomly drawn according to $S^*$. Lastly, the DM chooses  alternative $y\in \mathcal{Y}$ maximising $\mathbb{E}_{S^*, t}u(y, V)$.  

Note that our model nests the rational inattention framework because it remains completely silent on how agents choose information structures. In other words, we allow agents to choose information structures following the rational inattention protocol or any other procedure.

For an empirical counterpart of the rational inattention model, see \hyperlink{Csaba}{Csaba (2018)} and \hyperlink{Brown}{Brown and Jeon (2020)}.
A few theoretical papers on rational inattention and, more generally, stochastic choice also grapple with the identification problem. However, those papers typically require either to have data on choices for every possible realization of the state of the world (for example, \hyperlink{Caplin_Martin}{Caplin and Martin, 2015}), or to have data on choices for  multiple menus (for example, \hyperlink{Lu}{Lu, 2016}; \hyperlink{Lin}{Lin, 2022}).\footnote{According to our notation, having data on choices for multiple menus would consist of observing the choices of agents for multiple sets $\mathcal{Y}$.} These data are rarely available outside of laboratory experiments, especially for ``complex'' products as targeted here. Instead, in our framework the state of the world can be fully observed, partly observed, or fully unobserved by the researcher. Moreover, we assume to have data on choices for a single menu. In other words, we rely on less rich data. On one hand, this reduces the sources of identifying power and makes the task of deriving identification arguments more challenging; on the other hand, it allows for a wider applicability of our methodology.

Lastly,  \hyperlink{Hebert}{H\'ebert and Woodford (2018)} and \hyperlink{Morris}{Morris and Strack (2019)} consider continuous-time models of sequential evidence accumulation and show that the resulting choice probabilities are identical to those of a static rational inattention model with posterior-separable attentional cost functions. That is, there is an equivalence between the information that is ultimately acquired in some search models and the information acquired in a static model of rational inattention, under a particular parameterisation of the attentional costs. Therefore, our setting  also nests such search frameworks. 
\end{example}

%%%%%%%%%%%%%%%%%%%%%%%%%%%%%%%%%%%%%%%%%%%%%%%%%%%%%%%%%%%%%%%%%%
%%%%%%%%%%%%%%%%%%%%%%%%%%%%%%%%%%%%%%%%%%%%%%%%%%%%%%%%%%%%%%%%%%
%%%%%%%%%%%%%%%%%%%%%%%%%%%%%%%%%%%%%%%%%%%%%%%%%%%%%%%%%%%%%%%%%%
%%%%%%%%%%%%%%%%%%%%%%%%%%%%%%%%%%%%%%%%%%%%%%%%%%%%%%%%%%%%%%%%%%
\section{Proofs}
\label{proofs}

\paragraph{Proof of Theorem \ref{main}.} 
Fix $\theta \in \Theta$. When $\mathcal{V}$ and $\mathcal{T}$ are not finite sets, $P_V$ and $\{P_{T|V}(\cdot| v): v\in \mathcal{V}\}$ should be interpreted as densities. 
As a preliminary step, it is useful to equivalently rewrite Definition \ref{1BCE} using the distribution of $Y$ conditional on $V$ as unknown, $\mathcal{P}_{Y|V}\coloneqq \{P_{Y|V}(\cdot |v): v\in \mathcal{V}\}$.
 $\mathcal{P}_{Y|V}$ is a 1BCE of   $G(\theta)$ if:
\begin{enumerate}[1.]
\item It is {\it consistent}:
$$
\sum_{y\in \mathcal{Y}} P_{Y|V}(y|v) P_V(v; \theta_V)=P_V(v; \theta_V) \text{ }\forall v\in \mathcal{V}.
$$
\item It is {\it obedient}:
$$
\begin{aligned}
\int_{\mathcal{V}} P_{Y|V}(y| v )P_V(v; \theta_V)(u(y,v; \theta_u) - u(y', v; \theta_u ))dv \geq   0,\text{ } \forall y'\in \mathcal{Y}\setminus \{y\},\forall y\in \mathcal{Y}.
\end{aligned}
$$
\end{enumerate}
Observe that the {\it Consistency} requirement   reduces to $\sum_{y\in \mathcal{Y}} P_{Y|V}(y|v) =1$ for each $v\in \mathcal{V}$, which is  necessary for $P_{Y|V}(y|v) $ to be a well-defined distribution. Hence, in what follows, we just consider the obedience constraint.

Let $\mathcal{P}_{Y|V}$ be a 1BCE of $G(\theta)$. We now show that  there exists an information structure $S\coloneqq \{\mathcal{T}, \mathcal{P}_{T| V}\}\in \mathcal{S}$ and  an optimal strategy $\mathcal{P}_{Y| T}$ of   $\{G(\theta), S\}$ such that $\mathcal{P}_{Y| V}$ arises from $\mathcal{P}_{Y|T }$. 
Define $S\coloneqq \{\mathcal{T}, \mathcal{P}_{T| V}\}$ such that 
\begin{align}
&\mathcal{T}=\mathcal{Y}, \label{tv}\\
&P_{T|V}(y|v)=P_{Y|V}(y|v)\quad \forall (y,v)\in \mathcal{Y}\times \mathcal{V}. \label{tv2}
\end{align}
 Let $\mathcal{P}_{Y|T}$ be a strategy of    $\{G(\theta), S\}$ such that 
\begin{equation}
P_{Y|T}(y|t)=\begin{cases}
\label{yt}
1 & \text{if $y=t$}\\
0 & \text{if $y\neq t$}\\
\end{cases}\quad  \forall (y,t)\in \mathcal{Y}^2.
\end{equation}
Observe that 
\begin{equation}
\label{arise}
P_{Y|V}(y|v)=\sum_{t\in \mathcal{Y}} P_{Y|T}(y|t) P_{T|V}(t|v)= P_{Y|V}(y|v) \quad \forall (y,v)\in \mathcal{Y}\times \mathcal{V},
\end{equation}
where, in the first equality, we combine the formula of conditional probabilities with (\ref{tv}) and the fact that $Y$ is independent of $V$ conditional on $T$. In the second equality, we use (\ref{yt}) and (\ref{tv2}). From (\ref{arise}), we conclude that $\mathcal{P}_{Y| V}$ arises from $\mathcal{P}_{Y|T }$. We need to show that $\mathcal{P}_{Y|T }$ is an optimal strategy of $\{G(\theta), S\}$. This is the case if 
\begin{equation}
\label{theorem1_0}
\int_{\mathcal{V}} P_{V|T}(v|y) u(y,v; \theta_u)dv \geq \int_{\mathcal{V}}   P_{V|T}(v|y) u(y',v; \theta_u)dv,
\end{equation}
for each $ y'\in \mathcal{Y}\setminus \{y\}$ and for each $y\in \mathcal{Y}$. Using the formula of the posterior probability, (\ref{theorem1_0}) is equivalent to
\begin{equation}
\label{theorem1_1}
\int_{\mathcal{V}} P_V(v;\theta_V) P_{T|V}(y|v) u(y,v; \theta_u)dv \geq \int_{\mathcal{V}} P_V(v;\theta_V) P_{T|V}(y|v) u(y',v; \theta_u)dv,
\end{equation}
for each $ y'\in \mathcal{Y}\setminus \{y\}$ and $y\in \mathcal{Y}$. Further, 
by (\ref{tv2}), (\ref{theorem1_1}) is equivalent to
\begin{equation}
\label{theorem1_2}
\int_{\mathcal{V}} P_V(v;\theta_V) P_{Y|V}(y|v) u(y,v; \theta_u)dv \geq \int_{\mathcal{V}} P_V(v;\theta_V) P_{Y|V}(y|v) u(y',v; \theta_u)dv,
\end{equation}
for each $ y'\in \mathcal{Y}\setminus \{y\}$ and $y\in \mathcal{Y}$. Lastly, observe that (\ref{theorem1_2}) holds by the {\it Obedience} requirement. Therefore, $\mathcal{P}_{Y|T}$ is an optimal strategy of  $\{G(\theta), S\}$.

Let $\mathcal{P}_{Y| T}$ be an optimal strategy of   $\{G(\theta), S\}$ for some $S\in \mathcal{S}$. We now show that  $\mathcal{P}_{Y|V}$ arising from $\mathcal{P}_{Y| T}$ is a 1BCE of $G(\theta)$. By the definition of optimal strategy and using the formula of the posterior probability, it holds that 
\begin{equation}
\label{theorem1_3}
\int_{\mathcal{V}} P_V(v;\theta_V) P_{T|V}(t|v) u(y,v; \theta_u)dv \geq \int_{\mathcal{V}} P_V(v;\theta_V) P_{T|V}(t|v) u(y',v; \theta_u)dv,
\end{equation}
for each $t\in \mathcal{T}$, $ y'\in \mathcal{Y}\setminus \{y\}$, and $y\in \mathcal{Y}$ such that $P_{Y|T}(y|t)>0$. By pre-multiplying both sides of (\ref{theorem1_3}) by $P_{Y|T}(y|t)$ and integrating over $ \mathcal{T}$, we obtain
\begin{equation}
\label{theorem1_4}
\int_{\mathcal{T}} P_{Y|T}(y|t) \int_{\mathcal{V}} P_V(v;\theta_V) P_{T|V}(t|v) u(y,v; \theta_u)dvdt \geq\int_{\mathcal{T}} P_{Y|T}(y|t)  \int_{\mathcal{V}} P_V(v;\theta_V) P_{T|V}(t|v) u(y',v; \theta_u)dvdt,
\end{equation}
for each $ y'\in \mathcal{Y}\setminus \{y\}$. By rearranging integrals, (\ref{theorem1_4}) is equivalent to
\begin{equation}
\label{theorem1_5}
\int_{\mathcal{V}} P_V(v;\theta_V) \int_{\mathcal{T}} P_{Y|T}(y|t)  P_{T|V}(t|v) u(y,v; \theta_u)  dt dv\geq   \int_{\mathcal{V}} P_V(v;\theta_V) \int_{\mathcal{T}} P_{Y|T}(y|t) P_{T|V}(t|v) u(y',v; \theta_u)dt dv,
\end{equation}
for each $ y'\in \mathcal{Y}\setminus \{y\}$. Using the independence of $Y$ from $V$ conditional on $T$, (\ref{theorem1_5}) is equivalent to 
\begin{equation}
\label{theorem1_6}
\int_{\mathcal{V}} P_V(v;\theta_V) P_{Y|V}(y|v) u(y,v; \theta_u)dv \geq   \int_{\mathcal{V}} P_V(v;\theta_V) P_{Y|V}(y|v)  u(y',v; \theta_u)dv,
\end{equation}
for each $ y'\in \mathcal{Y}\setminus \{y\}$. Lastly, observe that (\ref{theorem1_6}) is just the {\it Obedience} requirement. Therefore, $\mathcal{P}_{Y|V}$ that arises from $\mathcal{P}_{Y| T}$ is a 1BCE of $G(\theta)$.
%%%%%%%%%%%%%%%%%%%%%%%%%%%%%%%%%%%%%%%%%%%%%%%%%%%%%%%%%%%%%%%%%%%%
\paragraph{Proof of Proposition \ref{main_ident}.} 
Fix $\theta \in \Theta$. We show that if $P_{Y}\in \mathcal{Q}(\theta)$, then $P_Y\in \mathcal{R}(\theta)$. If $P_Y\in \mathcal{Q}(\theta)$, then, by definition of $\mathcal{Q}(\theta)$, there exists 
$
P_{Y,V} \in\mathcal{W}(\theta)
$ such that $P_Y$ arises from $P_{Y,V}$. By Theorem \ref{main}, it follows that there exists 
$
S\in \mathcal{S}$ and  
$
\mathcal{P}_{Y| T}\in \mathcal{R}({\theta,S})$ such that $P_{Y,V} $ arises from  $\mathcal{P}_{Y| T}$.
Thus,  $P_Y$ arises from $\mathcal{P}_{Y| T}$. Therefore, by the  definition of $\mathcal{R}(\theta)$, $P_Y\in \mathcal{R}(\theta)$. 

Conversely, we show that $P_Y\in \mathcal{R}(\theta)$, then $P_Y\in \mathcal{Q}(\theta)$. First, let $\tilde{\mathcal{R}}(\theta) \subseteq \mathcal{R}(\theta)$ be the non-convexified set of model's predictions under $\theta$ while remaining agnostic about information structures. That is, 
\begin{equation*}
\begin{aligned}
\tilde{\mathcal{R}}(\theta) \coloneqq  \Big \{  P_{Y} \in \Delta(\mathcal{Y}):    P_Y(y)=\int_{\mathcal{V}} \int_{\mathcal{T}}      P_{Y|T}(y|t)   & P_{T|V}(t|v)   P_{V}(v; \theta_V)  dt dtv \text{ } \forall y \in \mathcal{Y}, \\
& \mathcal{P}_{Y| T} \in\mathcal{R}(\theta,S), S\coloneqq \{\mathcal{T}, \mathcal{P}_{T| V}\} \in \mathcal{S}\Big \},
\end{aligned}
\end{equation*}
Take $P_Y\in \tilde{\mathcal{R}}(\theta)$. Then, by the definition of $\tilde{\mathcal{R}}(\theta)$, there exists 
$
S \in \mathcal{S}
$ and 
$
\mathcal{P}_{Y| T} \in \mathcal{R}({\theta,S})
$
such that $P_Y$ arises from  $\mathcal{P}_{Y| T}$.  By Theorem \ref{main}, it follows that there exists  
$P_{Y,V}  \in\mathcal{W}(\theta)
$
 such that $\mathcal{P}_{Y| T}$ arises from $P_{Y,V} $. Thus,  $P_Y$ arises from $P_{Y,V}$. Hence, by the definition of $\mathcal{Q}(\theta)$, $P_Y\in \mathcal{Q}(\theta)$. Now, take any $K$ elements from $\tilde{\mathcal{R}}(\theta)$. Denote such elements by $P^1_Y\in \tilde{\mathcal{R}}(\theta),..., P^K_Y\in \tilde{\mathcal{R}}(\theta)$. Given the arguments above, it holds that $P^1_Y\in \mathcal{Q}(\theta),...,P^K_Y\in \mathcal{Q}(\theta)$. Moreover, any convex combination of $P^1_Y,...,P^K_Y$ belongs to $\mathcal{Q}(\theta)$ because  $\mathcal{Q}(\theta)$ is convex. Therefore, every $P_Y\in \mathcal{R}(\theta)$ is also contained in $\mathcal{Q}(\theta)$. 

We can conclude that $\mathcal{R}(\theta)= \mathcal{Q}(\theta)$ for each $\theta \in \Theta$. This implies $\Theta^*=\Theta^{**}$.

%%%%%%%%%%%%%%%%%%%%%%%%%%%%%%%%%%%%%%%%%%%%%%%%%%%%%%%%%%%%%%%%%%%%
\paragraph{Proof of Proposition \ref{id_power}.} 
 \hyperlink{HHK}{Haile, Horta\c{c}su, and Kosenok (2008)} investigate the identification power of the notion of quantile response
equilibrium (QRE) in $N$-player games with $N\geq 1$ and additively separable shocks. In their Theorem 1, they show
that QRE imposes no restrictions on behaviour. That is, for {any} systematic payoff and distribution of
observed behaviour, there exists a density  of the additively
separable unobserved terms  entering payoffs such that a QRE generates the distribution of
observed behaviour. When $N=1$, the QRE framework reduces to an additive random utility discrete choice model (Footnote 9, p.184, \hyperlink{HHK}{Haile, Horta\c{c}su, and Kosenok, 2008}). Hence, we can apply Theorem 1 of \hyperlink{HHK}{Haile, Horta\c{c}su, and Kosenok (2008)}  to our model and show that, when all DMs process the complete information structure, any $\tilde{u}$ rationalises the empirical choice distribution. Formally, for every  $P_Y \in \Delta(\mathcal{Y})$ and  $\tilde{u}\in \tilde{\mathcal{U}}$,  there exists  $P_V\in \tilde{\Delta}(\mathbb{R}^{|\mathcal{Y}|})$ such that $P_Y=P_{Y; \tilde{u}, P_V}$.
This implies that the projection of $\Theta^*$ on $\tilde{\Delta}(\mathbb{R}^{|\mathcal{Y}|})$ is equal to $\tilde{\mathcal{U}}$ and part (a) holds.\footnote{Note that part (a) of Proposition \ref{id_power} also  holds  in many-player games. Specifically, when $N\geq 2$, the QRE framework can be reinterpreted as a game with    incomplete information and Bayesian Nash equilibrium (Footnote 4, p.182, \hyperlink{HHK}{Haile, Horta\c{c}su, and Kosenok, 2008}). In that setting, Theorem 1 of \hyperlink{HHK}{Haile, Horta\c{c}su, and Kosenok (2008)} shows that, for any  systematic payoff and distribution of
observed behaviour, there exists a density  of the additively
separable unobserved terms  entering payoffs such that a Bayesian Nash equilibrium generates the distribution of
observed behaviour.}

 \hyperlink{GS2021}{Galichon and Salani\'e (2022)} study   identification in 2-sided matching models with additively separable shocks. In their Proposition 1, they show that 2-sided matching models are under-identified. 
Specifically, for {any} density of the additively separable unobserved terms entering payoffs and distribution of
observed behaviour, there exists a systematic payoff such that the model-implied stable match generates the distribution of
observed behaviour. The result applies also to the 1-sided case of an additive random utility discrete choice model. Hence, we can apply Proposition 1  of \hyperlink{GS2021}{Galichon and Salani\'e (2022)} to our model and show that, when all DMs process the complete information structure, any $P_V$ rationalises the empirical choice distribution. Formally, for  every $P_Y \in \Delta(\mathcal{Y})$ and $P_V\in \tilde{\Delta}(\mathbb{R}^{|\mathcal{Y}|})$, 
  there exists $\tilde{u}\in \tilde{\mathcal{U}}$ such that $P_Y=P_{Y; \tilde{u}, P_V}$. See also the Appendix   in \hyperlink{Berry}{Berry (1994)} for a similar result.
This implies that the projection of $\Theta^*$ on $\tilde{\mathcal{U}}$ is equal to $\tilde{\Delta}(\mathbb{R}^{|\mathcal{Y}|})$ and part  (b) holds.

%%%%%%%%%%%%%%%%%%%%%%%%%%%%%%%%%%%%%%%%%%%%%%%%%
\paragraph{Proof of Proposition \ref{count1}.} 
Fix $\theta\in \Theta$. We redefine the notions of  baseline decision problem and augmented decision problem. 
Let  $\{G(\theta),S\}$ be the {baseline decision problem} that the DM faces when processing the baseline information structure  ${S}\coloneqq \{{\mathcal{T}}, {\mathcal{P}}_{T|V}\}\in \mathcal{S}$ implemented by the policy program.
Let $\{G(\theta),S^\dagger\}$ be the {augmented decision problem} that the DM faces when processing the expanded information structure $S^\dagger\coloneqq \{\mathcal{T}^\dagger, \mathcal{P}_{T^\dagger|V}\}\in \mathcal{S}$ obtained as a combination of $S$ and some ${S}^\diamond \coloneqq \{{\mathcal{T}}^\diamond, {\mathcal{P}}_{T^\diamond |V}\} \in \mathcal{S}$. Let $Y^\dagger$ denote the DM's choice in the counterfactual scenario. 
  An optimal strategy in   $\{G(\theta),S^\dagger\}$  is a distribution of $Y^\dagger$ conditional on $(T, T^\diamond)$, $\mathcal{P}_{Y^\dagger|T, T^\diamond}\coloneqq \{P_{Y^\dagger|T, T^\diamond}(\cdot| t, t^\diamond): (t, t^\diamond)\in \mathcal{T}\times \mathcal{T}^\diamond\}$, such that, for each $(t, t^\diamond)\in \mathcal{T} \times \mathcal{T}^\diamond$, the DM maximises their  expected utility by choosing any alternative $y^\dagger\in \mathcal{Y}$  featuring $P_{Y^\dagger|T, T^\diamond}(y^\dagger|t, t^\diamond)>0$.

Next, we define a 1BCE of $\{G(\theta), S\}$. $P_{Y^\dagger,V, T}\in \Delta(\mathcal{Y}\times \mathcal{V}\times \mathcal{T})$ is a 1BCE of $\{G(\theta), S\}$ if it is consistent, i.e., $\sum_{y^\dagger\in \mathcal{Y}} P_{Y^\dagger,V,T}(y^\dagger,v,t)=P_{V}(v; \theta_V) P_{T|V}(t|v)$ $\forall (v,t)\in \mathcal{V}\times \mathcal{T}$, and  obedient, i.e., $\int_{\mathcal{V}} P_{Y^\dagger,V,T}(y^\dagger,v,t) (u(y^\dagger,v; \theta_u)- u(k^\dagger, v ; \theta_u)) dv \geq 0$ $\forall y^\dagger\in \mathcal{Y}$, $k^\dagger\in \mathcal{Y}\setminus\{y^\dagger\}$, and $t\in \mathcal{T}$.

We  now state Theorem 1 of \hyperlink{BM_2}{BM16}, which still applies to the new setting. This theorem shows  that $P_{Y^\dagger,V,T}$ is a 1BCE of $\{G(\theta), S\}$ if and only if, for some expansion $S^\dagger$ of $S$, there exists an optimal strategy of $\{G(\theta), S^\dagger\}$ that induces $P_{Y^\dagger,V,T}$.

It follows immediately that $ \mathcal{P}^*_{Y^\dagger}(h)$ can be characterised via the union across $\theta\in \Theta^*$ of  the set of 1BCE  $P_{Y^\dagger,V,T}$ of $\{G(\theta), S\}$. In particular, (\ref{con1}) is the {\it Consistency} constraint,  (\ref{con1a}) is the {\it Obedience} constraint, and (\ref{con2}) ensures that $P_{Y^\dagger,V,T}$ is a well-defined distribution. 

%%%%%%%%%%%%%%%%%%%%%%%%%%%%%%%%%%%%%%%%%%%%%%%%%
 \paragraph{Proof of Proposition \ref{count2}.} Fix $\theta\in \Theta$.  From Theorem 1  in \hyperlink{Bergemann_Brooks_Morris}{Bergemann, Brooks, and Morris (2022)}, $ \mathcal{Q}^*_{Y^\dagger|x, x^\dagger}(h)$ can be characterised via the union across $\theta\in \Theta^*$ of the set of 1BCE $P_{Y,Y^\dagger,V|x, x^\dagger}$ of  the linked baseline decision problem $\{G(\theta, x),G(\theta, x^\dagger)\}$. In particular, (\ref{con3}) is the {\it Consistency} constraint,  (\ref{con3a}) and  (\ref{con3b}) are the {\it Obedience} constraints for each decision problem, (\ref{con3c}) ensures that $P_{Y, Y^\dagger,V|x, X^\dagger}$ is a well-defined distribution, and (\ref{con4}) imposes that the factual choice probabilities are equal to the empirical ones.

%%%%%%%%%%%%%%%%%%%%%%%%%%%%%%%%%%%%%%%%%%%%%%%%%
%%%%%%%%%%%%%%%%%%%%%%%%%%%%%%%%%%%%%%%%%%%%%%%%%
%%%%%%%%%%%%%%%%%%%%%%%%%%%%%%%%%%%%%%%%%%%%%%%%%
%%%%%%%%%%%%%%%%%%%%%%%%%%%%%%%%%%%%%%%%%%%%%%%%%

 \section{Linear program with observed and unobserved heterogeneity}
 \label{epsilon}
In this section, we readapt (\ref{lin_pr_3}) to the case where   there are discrete covariates $X$ (observed by the researcher) and discrete  heterogeneity $\epsilon$ (unobserved by the researcher) which enter the DM's information set together with the signal. For simplicity, we assume that $X$ is independent of $(V, \epsilon)$. However, note that the computational procedure  goes through even when $X$ is not independent of $(V, \epsilon)$, and so the  distribution of $(V,\epsilon)$  is $X$-specific.  Let $\mathcal{X}$ and $\mathcal{E}$ be the supports of $X$ and $\epsilon$, respectively. Let $P_{\epsilon}(\cdot; \theta_{\epsilon})$ be the distribution of $\epsilon$, indexed by the structural parameter $\theta_\epsilon\in \Theta_{\epsilon}$.  For each $e\in \mathcal{E}$, let $P_{V|\epsilon}(\cdot| e; \theta_V)$ be the   distribution of $V$ conditional on $\epsilon=e$. For a given $\theta\coloneqq (\theta_U, \theta_V, \theta_\epsilon)\in \Theta \coloneqq \Theta_U\times  \Theta_V\times  \Theta_{\epsilon}$, the researcher needs to verify if the following finite-dimensional linear program has a solution with respect to $\lambda\coloneqq (\lambda_{k,K}^{y,x,e}: (k,y,x,e)\in \mathcal{K}\times \mathcal{Y}\times \mathcal{X}\times \mathcal{E})$:
\par\nobreak
\vspace{-0.7cm}
 {\small \begin{equation}
\label{lin_pr_3_epsilon}
\begin{alignedat}{3}
&\text{Obedience:}  \quad&&\sum_{k\in \mathcal{K}} \lambda_{k,K}^{y,x,e} \gamma^{y,y',x,e}_{1,k,K}(\theta_V, \theta_U) \geq 0 \text{ } \forall y\in \mathcal{Y}, \forall y'\in \mathcal{Y}\setminus\{y\}, \forall x \in \mathcal{X}, \forall e \in \mathcal{E},\\
&\text{Probability requirements:}  \quad&&\lambda_{k,K}^{y,x,e}\geq 0 \text{ }\forall k\in \mathcal{K}, \forall y \in \mathcal{Y},   \forall x \in \mathcal{X}, \forall e \in \mathcal{E},\text{ }\sum_{y\in \mathcal{Y}} \lambda_{k,K}^{y,x,e}= 1\text{ } \forall k\in \mathcal{K}, \forall x \in \mathcal{X}, \forall e \in \mathcal{E},\\
&\text{Data match:} \quad && \sum_{e\in \mathcal{E}} \sum_{k\in \mathcal{K}} \lambda_{k,K}^{y,x,e} \gamma^e_{2,k,K}(\theta_V; \theta_{\epsilon}) =\mathbb{P}_Y(y) \text{ }\forall y \in \mathcal{Y},\forall x \in \mathcal{X},
\end{alignedat}
\end{equation} }%
where 
$$ 
\begin{aligned}
&\gamma^{y,y',x,e}_{1,k,K}(\theta_V, \theta_U) \coloneqq \int_{\mathcal{V}}  a_{k,K}(v) P_{V|\epsilon}(v|e; \theta_V)({u(y,x,v,e; \theta_u)}-{u(y',x,v,e; \theta_u)}) dv,\\
& \gamma^e_{2,k,K}(\theta_V, \theta_{\epsilon}) \coloneqq 
 P_{\epsilon}(e; \theta_{\epsilon}) \int_{\mathcal{V}} a_{k,K}(v)  P_{V|\epsilon}(v|e; \theta_V)dv.
\end{aligned}
$$
Note that (\ref{lin_pr_3_epsilon}) is separable across $x\in \mathcal{X}$ and, therefore, can be solved separately across $x\in \mathcal{X}$.
 
%%%%%%%%%%%%%%%%%%%%%%%%%%%%%%%%%%%%%%%%%%%%%%%%%
%%%%%%%%%%%%%%%%%%%%%%%%%%%%%%%%%%%%%%%%%%%%%%%%%
%%%%%%%%%%%%%%%%%%%%%%%%%%%%%%%%%%%%%%%%%%%%%%%%%
%%%%%%%%%%%%%%%%%%%%%%%%%%%%%%%%%%%%%%%%%%%%%%%%%

\section{How consideration sets arise in our framework}
\label{optimal_appendix}
Let $\mathcal{P}_{Y|T}$ be an optimal strategy of the augmented decision problem $\{G, S\}$ as defined in Section \ref{model}. Following \hyperlink{Caplin_Dean_Leahy}{Caplin, Dean, and Leahy (2019)}, the DM's consideration set, $\mathcal{C}$, arises endogenously from $\mathcal{P}_{Y|T}$. In particular, $\mathcal{C}$ collects every alternative such that the subset of the signal's support inducing the DM to choose that alternative has positive measure:
$$\mathcal{C}\equiv \{y\in \mathcal{Y}: \int_{\mathcal{T}}P_{Y|T}(y|t) \int_{\mathcal{V}}P_{T|V}(t|v) P_{V}(v)dv dt >0\},$$
Observe that considerations sets can be heterogeneous across agents   as we leave the conditional signal densities  unrestricted.

Therefore, this paper also relates to the econometric literature on discrete choice models with consideration sets. For some recent contributions see, for example, \hyperlink{Cattaneo}{Cattaneo, Ma, Masatlioglu, and  Suleymanov (2020)};  \hyperlink{Manzini1}{Dardanoni, Manzini, Mariotti, and Tyson (2020)};  \hyperlink{Abaluck_Adams}{Abaluck and Adams (2021)}; \hyperlink{Molinari_2}{Barseghyan, Coughlin, Molinari, and  Teitelbaum (2021)};  \hyperlink{Molinari_1}{Barseghyan, Molinari, and Thirkettle (2021)};  \hyperlink{iaria}{Crawford, Griffith, and Iaria (2021)}; \hyperlink{Manzini2}{Dardanoni, Manzini, Mariotti, Petri, and Tyson (2022)}. Yet, there is an important difference between the literature on  consideration sets and this paper. The consideration set literature focuses on (partially) identifying the consideration
probabilities and payoffs under the assumption that the DM perfectly knows the payoff generated by each alternative in their  consideration set. Instead, our setting partially identifies the payoffs (and does not identify the consideration probabilities) while allowing the DM not to be fully aware of the payoff generated by each alternative in their  consideration set. Therefore, the two frameworks are non-nested and can answer different questions.

%%%%%%%%%%%%%%%%%%%%%%%%%%%%%%%%%%%%%%%%%%%%%%%%%
%%%%%%%%%%%%%%%%%%%%%%%%%%%%%%%%%%%%%%%%%%%%%%%%%
%%%%%%%%%%%%%%%%%%%%%%%%%%%%%%%%%%%%%%%%%%%%%%%%%
%%%%%%%%%%%%%%%%%%%%%%%%%%%%%%%%%%%%%%%%%%%%%%%%%

\section{Inference}
\label{inference}
We enrich the model of Section \ref{model} by assuming that there are discrete exogenous covariates $X$ with support $\mathcal{X}$ which enter the DM's information set together with the signal, as it is the case in the empirical application. We assume that the researcher has a sample of i.i.d. observations, $\{Y_i, X_i\}_{i=1}^n$. 
Given $\alpha\in (0,1)$, this section illustrates how to construct a uniformly  asymptotically valid $(1-\alpha)\%$ confidence region, $C_{1-\alpha}$, for any $\theta\in \Theta^{*}$. In particular, we suggest to reformulate our problem using conditional moment inequalities and apply the generalised moment selection procedure by \hyperlink{Andrews_Shi}{Andrews and Shi (2013)} (hereafter, AS13), as detailed in Appendix B.1 of \hyperlink{BMM}{Beresteanu, Molchanov, and Molinari (2011)} (hereafter, BMM11).

$C_{1-\alpha}$ is obtained  by running a test with null hypothesis $\text{H}_0: \theta_0=\theta$, for every $ \theta\in \Theta$, and then collecting all the values of $\theta$ which are not rejected. For a given $\theta$, the test rejects $\text{H}_0$ if $\text{TS}_n(\theta)>\hat{c}_{n,1-\alpha}(\theta)$, where $\text{TS}_n(\theta)$ is a test statistic and $\hat{c}_{n,1-\alpha}(\theta)$ is a corresponding critical value. Thus, 
\begin{equation}
\label{cr}
C_{1-\alpha}\equiv \{\theta \in \Theta\text{: }  \text{TS}_n(\theta) \leq \hat{c}_{n,1-\alpha}(\theta)\}.
\end{equation}
The remainder of the section explains how to compute $ \text{TS}_n(\theta)$ and $\hat{c}_{n,1-\alpha}(\theta)$ for any given $\theta\in \Theta$.

In order to define the test statistic, $\text{TS}_n(\theta)$, let us first rewrite the linear programming (\ref{lin_pr}) as a collection of conditional moment inequalities. To do so, we label  the elements of $\mathcal{Y}$ as $y^1,...,y^{|\mathcal{Y}|-1}, y^{|\mathcal{Y}|}$. We denote by $\mathbb{B}^{|\mathcal{Y}|-1}$ i  the unit ball in $\mathbb{R}^{|\mathcal{Y}|-1}$. For each $\theta\in \Theta$ and $x\in \mathcal{X}$, $ \mathcal{Q}(\theta,x)$ is the set of model-implied choice distributions, $P_{Y|X}(\cdot|x)$,  conditional on $X_i=x$. $\mathbb{P}_{Y|X}(\cdot|x)$ is the empirical choice distribution conditional on $X_i=x$.
\begin{prop}{\normalfont({\itshape Conditional moment inequalities)}}
\label{moment_ineq}
For each $\theta\in \Theta$, $\theta\in \Theta^*$ if and only if 
$$
\mathbb{E}[m(Y_i,X_i; b,\theta)|X_i=x]\leq 0  \text{ }\forall b \in \mathbb{B}^{|\mathcal{Y}|-1},\forall x \in \mathcal{X},
$$
where 
$$
m(Y_i,x;b,\theta)\equiv b^\top  \begin{pmatrix}
\mathbbm{1}\{Y_i=y^1\}\\
...\\
\mathbbm{1}\{Y_i=y^{|\mathcal{Y}|-1}\}\\
\end{pmatrix} - \max_{P_{Y|X}(\cdot|x)\in \mathcal{Q}(\theta,x) }b^\top  \begin{pmatrix}
P_{Y|X}(y^1|x)\\
...\\
P_{Y|X}(y^{|\mathcal{Y}|-1}|x)\\
\end{pmatrix}. 
$$
\end{prop}
Proposition \ref{moment_ineq} comes from the fact that, following \hyperlink{BMM}{BMM11}, one can express the condition $\mathbb{P}_{Y|X}(\cdot|x)\in \mathcal{Q}(\theta,x) $ as
\begin{equation}
\label{construct_ident_2}
 b^\top  \mathbb{P}_{Y|X}(\cdot|x) - \sup_{P_{Y|X}(\cdot|x)\in \mathcal{Q}(\theta,x)} b^\top  P_{Y|X}(\cdot|x)\leq 0\text{ }\forall b \in \mathbb{R}^{|\mathcal{Y}|}, 
\end{equation}
where the map $$b\in\mathbb{R}^{|\mathcal{Y}|}\mapsto  \sup_{P_{Y|X}(\cdot|x)\in \mathcal{Q}(\theta,x)} b^\top  P_{Y|X}(\cdot|x)\in \mathbb{R},$$ is the support function of $\mathcal{Q}(\theta,x)$. By exploiting the positive homogeneity of the support function and some algebraic manipulations, (\ref{construct_ident_2}) is equal to the  collection of conditional moment inequalities listed in Proposition \ref{moment_ineq}. 

Second, we rewrite the conditional moment inequalities in Proposition \ref{moment_ineq} as unconditional moment inequalities. Here, we use Lemma 2 in \hyperlink{Andrews_Shi}{AS13} which shows that conditional moment inequalities can be transformed into  unconditional moment inequalities by choosing  appropriate instruments, $h\in\mathcal{H}$, where $\mathcal{H}$ is a collection of instruments and $h$ is a function of $X_i$. Thus,
\begin{equation}
\label{inference2}
\theta\in \Theta^* \Leftrightarrow  \mathbb{E}[m(Y_i,X_i; b,\theta,h)]\leq 0 \text{ }\forall b\in \mathbb{B}^{|\mathcal{Y}|-1}, \forall h\in\mathcal{H} \text{ a.s.,}
\end{equation}
where 
$$
m(Y_i,X_i;b,\theta,h)\equiv m(Y_i,X_i; b,\theta) \times h(X_i). 
$$
Third, observe that $ \mathbb{E}[m(Y_i,X_i; b,\theta,h)]$ evaluated at $b\equiv 0_{|\mathcal{Y}|-1}$ is $0$. Therefore, (\ref{inference2}) is equivalent to
$$
\theta\in \Theta^* \Leftrightarrow \max_{b\in \mathbb{B}^{|\mathcal{Y}|-1}}\mathbb{E}[m(Y_i,X_i; b,\theta,h) ]=0  \text{ } \forall h\in\mathcal{H} \text{ a.s.}
$$
In light of the three steps above, following Appendix B.1 of \hyperlink{BMM}{BMM11}, we can use as test statistic
$$
\text{TS}_n(\theta)\equiv \int_{ \mathcal{H}} \Big[\sqrt{n} \max_{b\in \mathbb{B}^{|\mathcal{Y}|-1}} \bar{m}_n(b,\theta,h)\Big]^2 d\Gamma(h), 
$$
where $\Gamma$ is a probability measure on $\mathcal{H}$ as explained in Section 3.4 of \hyperlink{Andrews_Shi}{AS13}, and 
$$
\bar{m}_n(b,\theta,h)\equiv \frac{1}{n}\sum_{i=1}^n m(Y_i,X_i;b,\theta,h). 
$$
Intuitively, $\text{TS}_n(\theta)$ is built by imposing a penalty for each $h$ such that the maximum of $\mathbb{E}[m(Y_i,X_i; b,\theta,h) ]$ across $b\in \mathbb{B}^{|\mathcal{Y}|-1}$ is different from zero. Moreover, given that the support of $X_i$ is finite, the analyst can replace $\Gamma$ with the uniform probability measure on $\mathcal{X}$ as suggested by Example 5 in Appendix B of \hyperlink{Andrews_Shi}{AS13}. That is, 
\begin{equation}
\label{inf_discrete3}
\text{TS}_n(\theta)\equiv \frac{1}{|\mathcal{X}|}\sum_{x\in \mathcal{X}} \Big[\sqrt{n} \max_{b\in \mathbb{B}^{|\mathcal{Y}|-1}}   \bar{m}_n(b,\theta,x)\Big]^2,
\end{equation}
where $$
\bar{m}_n(b,\theta,x)\equiv \frac{1}{n}\sum_{i=1}^n m(Y_i, X_i; b,\theta)  \mathbbm{1}\{X_i=x\}. 
$$

Lastly, we compute the critical value, $\hat{c}_{n,1-\alpha}(\theta)$, by following \hyperlink{Andrews_Shi}{AS13}'s bootstrap method consisting of the following steps. First, we draw $W_n$ bootstrap samples using nonparametric i.i.d. bootstrap. Second, for each $w=1,...,W_n$, we compute the recentered test statistic
\begin{equation}
\label{boot}
\text{TS}^w_{n}(\theta)\equiv  \frac{1}{|\mathcal{X}|}\sum_{x\in \mathcal{X}} \Big[\sqrt{n} \max_{b\in \mathbb{B}^{|\mathcal{Y}|-1}} ( \bar{m}^w_{n}(b,\theta,x)-  \bar{m}_{n}(b,\theta,x))\Big]^2, 
\end{equation}
where  $\bar{m}^w_{n}(b,\theta,x)$ is calculated just as $\bar{m}_{n}(b,\theta,x)$, but with the bootstrap sample in place of the original sample. 
Third, $\hat{c}_{n,1-\alpha}(\theta)$ is set equal to the $(1-\alpha)$ quantile of $\{\text{TS}^w_{n}(\theta) \}_{w=1}^{W_n}$. 
Once $\text{TS}_n(\theta)$ and $\hat{c}_{n,1-\alpha}(\theta)$ are computed for each $\theta\in \Theta$ (or, in practice, for each $\theta$ belonging to a grid), the confidence region, $C_{1-\alpha}$, defined in (\ref{cr}) can be constructed. 

In Appendix \ref{computational} we provide more details on the computation of (\ref{inf_discrete3}) and (\ref{boot}). In particular, we show that computing (\ref{inf_discrete3}) and (\ref{boot}) amounts to solving some quadratically constrained linear programming problems. 

%We conclude by highlighting that, in addition to reporting a confidence region, it is often useful to report an estimated set, so as to reveal how much of the volume of the confidence region is due to randomness and how much is due to a large identified set. In this respect, \hyperlink{Andrews_Shi}{AS13} show that $C_{0.50}$ is an asymptotically half-median-unbiased estimated set. 

\subsection{Some computational simplifications}
\label{computational}
Following \hyperlink{Magnolfi_Roncoroni}{MR23}, we discuss  a way to simplify the computation of the test statistic, $\text{TS}_{n}(\theta)$,  as defined in (\ref{inf_discrete3}). Observe that, for each $x\in \mathcal{X}$ and $b\in \mathbb{B}^{|\mathcal{Y}|-1}$, 
\begin{equation}
\label{mean}
\bar{m}_n(b,\theta,x)= \mathbb{P}_X(x) b^\top  \Big(\tilde{\mathbb{P}}_{Y|X}(\cdot|x) -\max_{P_{Y|X}(\cdot|x)\in \mathcal{Q}(\theta,x)}\tilde{P}_{Y|X}(\cdot|x)\Big),
\end{equation}
where $\mathbb{P}_X$ is the empirical distribution of $X_i$, $\tilde{\mathbb{P}}_{Y|X}(\cdot|x)\equiv \begin{pmatrix}
\mathbb{P}_{Y|X}(y^1|x)\\
\vdots\\
\mathbb{P}_{Y|X}(y^{|\mathcal{Y}|-1}|x)\\
\end{pmatrix}$, and $\tilde{P}_{Y|X}(\cdot|x)\equiv \begin{pmatrix}
P_{Y|X}(y^1|x)\\
\vdots\\
P_{Y|X}(y^{|\mathcal{Y}|-1}|x)\\
\end{pmatrix}$. 

Therefore, (\ref{inf_discrete3}) is equal to
\begin{equation}
\label{inf_discrete}
\begin{aligned}
\text{TS}_n(\theta)& \equiv \frac{1}{|\mathcal{X}|}\sum_{x\in \mathcal{X}}\Big[\sqrt{n}  \max_{b\in \mathbb{B}^{|\mathcal{Y}|-1}}  b^\top   \Big(\mathbb{P}_X(x)\tilde{\mathbb{P}}_{Y|X}(\cdot|x) -\mathbb{P}_X(x) \max_{P_{Y|X}(\cdot|x)\in \mathcal{Q}(\theta,x)}\tilde{P}_{Y|X}(\cdot|x)\Big) \Big]^2.
\end{aligned}
\end{equation}
To compute (\ref{inf_discrete}), the researcher should calculate, for each $x\in \mathcal{X}$, 
$$
\max_{b\in \mathbb{B}^{|\mathcal{Y}|-1}}  b^\top   \Big(\mathbb{P}_X(x) \tilde{\mathbb{P}}_{Y|X}(\cdot|x) -\mathbb{P}_X(x)\max_{P_{Y|X}(\cdot|x)\in \mathcal{Q}(\theta,x)}\tilde{P}_{Y|X}(\cdot|x)\Big),
$$
which is equivalent to 
\begin{equation}
\label{inf_discrete2}
\max_{b\in \mathbb{B}^{|\mathcal{Y}|-1}}  \min_{P_{Y|X}(\cdot|x)\in \mathcal{Q}(\theta,x)} b^\top   \Big(\mathbb{P}_X(x)\tilde{\mathbb{P}}_{Y|X}(\cdot|x) -\mathbb{P}_X(x)\tilde{P}_{Y|X}(\cdot|x)\Big). 
\end{equation}
(\ref{inf_discrete2}) is a max-min problem which can be simplified as follows. Note that the inner constrained minimisation problem in (\ref{inf_discrete2}) is linear in $P_{Y|X}(\cdot|x)$. Thus, it can be replaced by its dual, which consists of a linear constrained maximisation problem. Moreover, the outer constrained maximisation problem  in (\ref{inf_discrete2}) has a quadratic constraint, $b^\top b\leq  1$. Therefore,  (\ref{inf_discrete2}) can be rewritten as a quadratically constrained linear maximisation problem which is a tractable exercise.  This is described in detail below.

By (\ref{lin_pr_3}), (\ref{inf_discrete2})  is equivalent to 
\begin{equation}
\label{main_ident10}
{\footnotesize
\hspace{-1cm}
\begin{alignedat}{3}
 & \max_{b\in \mathbb{R}^{|\mathcal{Y}|-1}} &&  \min_{\substack{\text{$P_{Y|X}(\cdot|x), \lambda$}}} &&  b^\top  [\mathbb{P}_X(x) \tilde{\mathbb{P}}_{Y|X}(\cdot|x) -\mathbb{P}_X(x) \tilde{P}_{Y|X}(\cdot|x)],\\
  & \text{s.t.} && \text{$b\in \mathbb{B}^{|\mathcal{Y}|-1}$: }  && b^\top b\leq 1, \\
&&&\text{Obedience:}  &&\sum_{k\in \mathcal{K}} \lambda_{k,K}^{y,x} \gamma^{y,y',x}_{1,k,K}(\theta) \geq 0 \text{ } \forall y\in \mathcal{Y}, \forall y'\in \mathcal{Y}\setminus\{y\},\\
&&&\text{Probability requirements:}  && \lambda_{k,K}^{y,x}\geq 0 \text{ }\forall k\in \mathcal{K}, \forall y \in \mathcal{Y},  \text{ }\sum_{y\in \mathcal{Y}} \lambda_{k,K}^{y,x}= 1\text{ } \forall k\in \mathcal{K},\\
&&&\text{Choice probability:} &&P_{Y|X}(y|x)= \sum_{k\in \mathcal{K}} \lambda_{k,K}^{y,x} \gamma_{2,k,K}(\theta_V) \text{ }\forall y \in \mathcal{Y}.
\end{alignedat}}
\end{equation}

We simplify (\ref{main_ident10}) by introducing new variables.  Let  $W_1\equiv \mathbb{P}_X(x)(\mathbb{P}_{Y|X}(\cdot|x)-P_{Y|X}(\cdot|x))$.  
Let $W$ be the $(|\mathcal{Y}|+|\mathcal{Y}|\cdot K)\times 1$ vector
collecting $W_1$ and $\lambda$.
 (\ref{main_ident10}) can be rewritten as
\begin{equation}
\label{main_ident11}
\begin{alignedat}{3}
 \max_{b\in \mathbb{R}^{|\mathcal{Y}|-1}} &&&  \min_{\substack{\text{$W$}}} 
\Big[b^\top  \text{ } \text{ } \text{ }0_{1+|\mathcal{Y}|\cdot K}^\top \Big] W,\\
 \text{s.t. }   &&&    b^\top b\leq  1, \\
  &&&   A_{\text{eq}}\text{ }W = B_{\text{eq}},\\
 &&&   A_{\text{ineq}} \text{ }W \leq 0_{d_{\text{ineq}}},  
\end{alignedat}
\end{equation}
where $A_{\text{eq}}$ is the matrix of  coefficients multiplying $W$ in the equality constraints of (\ref{main_ident10}) with $d_{\text{eq}}$ rows, $B_{\text{eq}}$ is the vector of constants appearing in the equality constraints  of (\ref{main_ident10}), and $A_{\text{ineq}}$ is the matrix of  coefficients multiplying $W$ in the inequality constraints  of (\ref{main_ident10}) with $d_{\text{ineq}}$ rows.

Further, the inner  constrained minimisation problem in (\ref{main_ident11}) is linear. Hence, by strong duality, it can be replaced with its dual. This allows us to solve one unique maximisation problem. Precisely, the solution of (\ref{main_ident11}) is equivalent to the solution of 
\begin{equation}
\label{main_ident12}
\begin{alignedat}{3}
 \max_{\substack{\text{$b\in \mathbb{R}^{|\mathcal{Y}|-1}$} \\ \text{$\tau_{\text{eq}}\in \mathbb{R}^{d_{\text{eq}}}$} \\ \text{$\tau_{\text{ineq}}\in \mathbb{R}^{d_\text{ineq}}_{+}$}}} &&& \Big[-B_{\text{eq}}^\top  \text{ } \text{ } \text{ }0_{d_{\text{ineq}}}^\top \Big] \tau,\\
 \text{s.t. }   &&&    b^\top b\leq  1, \\
  &&&   [A^\top ]_{1:|\mathcal{Y}|} \cdot \tau = \begin{pmatrix}-b\\ 0\end{pmatrix},\\
 &&&   -[A^\top ]_{|\mathcal{Y}|+1:|\mathcal{Y}|+|\mathcal{Y}|\cdot K} \cdot \tau \leq 0_{|\mathcal{Y}|\cdot K},  
\end{alignedat}
\end{equation}
where $\tau$ is the $(d_{\text{eq}}+d_{\text{ineq}})\times 1$ vector collecting $\tau_{\text{eq}}$ and $\tau_{\text{ineq}}$,  $A$ is the  $(d_{\text{eq}}+d_{\text{ineq}})\times( |\mathcal{Y}|+|\mathcal{Y}|\cdot K )$ matrix obtained by stacking one on top of the other the matrices $A_{\text{eq}}$ and $A_{\text{ineq}}$, and $[A]_{i:j}$ denotes the sub-matrix of $A$ containing the rows $i,i+1,...,j$ of $A$. 

Note that (\ref{main_ident12}) is a quadratically constrained linear maximisation problem. In particular, the first constraint in (\ref{main_ident12}) is quadratic. The objective function and the remaining constraints in (\ref{main_ident12}) are linear. 
Close derivations are discussed in \hyperlink{Magnolfi_Roncoroni}{MR23} for an entry game setting.

We now discuss  a way to simplify the computation of bootstrap test statistic, $\text{TS}^w_{n}(\theta)$, as defined in (\ref{boot}). 
Similarly to (\ref{mean}), by rearranging terms it holds that 
$$
\begin{aligned}
& \bar{m}^w_{n}(b,\theta,x)= \mathbb{P}^w_X(x) b^\top  \Big(\tilde{\mathbb{P}}^{w}_{Y|X}(\cdot|x) -\max_{P_{Y|X}(\cdot|x)\in \mathcal{Q}(\theta,x)}\tilde{P}_{Y|X}(\cdot|x)\Big),
\end{aligned}
$$
where the superscript ``$w$'' denotes the bootstrap probabilities. 
Therefore, 
$$
\begin{aligned}
& \bar{m}^w_{n}(b,\theta,x)- \bar{m}_n(b,\theta,x)\\
&= \mathbb{P}^w_X(x) b^\top  \Big(\tilde{\mathbb{P}}^{w}_{Y|X}(\cdot|x) -\max_{P_{Y|X}(\cdot|x)\in \mathcal{Q}(\theta,x)}\tilde{P}_{Y|X}(\cdot|x)\Big) - P^{0}_X(x) b^\top  \Big(\tilde{P}^{0}_{Y|X}(\cdot|x) -\max_{P_{Y|X}(\cdot|x)\in \mathcal{Q}(\theta,x)}\tilde{P}_{Y|X}(\cdot|x)\Big), \\
&= b^\top  \Big[  \mathbb{P}^w_X(x) \tilde{\mathbb{P}}^{w}_{Y|X}(\cdot|x) -P^{0}_X(x) \tilde{P}^{0}_{Y|X}(\cdot|x)- (\mathbb{P}^w_X(x)-P^{0}_X(x) )\max_{P_{Y|X}(\cdot|x)\in \mathcal{Q}(\theta,x)}\tilde{P}_{Y|X}(\cdot|x)\Big]. 
\end{aligned}
$$
To simplify the notation, let us rename 
$$
A^w_x\equiv \mathbb{P}^w_X(x) \tilde{\mathbb{P}}^{w}_{Y|X}(\cdot|x) -P^{0}_X(x) \tilde{P}^{0}_{Y|X}(\cdot|x),
$$
and
$$
C^w_x\equiv \mathbb{P}^w_X(x)-P^{0}_X(x). 
$$
Therefore, (\ref{boot}) is equal to
\begin{equation}
\label{boot2}
\text{TS}^w_{n}(\theta)\equiv  \frac{1}{|\mathcal{X}|}\sum_{x\in \mathcal{X}} \Big[\sqrt{n} \max_{b\in \mathbb{B}^{|\mathcal{Y}|-1}} b^\top  \Big(  A^w_x- C^w_x \max_{P_{Y|X}(\cdot|x)\in \mathcal{Q}(\theta,x)}\tilde{P}_{Y|X}(\cdot|x)\Big) \Big]^2. 
\end{equation}
To compute (\ref{boot2}), the researcher should calculate, for each $x\in \mathcal{X}$, 
$$
\max_{b\in \mathbb{B}^{|\mathcal{Y}|-1}} b^\top  \Big ( A^w_x- C^w_x \max_{P_{Y|X}(\cdot|x)\in \mathcal{Q}(\theta,x)}\tilde{P}_{Y|X}(\cdot|x)\Big),
$$
which is equivalent to 
\begin{equation}
\label{boot3}
\max_{b\in \mathbb{B}^{|\mathcal{Y}|-1}}  \min_{P_{Y|X}(\cdot|x)\in \mathcal{Q}(\theta,x)} b^\top   \Big(A^w_x-C^w_x\tilde{P}_{Y|X}(\cdot|x)\Big). 
\end{equation}
(\ref{boot3}) can be rewritten as a quadratically constrained linear maximisation problem as done for (\ref{inf_discrete2}). Once (\ref{boot3}) is computed for each $x\in \mathcal{X}$, the analyst easily obtains $\text{TS}^w_{n}(\theta)$.

\end{appendix}

\end{document}